\documentclass[12pt,preprint]{aastex}

\newcommand{\Msun}{{\rm M}_\odot}
\newcommand{\Lsun}{{\rm L}_\odot}
\newcommand{\spit}{{\it Spitzer}}
\usepackage{subfigure}

\shorttitle{NGC~602}
\shortauthors{Carlson et al}

\begin{document}

%% You may use \\ to force a line break if you desire.
\title{A PANCHROMATIC VIEW OF NGC~602: \\Time-Resolved Star Formation with the Hubble and Spitzer Space Telescopes}

\author{Lynn Redding Carlson\altaffilmark{1}}
\affil{Leiden Observatory, Leiden University, Leiden, Netherlands}
\email{carlson@strw.leidenuniv.nl}
\author{Marta Sewilo}
\affil{Johns Hopkins University, Baltimore, 3400 N. Charles St., Baltimore, MD, USA} 
\email{mmsewilo@stsci.edu}
\author{Margaret Meixner}
\affil{Space Telescope Science Institute, 3700 San Martin Drive, Baltimore, MD, USA}  
\email{meixner@stsci.edu}
\author{Krista Alexandra Romita}
\affil{University of Florida, Gainesville, FL, USA} 
\email{k.a.romita@gmail.com}
\author{Barbara Whitney}
\affil{Space Science Institute, Boulder, CO, USA} 
\email{bwhitney@spacescience.org}
\author{J. L. Hora}
\affil{Harvard/CfA, Cambridge, MA, USA}
\email{jhora@cfa.harvard.edu}
\author{M. Cignoni\altaffilmark{2}}
\affil{Universit\'a degli Studi di Bologna, Bologna, IT}
\email{michele.cignoni@unibo.it}
\author{Elena Sabbi}
\affil{Space Telescope Science Institute, 3700 San Martin Drive, Baltimore, MD, USA} 
\email{sabbi@stsci.edu}
\author{Antonella Nota\altaffilmark{3}}
\affil{Space Telescope Science Institute, 3700 San Martin Drive, Baltimore, MD, USA} 
\email{nota@stsci.edu}
\author{Marco Sirianni\altaffilmark{3}} 
\affil{ESA/ESTEC Noordwijk, Netherlands}
\email{marco.sirianni@esa.int}
\author{Linda J. Smith\altaffilmark{4}}
\affil{University College London, London, England}
\email{lsmith@stsci.edu}
\author{Karl Gordon}
\affil{Space Telescope Science Institute, 3700 San Martin Drive, Baltimore, MD, USA}
\email{kgordon@stsci.edu}
\author{B. Babler}
\affil{University of Wisconsin, Madison, WI, USA}
\email{brian@sal.wisc.edu}
\author{S. Bracker}
\affil{University of Wisconsin, Madison, WI, USA}
\email{s\_bracker@hotmail.com}
\author{J. S. Gallagher, III}
\affil{University of Wisconsin, Madison, WI, USA}
\email{jsg@astro.wisc.edu}
\author{M. Meade}
\affil{University of Wisconsin, Madison, WI, USA}
\email{meade@astro.wisc.edu}
\author{K. Misselt}
\affil{Steward Observatory, University of Arizona, Tucson, AZ, USA} 
\email{kmisselt@as.arizona.edu}
\author{A. Pasquali}
\affil{Max-Planck-Institut f$\ddot{u}$r Astronomie, Heidelberg, Germany}
\email{pasquali@mpia-hd.mpg.de}
\author{B. Shiao}
\affil{Space Telescope Science Institute, 3700 San Martin Drive, Baltimore, MD, USA}
\email{shiao@stsci.edu}
\altaffiltext{4}{Space Telescope Science Institute, 3700 San Martin Drive, Baltimore, MD, USA} 
\altaffiltext{3}{Space Telescope Operation Division, ESA, Baltimore, MD, USA} 
\altaffiltext{2}{INAF-Osservatorio di Bologna, IT}
\altaffiltext{1}{Johns Hopkins University, Baltimore, 3400 N. Charles St., Baltimore, MD, USA}

% ABSTRACT
\begin{abstract}

We present the photometric catalogs for the star-forming cluster NGC~602 in the wing of the Small Magellanic Cloud  covering a range of wavelengths from optical  ({\it HST/ACS} F555W, F814W and SMARTS/ANDICAM {\it V}, {\it I})  to infrared  (\spit\ /IRAC 3.6, 4.5, 5.8, and 8~$\mu$m and MIPS 24~$\mu$m).  Combining this with IRSF (InfraRed Survey Facility) near-infrared photometry ({\it J}, {\it H}, $K_s$), we  compare the young main sequence (MS) and pre-main sequence (PMS)  populations prominent in the optical with  the current young stellar object (YSO) populations  revealed by the infrared (IR).   We analyze the MS and PMS population with isochrones in color-magnitude diagrams to derive ages and masses.  The optical data reveal $\sim 565$ PMS candidates, low mass Stage~{\sc iii} YSOs.  We  characterize $\sim 40$ YSOs by fitting their spectral energy distributions (SEDs) to a grid of models \citep{robitaille07} to derive luminosities, masses and evolutionary phase (Stage~{\sc i}-{\sc iii}). The higher resolution {\it HST} images reveal that $\sim70\%$ of the YSO candidates are either multiples or protoclusters.  For YSOs and PMS sources found in common, we find a consistency in the masses derived.  We use the YSO mass function to derive a present-day star-formation rate of $\sim 0.2-1.0  ~\Msun$ yr$^{-1}$ kpc$^{-2}$, similar to the rate derived from the optical  star formation history suggesting a constant star formation rate for this region.  We demonstrate a progression of star formation from the optical star cluster center to the edge of the star forming dust cloud.  We derive lifetimes of a few $10^5$ years for the YSO Stages~{\sc i} and {\sc ii}.  

Please note that color images are compressed for space.  Please contact the first author for superior, high resolution versions.

\end{abstract}

%% subject header.  Objects should be in the appropriate "individual"
%% headers (e.g. quasars: individual, stars: individual, etc.) with the
%% additional provision that the total number of headers, including each
%% individual object, not exceed six.  The \objectname{} macro, and its
%% alias \object{}, is used to mark each object.  The macro takes the object
%% name as its primary argument.  This name will appear in the paper
%% and serve as the link's anchor in the electronic edition if the name
%% is recognized by the data centers.  The macro also takes an optional
%% argument in parentheses in cases where the data center identification
%% differs from what is to be printed in the paper.


\section{INTRODUCTION}
\label{intro}

The Small Magellanic Cloud (SMC) is a valuable astrophysical laboratory for understanding the processes of star formation in a galaxy that is extremely different from the Milky Way.  
In particular, it has a subsolar chemical abundance of Z$\sim 0.004$ \citep{rolleston99,lee05} and a low dust-to-gas ratio of $\sim$~1/30 Milky Way in the diffuse interstellar medium \citep{stanimirovic00} and $\sim1/6$ in star-forming regions \citep{bot07}.  The SMC's lack of organized rotation means that star formation is predominantly driven by a combination of tidally-induced cloud-cloud interactions \citep{zaritsky00,zaritsky04} and shell formation \citep{hatzidimitriou05}.   With its close proximity \citep[60.6~kpc;][]{ hilditch05}, the SMC is uniquely suited to detailed investigation of the stellar content, down to the sub-solar mass regime, in the youngest and most compact star clusters.
  
The small, young star cluster NGC~602, associated with the highly structured H~{\sc ii} region N90 \citep{henize56}, is one of the most interesting star forming regions in the SMC.  Located at the boundary between the SMC wing and the Magellanic Bridge, at the intersection of three H{\sc i} shells, it is probably the result of the interaction of two expanding shells of H{\sc i} that occurred approximately $\sim 7\, {\rm Myr}$ ago \citep{nigra08}.  \citet{nigra08} also show that the N90 region is quiescent, with negligible H$\alpha$ shell expansion velocities.  Propagation of star formation is most likely driven by radiation with stars forming along the edges of the photodissociation region.   Star formation started approximately 4~Myr ago, with the formation of the central cluster, and gradually propagated towards the outskirts, where star formation continues  \citep{carlson07}.  \citet{gouliermis07} provide a list of 22 candidate Young Stellar Objects (YSOs) in the outskirts of N90.  The optical cluster's Mass Function (MF) is consistent with a standard Salpeter~\citep{salpeter55} Initial Mass Function \citep{schmalzl08}.  \citet{cignoni09} reconstruct a complete star formation history for the optical population and find that the pre-main sequence (PMS) is not more than $\sim$5~Myr, and the star formation rate (SFR) has reached a maximum in the last 2.5~Myr.  

\begin{figure}
\epsscale{1.}
\plotone{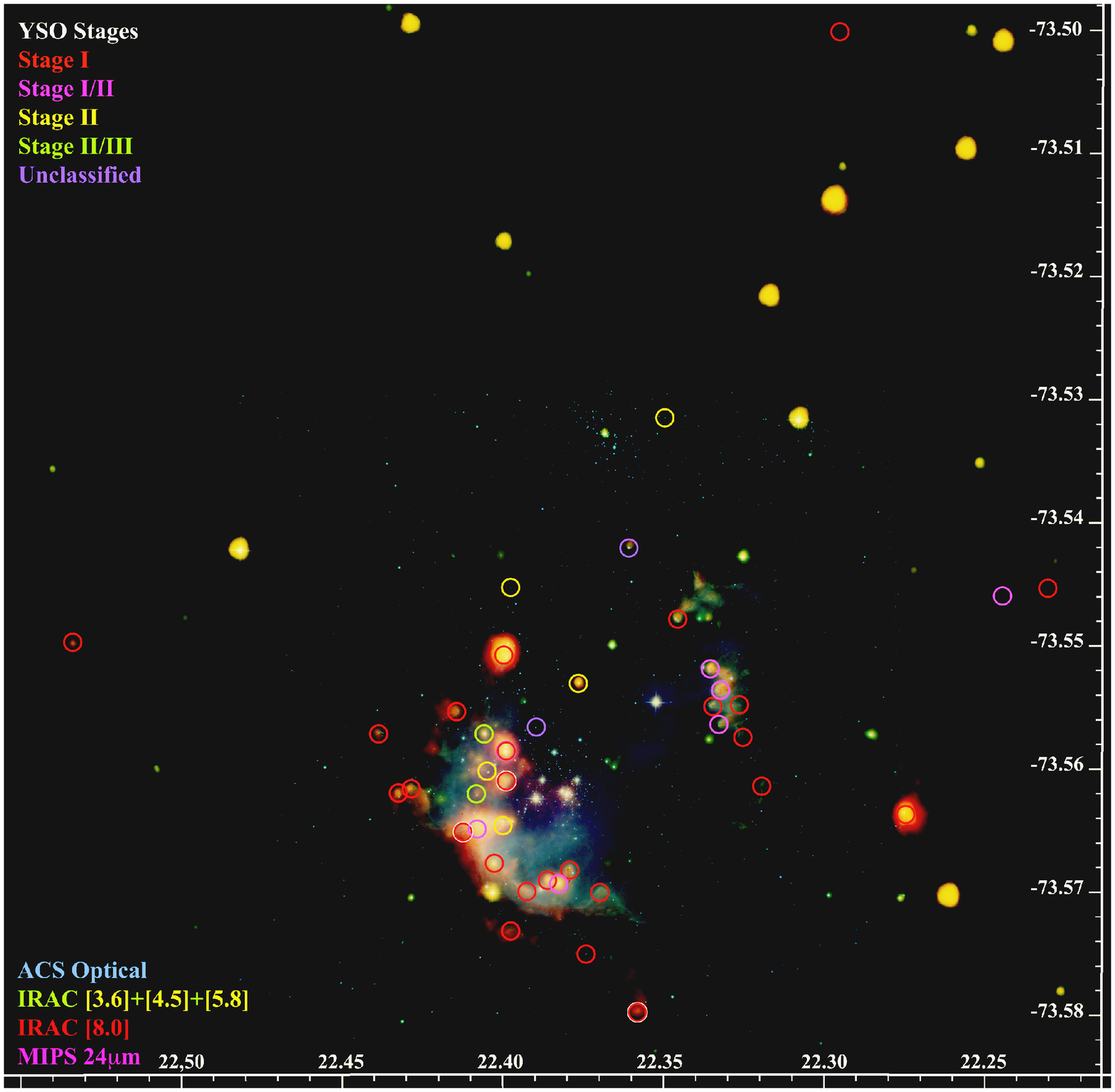}
\caption{\label{8col} Image of NGC~602 in optical through 24~$\mu$m.  Circle of radii $1\farcs 6$ indicate the positions of YSO candidates, color-coded by evolutionary stage as in the legend.  The addition of a white circle indicates that we do not consider the fit good.  North is up, east to the left.}
\end{figure}

We employ a new panchromatic approach to extend the star formation history analysis to the present day.  Population identification via Color-Magnitude Diagram (CMD) analysis informs us of cluster-scale star formation, with optical revealing a bright main sequence (MS) and a young PMS and infrared (IR) CMD highlighting the youngest embedded sources.  Using spectral energy distribution (SED) analysis, for which we combine Hubble Space Telescope ({\it HST}) optical to probe central stellar sources and Spitzer Space Telescope (\spit) IR to probe circumstellar disks and envelopes, we characterize star formation in the NGC~602 region (Figure~\ref{8col}) on the scale of single sources and protoclusters. 

The structure of this paper is as follows.  We describe observations and data reduction in Section \ref{obs}.  In Section~\ref{cmdanal}, we discuss what can be gleaned from the IR and optical data separately via CMD analysis.  We describe the application of the SED fitter in Section~\ref{FIT} and present the results of our SED analysis of 77 sources, combining optical to mid-IR data, for YSOs (\ref{PAHs}, \ref{yseds0}, \ref{yseds1}, and \ref{yseds+}) and other types of IR sources (\ref{nseds}).  We address the YSO Mass Function and Star Formation Rate (SFR) in Section \ref{mf} and the spatial and temporal distribution of young sources in Section~\ref{SpaceTime}.

\section{OBSERVATIONS AND DATA REDUCTION}
\label{obs}

\begin{figure}
\epsscale{1.}
\plotone{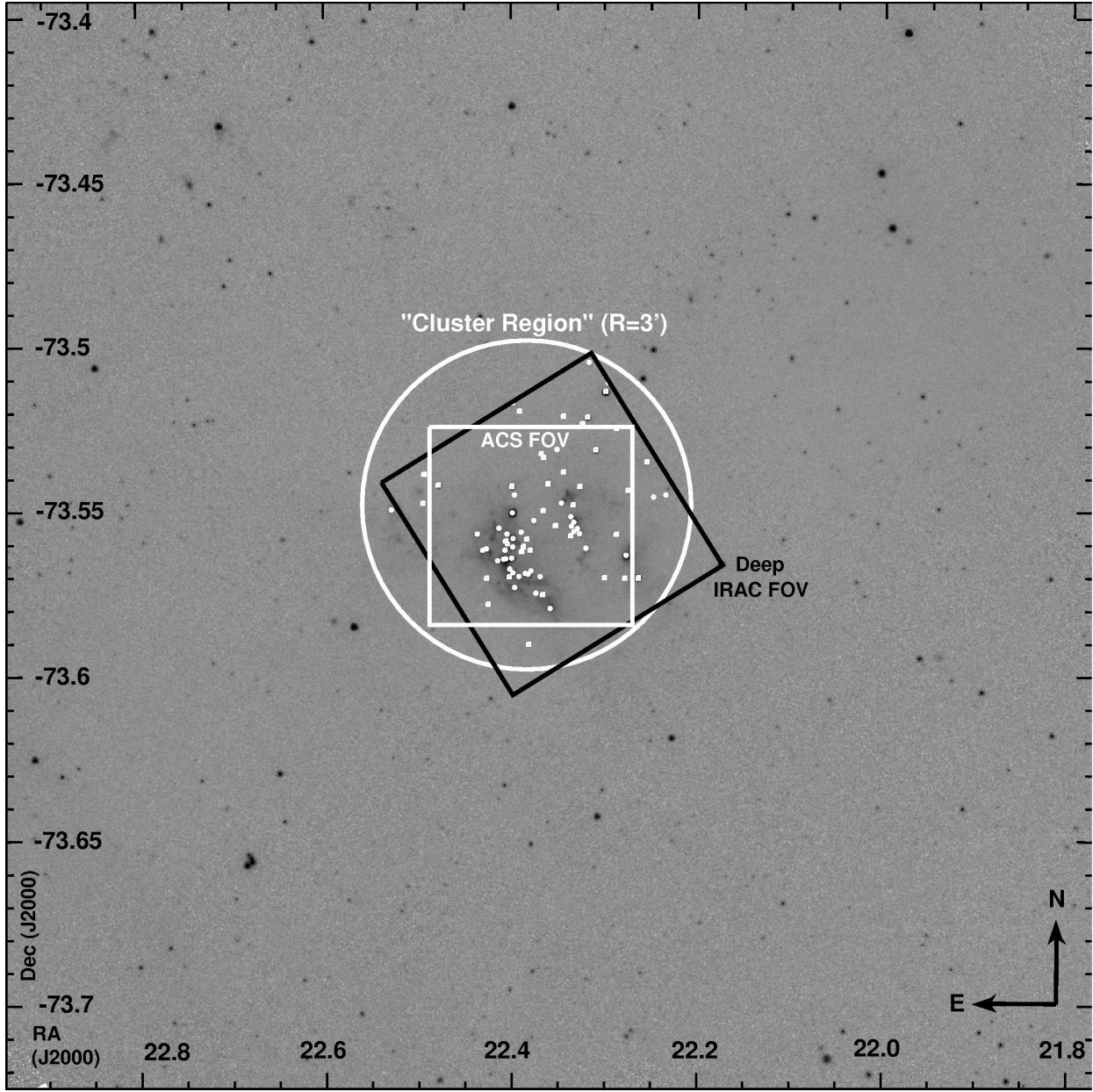}
\caption{\label{FOV} Large 8.0~$\mu$m image of the NGC~602 region.  Black Box: Deep \spit/IRAC FOV;  White Box: {\it HST/ACS/WFC} FOV;  Large White Circle: $3\arcmin$ radius region studied here.  Small white markings: 77 ``fitter sources."  The full area shown is 18$\arcmin \times 18\arcmin$.  Sources outside the Cluster Region are marked in grey in Figure~\ref{iracc} to represent the background population.} 
\end{figure}

Observations of NGC~602 (R.A. = $01^{\rm h}29^{\rm m}31^{\rm s}$, Dec. = $-73^{\circ}33\arcmin15\arcsec$, J2000) were taken from the optical to the mid-infrared (Figure~\ref{FOV}).  Optical photometry (Section~\ref{optdat}) reveals the naked MS and PMS populations, and we determine their ages through CMD analysis.  Accompanying {\it HST} Advanced Camera for Surveys ({\it ACS}) imaging offers a high-resolution view of both the nebular structure and the physical distribution of the stars.  \spit\ Infrared Array Camera (IRAC) data (Section~\ref{IRACdat}) reveal the YSOs. Relying completely on \spit\ photometry and imaging, however, would introduce significant contamination from background galaxies, contamination which can be avoided through comparison to ACS images \citep[See Section~\ref{gals} and][]{nigra08}.  Likewise, optical data alone miss the youngest sources.  We characterize YSOs and other infrared sources by combining optical, near infrared (NIR), and IR photometry (Section~\ref{SEDdata}) and fitting the resultant SEDs to those of numerically modeled YSOs, catalog naked stars, and galaxies.  These sources are referenced below with prefixes Y~(YSO candidate), G~(galaxy), K~(naked star), S~(stellar in images but not fit), and U~(unclassified).

\subsection{Optical}
\label{optdat}

We use optical data primarily from the {\it HST/ACS} Wide Field Channel ({\it WFC}) in F555W~({\it V}), F658N~(H$\alpha$), and F814W~({\it I}).  Some bright Red Giant and upper-MS stars are saturated in our {\it ACS} images.  We correct for this by adding ground-based observations from the SMARTS (Small and Moderate Aperture Research Telescope System) ANDICAM instrument on a 1.3~m telescope in Johnson-Cousins {\it V} and {\it I} bands.  The log-book of optical observations is presented online in Table~\ref{t:obs}.  The final optical catalog \citep[Table~\ref{t:optphot}, also used in ][]{cignoni09} includes 4535 stars, from a combination of these {\it HST} and ground-based data. 

\subsubsection{HST Data}
\label{HSTdat}

Five long exposures were taken through filters F555W~($\sim${\it V}) and F814W~($\sim${\it I}) using {\it HST/ACS/WFC} (Table~\ref{t:obs}; {\it HST} Proposal 10248, P.I. Nota).  The dither pattern allowed hot pixel removal and filled the gap between the two $2048\times4096$ pixel detectors. This dithering technique also improved both Point Spread Function (PSF) sampling and photometric accuracy by averaging flat-field errors, and by smoothing over the spatial variations in detector response.   The entire data set was processed with the CALACS calibration pipeline, and the long and short exposures for each filter were separately co-added with the MULTIDRIZZLE package.  The final exposure times of the deep combined F555W and F814W are $2150\, {\rm s}$ and $2265\, {\rm s}$ respectively. The images cover a region of $200\arcsec\times 200\arcsec$ corresponding to linear size of $58\, {\rm pc} \times 58\, {\rm pc}$ at the SMC distance.

The photometric reduction of all optical images has been performed within the IRAF\footnote{IRAF is distributed by the National Optical Observatories, which are operated by AURA Inc., under cooperative agreement with the National Science Foundation.} environment via the same method used for NGC~330 \citep{sirianni02} and NGC~346 \citep{sabbi07}.  We use PSF fitting  and aperture photometry routines provided within IRAF's DAOPHOT package \citep{stetson87} to derive accurate photometry of all stars in the field.   The photometry is calibrated into the {\it HST} Vegamag photometric system, following the prescription of \citet{sirianni05}.  Charge Transfer Efficiency (CTE) corrections are included, following the procedure described in \citet{sabbi07}.  We consider a source photometric in F555W (in F814W) only if its detection is at least $4 \sigma$ ($3.5 \sigma$) above the background level, and its DAOPHOT sharpness parameter is $-0.5 <$~{\it sharpness}~$ < 0.5$ ($-0.3 <$~{\it sharpness}~$ < 0.3$).  This allows us to maximize inclusion of real point sources while avoiding inclusion of ``fuzzy" background galaxies.  Our final catalog from WFC includes 4496 sources.  For further specifics of our photometry, including completeness testing and detailed selection criteria, see \citet{cignoni09}.

\subsubsection{SMARTS Ground-Based Data}
\label{SMARTSdat}

A few of bright stellar sources in NGC~602 are saturated in the {\it ACS} images.Photometric observations were taken the night of 14~October~2006 (JD: 2454023.67) and include three 45 second {\it V}-band exposures and three 30 second {\it I}-band exposures\footnote{{\it B}-band imaging and photometry were also performed but are not presented here.  Please contact authors for further information.}.  Some dithering was applied to cover bad pixels.  

We employ IRAF packages XREGISTER, IMSHIFT, and IMCOMBINE to properly align and combine the three images in each band.  We perform aperture photometry on a subset of stars across the FOV, finding the magnitude of each star at aperture radii 3,~4,~...~,~19,~and~20~pixels and select an aperture size of 13 pixels ($4\farcs 8$) for use in all further photometry.  This aperture captures the most stellar flux with the least background contamination.  We use DAOFIND to construct a photometry list with DAOPHOT parameters $ 0.1<{\it sharpness}<0.9 $ and  $ -0.8<{\it roundness}<0.8$, thus eliminating extended sources, bad pixels, and cosmic rays from our list.  We complete PSF photometry using packages PHOT, PSF, and ALLSTAR, requiring $\sigma$ values of 2.3 (in {\it V}) and 1.6 (in {\it I}).  We use SMARTS standard-star observations, air mass, and air mass coefficients  from the same night to calibrate our cluster photometry into the Landolt Vegamag system.  See the SMARTS/ANDICAM website\footnote{http://www.astronomy.ohio-state.edu/ANDICAM} for more information.  

Measured magnitudes are then converted from the Landolt to the HST Vegamag system using the equations from \citet{sirianni05} and used to fill in for sources saturated in our HST observations.  These are optically bright sources; none are YSO candidates.  We replace HST data photometry with SMARTS for $m_{F555W} < 18$ and for $m_{F814W}< 17.5$.  In Table~\ref{t:optphot}, 15 sources have SMARTS photometry only and twenty-six have a combination of SMARTS and HST photometry. 

\subsection{IRAC Mid-Infrared Data}
\label{IRACdat}
Five mid-IR images were taken through each of the four \spit/IRAC bands (3.6, 4.5, 5.8, and 8~$\mu$m) described in \citet{fazio04}.  The data were taken in $30\, {\rm s}$ High Dynamic range mode, using five dithers in the medium cycling pattern (Program ID 125, P.I. Fazio, AOR 12485120).  For each band the total integration time is $134\, {\rm s}$.  The Basic Calibrated Data (BCD; version S11.4.0) were used.

For each channel the IRAC mosaics have been constructed using the \spit\ Mosaicker software, along with the IRACproc package \citep{schuster06}. The overlap correction module was used to minimize the instrumental offsets between the frames.  The total field of view (FOV) is approximately $5\arcmin \times 5\arcmin$, corresponding to a linear size of 88~pc$ \times 88\, {\rm pc}$ at the distance of the SMC.  The native IRAC pixel scale is $1\farcs2$; through drizzling dithered images, we reach a pixel scale of $0\farcs 4/$pixel.

We use IRAF tasks DAOFIND and PHOT to locate and perform photometry on sources in each of the mosaics.  A 4$\sigma$ cutoff was used in source identification in DAOFIND.  Photometry was performed with PHOT, using a $1\farcs 6$ (4 drizzled pixels) aperture radius.  This is close to the IRAC PSF at [3.6] and encircles most of the flux without including nearby sources in crowded fields.  Background levels are determined and subtraction performed based on the background levels in the annuli with inner radius 14.1~pixels and outer radius 28.2~pixels (or $5\farcs 64$ and $11\farcs 28$), chosen through image examination. Zero-points \citep{reach05} were applied to normalize photometry to the nominal $12\arcsec$ (30 drizzled pixels, 10 IRAC native pixels) radius and to calibrate to the Vega magnitude system.  For SED fitting (See Section~\ref{SEDdata}), fluxes are calculated according to \citet{reach05}.  

Our deep IRAC photometric catalog (Table \ref{t:iracphot}) includes 497 sources, each detected in at least one IRAC band.  Of these forty-eight are photometric in all four IRAC bands and forty-one in three bands, using aperture photometry.  The other 408 sources are photometric in only one (164 sources) or two (244 sources) bands.  \citet{gouliermis07} use the same IRAC data set and apply PSF photometry to produce a list of 103 sources detected in at least three IRAC bands.  In our online photometric catalog, we include cross identification with the 22 YSO candidates reported by \citet{gouliermis07}.

For the purposes of image examination, we supplement our deeper IRAC data with photometry from 12~s exposures from the \spit\ Legacy Program SAGE-SMC \citep[Surveying the Agents of Galaxy Evolution in the Small Magellanic Cloud; ][]{gordon10}.  While this does not significantly increase the depth, the mosaicing of images taken with differing camera orientations allows us to create cleaner images with rounder point spread functions.  We also use SAGE-SMC IRAC photometry for 3 sources within our $3\arcmin$ radius but outside of our deeper IRAC FOV.  

\subsection{Supplemental Photometry for SED Construction} 
\label{SEDdata}

We characterize individual sources by constructing SEDs and comparing them to theoretical YSO model SEDs and SEDs of catalog naked stars and galaxies \citep[See Section~\ref{fitter};][]{robitaille06,robitaille07}.  We need to include as many photometric data points as possible to best constrain fits.  For this analysis, therefore, we require sources to have at least 4 bands of photometry, including [8.0] and at least 2 other IRAC bands.  We examine the 77 IRAC ``fitter sources" meeting these criteria.  We add optical and NIR data where available, but differences between instrument resolution, particularly at extragalactic distances, complicate this.  Of our 77 fitter sources, 61 lie within the {\it ACS} FOV, 6 with no apparent optical (point-source) counterpart, 29 corresponding to a single optical source, and 26 with multiple optical counterparts.  We must handle our optical fluxes carefully.

Further complication results from the fact that the SED fitter allows input only in specific photometric bands, including Landolt {\it V} and {\it I} and Two-Micron All Sky Survey \citep[2MASS;][]{skrut06} {\it J}, {\it H}, and $K_s$.  We therefore convert magnitudes from similar native bands, introducing an additional element of uncertainty, as conversions are generally calibrated for MS stars rather than PMS or YSO sources.  Our treatment of these uncertainties is discussed below.  
% Visual inspection of SEDs e.g., Figures~\ref{SED283}, \ref{seds28788}, \ref{starseds}, and \ref{galseds}) indicates that 
Our applied error estimates are appropriate with fluxes  tracing continuous curves and no discontinuities between fluxes from different instruments except where fluxes are used as upper limits for the physical reasons explained below.

\subsubsection{Optical}
\label{sed_optdat}

Comparing deep, high-resolution {\it ACS} optical and IRAC images and photometry requires special care.  Several physical sources at the distance of the SMC may fall within the $1\farcs 6$ IRAC [3..6] aperture.  Something that appears to be a single source in the IR images, may be resolved into ten or more sources in {\it ACS} images, with no obvious way to tell which one is or which ones are the primary IR contributer(s).  Likewise, faint or embedded sources may have no resolvable optical point source within the IR aperture.  

We consider four possible types of IR-optical matches:

\begin{enumerate}
\item{\it Single Optical Point Sources:}\label{one}  These are unsaturated in the ({\it HST}) optical and have only one obvious optical point source.
\item{\it No Optical:}\label{zero}  There is no apparent optical counterpart to the IR source.  These often correspond to dusty structures remmeniscent of the famous pillars and mountains of creation \citep[e.g.,][]{thompson02}.
\item{\it Multiple Optical Sources:}\label{many} Some of these are probable protoclusters, while others are superpositions.  These may be faint or bright.
\item{\it Saturated Sources:}\label{sat} Optically saturated sources mostly seem to be bright Main Sequence stars.  In some cases, faint neighbors with poor photometry (often because of diffraction spikes) are distinguishable in optical images.
\end{enumerate}

We want to include light from the same sources in each of the photometric bands to the greatest extent possible to produce the most physically meaningful SEDs (see Section \ref{FIT}).  We perform aperture photometry with DAOPHOT's PHOT task at the exact IRAC coordinates of the 43 fitter sources which lie within our {\it ACS} FOV and are not saturated in the optical and use a 32 {\it ACS} pixel aperture radius to measure the integrated optical flux over the $1\farcs 6$ IRAC aperture.  For 13 saturated sources, we use the SMARTS ground-based photometry (aperture radius $4\farcs 8$) and calculate the flux as with other sources and apply a 40\% uncertainty.  One optically saturated source (S456) has no good optical photometry.

The SED fitter requires the input of fluxes in specific photometric bands, including Landolt {\it V} and {\it I} but not {\it ACS} F555W or F814W.  We must convert from the {\it ACS} to the Landolt system and calculate fluxes from magnitudes.  To this end, photometry is performed in the {\it ACS} OBMAG system (PHOT's ABMAG with no zero point).   We then convert to the Landolt photometric system according to the description of \citet{sirianni05}.  We calculate the fluxes in mJy, using the zero point fluxes $f_{0V}=3.789\times10^6 $~mJy and $f_{0I}=2.635\times10^6$~mJy adapted from \citet{crawford75}.  The conversions from {\it ACS} to Landolt bands are not designed for dusty, young, or low-metalicity sources.  We therefore calculate the errors in flux by squaring the errors described in the conversion.  Optical fluxes are treated as upper limits for several YSO candidates, if their high resolution optical morphology indicates the presence of multiple optical or zero optical counterparts.  Source Y217 has indefinite OBMAG in F555W but a measurable magnitude in F814W.  We calculate the flux in F814W using the equation specified in \citet{sirianni05} and assume that the real {\it I} flux is not more than a factor of two greater.  We thus double calculated flux as an upper limit in the SED fitter input (Sections~\ref{fitter} and \ref{quality}).

In multiple optical sources, it is unlikely that all optical sources are significant IR emitters \citep[i.e.][L$_{YSO} \propto $~M$_{YSO}^3$]{bernasconi96}.  It is instead probable that one, two, or even zero of the optical sources is ``the" infrared emitter.  These sources should be treated with extra care.  The approximate number of optical sources corresponding to a single IR source is determined visually, as some optical sources are non-photometric (See Section~\ref{optdat}).  Optical fluxes are treated as upper limits for several YSO candidates, based on their high resolution optical morphology.   For sources that have no optical or are saturated in the optical but have 4 IR bands to constrain the SED fit, we set the calculated {\it V} and {\it I} fluxes as upper limits rather than assigning errors bars.  They are noted in Table~\ref{t:ysoresult}, which provides details of the fitter results as well as the number of optical sources.  Optical matches to IRAC sources are also noted in online photometry Tables~\ref{t:optphot} and \ref{t:iracphot}.

\subsubsection{IRSF Near Infrared Data}
\label{NIRdat}
We use Near-IR photometry from the Magellanic Clouds Point Source Catalog of \citet{kato07}.  This imaging survey was performed with the Simultanious three-color InfraRed Imager for Unbiased Surveys (SIRIUS) aboard the InfraRed Survey Facility (IRSF) 1.4~m telescope at the South African Astronomical Observatory's Southland.  Photometry covers the {\it J} ($\lambda_c=1.25~\mu$m), {\it H} (1.63~$\mu$m), and $K_s$ (2.4~$\mu$m) bands.  We choose to use this IRSF/SIRIUS data for its depth ($\la 16$~mag in $K_s$ compared to $\la 14$~mag in 2MASS) and high spatial resolution ($FWHM\sim1\farcs 1-1\farcs 3$, compared to 2MASS's $2\arcsec-3\arcsec$).  \citet{kato07} performed PSF photometry for aperture radius 1$\arcsec$.35 to construct their catalog.  

Sixty-four fitter sources have IRSF matches.  For use in the SED fitter, we convert IRSF magnitudes to 2MASS fluxes.  We apply magnitude conversions given in \citet{kato07}, which require at least 2 bands of IRSF data, then apply 2MASS zero-points of 1.594, 1.024, and 0.6668~mJy in {\it J}, {\it H}, and $K_s$ \citep{cohen03} respectively to obtain 2MASS fluxes.  Unfortunately, uncertainty is always introduced when converting between photometric bands, particularly for non-photospheric objects.  We estimate a nominal 20\% uncertainty in flux.  Eight of the 64 matched sources have only one band of IRSF photometry, making the usual conversion to 2MASS magnitudes impossible.  We use the IRSF magnitude as though it were a 2MASS magnitude and account for extra uncertainty introduced by lack of color information by increasing our applied error bars from 20\% to 40\% in flux.  Another 13 of the 64 matched fitter sources have multiple matches within the IRSF catalog.  For these, we take the brightest IRSF measurement in each band and treat the flux as an upper limit in our SED fitting.

\subsubsection{MIPS Data}
\label{MIPSdat}
Multiwavelength Imaging Photometer for \spit\ \citep[MIPS; ][]{rieke04} 24~$\mu$m data are taken from the first epoch observations of SAGE-SMC \citep{gordon10}.  These observations were performed in October 2007.  Post-processing is performed as it was for the SAGE-LMC MIPS 24~$\mu$m data \citep[as in ][]{meixner06}.  Photometry is performed using the shape-fitting algorithm STARFINDER \citep{diolaiti00}.  Sources for which we include 24~$\mu$m fluxes have been selected through automated matching to our known IRAC sources and verified through careful visual inspection of the 24~$\mu$m image.  We match eleven 24~$\mu$m sources to sources fit as YSO candidates and seven more to other types of sources (See Section~\ref{FIT}).

\section{COLOR MAGNITUDE ANALYSIS - STELLAR AND PROTO-STELLAR POPULATIONS} 
\label{cmdanal}

We examine optical and IR data independently before combining them.  This strategy results in a complete picture of stellar and proto-stellar populations in the region with the more evolved populations revealed by optical alone and the least evolved sources appearing only in the IR. 

\subsection{The Optical CMD}
\label{opt_pop}
Examination of the optical CMD (Figure~\ref{OptCMD}) reveals a quantitative picture of the optical populations.  In NGC~602, optical point sources comprise MS, Red Giant Branch (RGB) evolved, and PMS stars.  Old (lower) MS and their RGB counterparts are found to be primarily background sources \citep{cignoni09}.  More important for this study, the optical CMD reveals young ($\la 5$~Myr) MS and PMS stars.

Our optical catalog of over 4500 sources is a combination of {\it HST/ACS} and SMARTS data (Table~\ref{t:optphot}).  The applied PMS evolutionary tracks (and isochrones) were calculated for NGC~602 by \citet{cignoni09} working from FRANEC stellar evolutionary code \citep[cf.][]{chieffi89,deglin08} and using $Z=0.004$ and distance modulus $(m-M)_0=18.9$ and reddening $E(B-V)\sim0.08$.  Padua stellar evolutionary tracks \citep{fagotto94} are used for MS and RGB populations.  The RGB and lower-MS populations are best fit with $Z=0.001$, consistent with the Magellanic Bridge/Tail field population \citep{zaritsky04}, supporting the supposition that these are background sources.

We identify PMS stars via the color-magnitude selection (yellow line in Figure~\ref{isoCMD}) from \citet{carlson07}: 
\begin{equation}
\label{pmseq}
(28-4.3\times[m_{\rm F555W}-m_{\rm F814W}])  <  m_{\rm F555W}  <  (5.5\times[m_{\rm F555W}-m_{\rm F814W}]+17)
\end{equation}
In Figure \ref{OptCMD}, most of the PMS population lies between the example 1~Myr and 10~Myr isochrones.  In \citet{cignoni09}, it is shown that most PMS stars in NGC~602 are less than 5~Myr in age, and the star formation rate has been increasing for the past few Myr.  The star formation rate was $\sim1.5 \times 10^{-4} ~ \Msun$~yr$^{-1}$ from 5 to 2.5~Myr ago and has increased to $\sim 3 \times 10^{-4} ~ \Msun$~yr$^{-1}$ in the last 2.5~Myr.  Further detailed analysis of the star formation history of the optical population can also be found in \citet{cignoni09}. 

\subsection{The Infrared Population}
\label{IR_pop}

Figure~\ref{iracc} shows both a CMD and a Color-Color Diagram (CCD) in IRAC featuring the 77 sources within $3\arcmin$ of cluster center for which we perform detailed SED analysis below.  SAGE-SMC IRAC sources for the larger $18\arcmin \times 18\arcmin$ FOV (317~pc~$ \times $~317~pc; Figure~\ref{FOV}) are plotted in grey.  This background population is quite sparse and is concentrated near the zero-color MS.

Infrared sources in the NGC~602 region may be classified into 3 groups: YSOs, other stars, and background galaxies.
Significant features are the colorless MS and the primary YSO color-magnitude space, centered around $[3.6]-[8.0] = 3$ (Figure~\ref{iracCMD}).  In the color-range $0.3 \la [3.6]-[8.0] \la 3$, only 6 sources are brighter than about $[3.6]=13.5$, indicating that there are virtually no IR-bright evolved sources in the entire region.  Six fit stellar sources (K181, S213, K225, K232, K441, and K444) have colors $0.6 \la [3.6]-[8.0]$, redward of the MS, and are discussed further in Section~\ref{star}.  Five of the 7 identified galaxies on the CMD lie in the color range $2.5\la ([3.6]-[8.0])\la 2.9$.  The YSO population is on average redder and fainter than the MS population, with most sources redder than $\sim2.8$ in [3.6]-[8.0].  Examination of IRAC images of sources from the larger FOV in this YSO CMD region suggests that many of them are background galaxies; a dense population of these background sources falls within the same $2.5\la ([3.6]-[8.0])\la 2.9$ range.
 
The CCD (Figure~\ref{iracCCD}) illustrates that the YSOs are significantly redder than both the cluster and the background populations.  This is particularly true in [4.5]-[5.8] where only two YSO candidates have colors $[4.5]-[5.8] < 0.5$.  Twenty-three of the 26 YSO candidates with these three bands of data lie in two sections of color-color space, outlined in green and discussed as indicators of strong and weak emission from polycyclic aromatic hydrocarbons (PAHs) in Section~\ref{PAHs}.  Non-YSOs are strongly clustered around zero on both axes, with no non-YSO cluster sources and very few background (``outside cluster") sources falling within the YSO color-color spaces.  {\it Zero} non-YSOs have $1.5 \la [4.5]-[5.8]$.
  
\section{Combining Optical and IR: Fitting Spectral Energy Distributions}
\label{FIT}

We apply the SED fitting tool to a total of 77 ``fitter sources," two of which are multiple sources with sufficient wavelength coverage to be fit as two separate sources (Sections~\ref{340} and \ref{270}).  Of these 79 sources then, forty-one are YSO candidates, twenty-eight are stellar, eight are background galaxies, and two are of unknown nature.  Fluxes used are given in Table~\ref{t:flux}.  Images of all YSO candidates are shown in Figure~\ref{minis} and cover {\it ACS} optical, IRAC, and MIPS 24~$\mu$m data.

We characterize 37 sources as well-fit YSOs.  We name these with their IRAC photometric IDs (used in Table~\ref{t:iracphot}) preceded by Y.  The evolutionary stages for the YSOs are concentrated in the less evolved Stage~{\sc i} (21 {\sc i} YSOs) and Stage~{\sc i} or {\sc ii} (6  {\sc i}/{\sc ii} YSOs) with a few Stage {\sc ii} (7  {\sc ii} YSOs).  Of the three remaining well-fit YSO candidates, two may be Stage~{\sc ii} or Stage~{\sc iii}, and one can be fit with YSO models of all evolutionary stages and is considered unclassified.  Sources Y090, Y227, Y240, and Y251 are probable YSOs based on color, environment, morphology, and best fits.  They are Stages {\sc i} (Y090, Y240, Y251) and {\sc ii} (Y227), but their SED fits are not as good (${\chi^2/pt}_{min} > 3.00$).  

Table \ref{t:ysoresult} gives the average fitter results for all YSO candidates, including the number of optical sources, the optical and IR masses, the YSO's luminosity and accretion rate, and the estimated evolutionary stage.  Where one or more optical sources correspond to a YSO candidate, we attempt to estimate the optical mass in addition to the mass given by the fitter.  We determine the location of optical sources on the CMD and compare to evolutionary tracks (Figure~\ref{massCMD}).  Only a small percentage of the optical counterparts have good photometry and fall within the color-magnitude space covered by evolutionary tracks; most are too young, too red, or of too small a mass for these estimates.  Intrinsic reddening will also cause some optical sources to appear to have somewhat lower masses than they do.  The optical masses we do calculate (summing over the entire IRAC source) are consistent as lower limits with mass estimates from SED fits.

We fit an additional 38 sources as non-YSO candidates.  Nineteen are classified as single, naked stars with Kurucz models \citep{kurucz93}, and we name them by their IRAC photometric ID preceded by K.  Nine other sources are clearly stellar (named S- - -) based on imaging but do not have conclusive SED fits.  Eight sources are  visually identified as galaxies (G- - -).  Sources U346 and U703 are the only sources we are unable to classify.  They fall outside the {\it ACS} optical FOV, are poorly fit as YSOs, and neither their positions nor their morphologies provide clues to their natures.

\subsection{The Fitter}
\label{fitter}

We use the model SED fitting tool described by \citet{robitaille07}, which compares fluxes in specific photometric bands to a grid of 200,000 YSO model SEDs  \citep{robitaille06}.  The model grid consists of model SEDs computed using the radiation transfer codes of \citet{whitney03a, whitney03b, whitney04} for 20,000 sets of physical parameters, 10 viewing angle (from edge-on to pole-on), and in 50 different photometric apertures (from 100 to 100,000~AU).  Among the physical parameters varied in the production of model SEDs are stellar mass, radius, and temperature, envelope mass and accretion rate, and disk mass, flaring angle, and accretion rate.  Ranges for these parameters are determined from observational data.

A limited number of naked photosphere, asymptotic giant branch (AGB), and galaxy SEDs are also fit to the photometric data for comparison.  All sources are allowed to be fit with YSO model SEDs and Kurucz naked star SEDs \citep{kurucz93}, including stellar photospheres with extinction $0<A_V<15$ in addition to internal reddening for YSOs.  For AGBs, we consider extinction in the range $2<A_V<20$, while galaxy extinction factors are  $3<A_V<30$.  

Synthetic YSO SEDs are computed for five distances in the prescribed range of 60-65~kpc, appropriate for the SMC.  At a distance of 60.6~kpc, our aperture radii of $1\farcs 6$ in IR and optical and $10\arcsec$ in MIPS 24~$\mu$m correspond to $9.48 \times 10^7 $~AU and $5.92 \times 10^8 $~AU respectively.  The largest aperture available for the computed SEDs is $10^5 $~AU, so this is the aperture radius used for our fits.  

The current models are calculated for Solar metallicities and do not include effects from PAH emission, external illumination, or protoclusters.  Partially as a consequence of the physical aperture size at SMC distances, there are often multiple optical sources within the aperture of a single IR source.  Emission from PAHs can contribute significantly the flux in some IRAC bands.  Metallicity is expected to play a role in the timescales of star formation, because metals help carry heat away from collapsing gas and dust, (possibly) increasing the speed at which a star can form.  Models are being developed to include these effects \citep[cf.][]{sewilo10}.  To account for multiplicity and PAHs, optical fluxes as well as 5.8, 8.0~$\mu$m fluxes must be treated with care, as discussed in Sections~\ref{sed_optdat}, \ref{PAHs}, \ref{340}, and \ref{270}.    

\subsection{Quality of Fits}
\label{quality}
We quantify how well a source is fit by a given model SED by considering the value of $\chi^2$/point \citep[as in][]{robitaille07}.  In general, the probability that a given model reproduces the input data is ${\rm p} \sim {\rm e}^{-\chi^2/2}$ (or $\chi^2 \sim -2 \ln {\rm p}$).  We consider p to be a  Gaussian for photometric data points, giving the standard $\chi^2$ in equation (2).  In SED fitting, we include some fluxes as upper limits only.  Our confidence (${\rm c} = 1 - {\rm p}$) that the source is not brighter than this limit is input to the fitter.  $\chi^2$ values for these points are then calculated as in equation (3).
\begin{eqnarray}
\chi^2_{data} & = & \sum_{i=1}^n \left ( \frac{F_i-M_i}{\sigma_i} \right )^2 \\
\chi^2_{upper} & = & \sum_{i=1}^n 
\left \lbrace
\begin{array}{lc}
-2 \ln (1- {\rm c}) & \mbox{if }M_i > F_i  \nonumber \\
0 & \mbox{if }M_i \le F_i  \nonumber
\end{array}     \right \rbrace \\
\end{eqnarray}
Where $F_i$ is the flux at wavelength $\lambda_i$ with error bar $\sigma_i$ describing the gaussian probability distribution.  $M_i$ is the model flux value (which includes extinction and scaling for distance) at $\lambda_i$, and n is the total number of data points.
\begin{eqnarray}
\begin{array}{cccc}
\chi^2_{total} & = & \chi^2_{data} + \chi^2_{upper} &  \nonumber \\
\chi^2 /pt & = & \chi^2_{total}/n & \mbox{\small{(i.e., reduced $\chi^2$)}} \nonumber 
\end{array}
\end{eqnarray}
Based on visual examination of both SEDs and images, sources with at least one fit of $\chi^2/pt \le 3.00$ are considered "well-fit".  We average the parameters of all models with ${\chi^2/pt} \le ({\chi^2/pt}_{min} + 0.50)$ to characterize each YSO candidate.  Table~\ref{t:ysoresult} lists the number of models considered.  In all example YSO SEDs (Figures~\ref{noopt}, \ref{sed700}, \ref{oneopt}, \ref{seds28788}, \ref{sed312}, \ref{340set}, and \ref{270set}), we show the best fit (that with $\chi^2_{min}$) as a dark solid line with grey lines indicating all fits considered.  SEDs of non-YSOs (Figures~\ref{340set}, \ref{starseds}, and \ref{galseds}) are shown with only the top twenty-five fits (which are representative of the whole set).

\subsection{Characterizing YSOs}
\label{YSOfit}

After applying the fitter to all 77 fitter sources, we select YSO candidates.  To calculate their physical characteristics, we take the well-fit-model parameters and apply a dust-to-gas ratio of 1/6 Galactic, appropriate for star-forming clusters in the SMC \citep{bot07}.

YSO sources can then be classified according to evolutionary stage, based on the fractional disk mass and the envelope accretion rate.   We use the physical Stages {\sc i}, {\sc ii}, and {\sc iii} from \citet{robitaille06} which are roughly equivalent to the the observational Class {\sc i}, {\sc ii}, and {\sc iii} classifications of \citet{lada87}.  
\begin{enumerate}
\item{\it Stage {\sc i}:} Embedded Source ($\dot{M}_{env} > 10^{-6} M_{\star} $~yr$^{-1}$)
\item{\it Stage {\sc ii}:} Source with Disk ($\dot{M}_{env} < 10^{-6} M_{\star} $~yr$^{-1}$  \&  $M_{disk}/M_{\star} > 10^{-6}$)
\item{\it Stage {\sc iii}:} Source with Optically Thin Disk or No Disk ($\dot{M}_{env} < 10^{-6} M_{\star} $~yr$^{-1}$ \&  $M_{disk}/M_{\star} < 10^{-6}$)
\end{enumerate} 
In Table~\ref{t:ysoresult}, we list the parameters resultant from both the best fit model and the average of all models in the ${\chi^2/pt} \le ({\chi^2/pt}_{min} + 0.50)$ range.

\subsection{Accounting for Polycyclic Aromatic Hydrocarbons}
\label{PAHs}

Emission from PAHs is not incorporated into the SED models in the current version of the fitter.  IR flux can be significantly increased by PAH emission.  Of the IRAC bands, only [4.5] should be unaffected by PAH contributions \citep{churchwell04}.  The sources affected by these PAH emissions are readily identifiable by the characteristic dip in $\lambda log_{10}({\rm flux})$  at $\sim 4.5 \mu$m in the SED, which can be quantified using color-selection such as that shown in Figure~\ref{iracCCD}.  We must weight the uncontaminated [4.5] band flux more heavily than the measured fluxes in the other three IRAC bands.  Most appropriately, the heavily contaminated [5.8] and [8.0] fluxes would be treated as upper limits (Robitaille, Private communications, 2009); however, in the absence of longer wavelength measurements, this is impractical.  We therefore adjust the error bars to 20, 10, 30, and 40\% in the 3.6~$ \mu$m, 4.5~$ \mu$m, 5.8~$ \mu$m, and 8.0~$ \mu$m bands, respectively. 

We perform these PAH corrections on the 13 sources falling within the ``Strong PAHs" area given in Figure~\ref{iracCCD}.   The relatively blue slope of 3.6 to 4.5~$ \mu$m, combined with the relatively red slope of 4.5 to 5.8~$ \mu$m, is indicative of a strong dip in the SED at 4.5~$ \mu$m and thus of significant PAH ``contamination" in the other IRAC bands.  
We note sources to which we apply PAH corrections in Table~\ref{t:ysoresult}. 

Five sources with probable significant PAH contributions (Y217, Y227, Y270, Y287, and Y340) do have 24~$\mu$m flux measurements; we consider both methods of PAH correction.  We perform SED fits considering 5.8~$ \mu$m and 8.0~$ \mu$m fluxes as upper limits, relying upon the additional SED constraint of  the longer wavelength measurements.  These are the fits we use in our calculation of the mass function and in the determination of YSO stages as reported in all figures.  The difference in fitting PAHs with 5.8~$\mu$m and 8.0~$\mu$m fluxes as upper limits rather than as having error bars of 30\% and 40\% is shown in Table~\ref{t:ysoresult}.  For sources Y217, Y227, and Y287, fits with increased error bars suggest higher masses and later stages of evolution in comparison to fits with 5.8 and 8.0~$\mu$m treated as upper limits.  Source Y340 is unclassified using the upper limits method but classified as Stage~{\sc iii} using increased error bars; the estimated mass is higher using the error bar method.  Y270i (Section~\ref{270}) is the only source which breaks the trend and is fit as Stage~{\sc i} with the error bar method but Stage~{\sc i}/{\sc ii} using the upper limits method.  Results using the error bar method are reported in Table~\ref{t:ysoresult} as Y~-~-~-e.

\subsection{YSOs without Optical Counterparts} 
\label{yseds0}
Sources Y142, Y143, Y174, Y179, and Y217 (Figure \ref{mini142}, \ref{mini143}, \ref{mini162170171174}, \ref{mini179196197206}, and \ref{mini217237} respectively) have no readily identifiable optical counterparts.  Y217, which lies at the tip of one of the pillars along the western molecular ridge, is classified as a Stage {\sc i} or {\sc ii} source.  The rest are fit as Stage {\sc i} and correspond to distinctive dust features.   Each is an optically dark bump along a molecular ridge.  Y142 lies within the ``mountain of creation" south of the central cluster.  All five have masses estimated between 6 and 7 $\Msun$.  Y149, Y163, Y493, and Y700 lie outside of the ACS FOV and have no known optical counterparts for this reason.  

The SEDs of Y142, Y217, and Y700 are shown in Figures~\ref{noopt} and \ref{sed700} and are markedly different.  The fit to Y142 is poorly constrained at wavelengths greater than 8~$\mu$m and may peak anywhere between 8~and~70~$\mu$m.  Y217 has a double peak SED but is unconstrained at wavelengths greater than 24~$\mu$m.  Y700's SED is remarkably flat from about 1~$\mu$m to at least 24~$\mu$m.

\subsection{YSOs with Single Optical Counterparts} 
\label{yseds1}
Sources Y118, Y148, Y251, and Y283 (Figure~\ref{mini118}, \ref{mini148}, \ref{mini251255264270}, and \ref{mini283}) correspond to single optical point sources, but each is in a different environment.  Y118 matches a relatively isolated PMS candidate in a region of apparently thin diffuse dust and is fit as a Stage {\sc i} YSO (Figure~\ref{SED118}).  The mass of the optical source ($\sim 0.8 ~\Msun$) as determined from the optical CMD (Figure~\ref{massCMD}) is significantly smaller than the $\sim 7.4~ \Msun$ of the YSO fit, likely indicating the presence of a separate IR-bright source.  It is also possible that intrinsic reddening causes and underestimate of the optical source's mass; with A$_V \sim 3$, its mass could be as high as 1.5~$\Msun$, still leaving $\sim 6 \Msun$ unaccounted for.  Stage {\sc i} source Y148 appears as a very faint optical source in a region of optically thick dust.  Y251 is poorly fit as a Stage~{\sc i} YSO but corresponds to a PMS candidate in a dust peak near cluster center.  Y283 is a well-fit Stage {\sc ii} source with an SED peaking around 10~$\mu$m (Figure~\ref{SED283}) and an estimated mass of $\sim 4.6 ~\Msun$ in fair agreement with its optical mass of $3-4 ~\Msun$  .

\subsection{YSOs with Multiple Optical Components}
\label{yseds+}
Approximately 70\% of our YSO candidates have multiple optical counterparts.  Optical sources corresponding to YSO candidates are marked on the CMDs in Figure~\ref{OptCMD}.  Most of these are fainter or redder than the 1~Myr isochrone, indicating that they are very young and probably embedded, as we would predict for sources related to the earliest stages of star formation.  Where possible we determine the masses of YSOs' optical counterparts through comparison the evolutionary tracks (Figure~\ref{massCMD}); we sum the apparent masses of these optical sources to determine the optical masses (M$_{\rm opt}$) given in Table~\ref{t:ysoresult}.  We find that YSO model masses are systematically higher that optical masses.  Many masses are probably underestimated, because their embedded nature makes them look fainter in the optical.  Looking at Figure \ref{minis} there are far more optical sources corresponding to YSO candidates than appear on the optical CMD.  In particular, sources Y090, Y096, Y223, Y270, Y326, and Y327 (Figure \ref{mini90}, \ref{mini96}, \ref{mini223227240}, \ref{mini251255264270}, \ref{mini326}, and \ref{mini327}) have more than ten optical counterparts each (see also Table~\ref{t:ysoresult}).  Some of these lie redward of evolutionary tracks on CMDs, preventing mass estimates.  Others are non-photometric as a result of confusion from source multiplicity and high background, as well as their faint nature, and they are not in our catalog (Table~\ref{t:optphot}).  There may be embedded IR-dominant sources, likely representing ongoing star-formation (as Y340 and Y270), as massive YSOs are expected to evolve more quickly than their low-mass counterparts.  It is also possible that multiple optical sources are heating more of the surrounding ISM than is truly circumstellar matter, causing additional infrared excesses.  SEDs of distinct multiple sources Y287, Y288, and Y312 are shown in Figures~\ref{seds28788} and \ref{sed312}. 

\subsubsection{Source 340}
\label{340}
Optical imaging of Source 340 reveals exactly two optical sources, one optically bright, the other faint (Figure~\ref{340set}).  They lie at the tip of a dusty pillar highlighted by H$\alpha$ emission.  Source 340 is also a PAH source with 24~$\mu$m flux.  Fitting all data points with either PAH correction method results in a poor Stage~{\sc iii} fit (A340; ${\chi^2/pt}_{min} \sim 3.5$).  Fitting different wavelength regimes separately results in two good fits, which add together to describe the SED nicely.  A main sequence fit (K340; ${\chi^2/pt}_{min}=0.03$) accounts for the majority of the optical and NIR flux, while the longer wavelengths are well-fit as an unclassified YSO (Y340; ${\chi^2/pt}_{min}=0.62$).  As with Y217, Y227, and Y287, treating 5.8 and 8~$\mu$m fluxes as upper limits for Y340 results in a lower mass estimate than if we treat them as concrete data points.  We show the SEDs for A340, K340, Y340, and all three overlaid, demonstrating the improved (and more physically meaningful) fit that results from treating K340 and Y340 separately.  The brighter of the two optical sources distinguishable in Figure~\ref{340Halpha}, has a mass of $\sim4~\Msun$ based on its position on the optical CMD (Figure~\ref{massCMD}) and likely corresponds to the K340 fit.  The fainter optical source may be dominant in the longer wavelengths, and it is probable that one or more protostars are still embedded in the dusty pillar.  Physical characterizations of each fitting method are given in Table~\ref{t:ysoresult}.

\subsubsection{Source 270}
\label{270}
Optically, Source 270 appears to be a protocluster emerging from a pillar of dust (Figure~\ref{270set}).   At least ten sources are discernible in the optical image, including one B2 star classified by \citet[star 12]{hutchings91}.  We fit source 270 in the three ways shown in Figure~\ref{270set}: (A270) fitting all data points, (Y270o) fitting optical plus NIR, and (Y270i) fitting IRAC plus MIPS~24~$\mu$m.  A270 is poorly fit as Stage~{\sc ii} (${\chi^2/pt}_{min}=3.36$), while Y270o and Y270i are well-fit as Stages~{\sc i} and {\sc i}/{\sc ii} YSOs of mass $\sim 5.6~\Msun$ (${\chi^2/pt}_{min}=1.79$) and $\sim9.3~\Msun$ (${\chi^2/pt}_{min}=1.05$), respectively.  Using the error bar method of PAH correction for Y270i results in evolutionary Stage~{\sc i} and a slightly lower mass of 8.2~$\Msun$ compared to the upper limit method.
  
Eight of the optical sources corresponding to source 270 have good photometry, and we are able to approximate masses for five of these (see Figure~\ref{massCMD}).  Two lie near the MS and have masses greater than $\sim 5.5~ \Msun$; a third falls on the PMS $3~\Msun$ evolutionary track and is much younger/redder than 1~Myr, and the other two are PMS stars with probable masses $<1~\Msun$.  Other optical sources are too red or too faint for a mass determination.  The sum of these optical masses is $\sim 15~ \Msun$, which is also the sum of the Y270o and Y270i masses given by the fitter.
 
\subsection{Non-YSOs}
\label{nseds}

\subsubsection{Stellar Fits}
\label{star}
We are able to fit 18 of the 77 fitter sources (plus K340) as naked stars (MS or Giant stars).  Four of these correspond to O or B type stars defined by \citet{hutchings91} on the basis of optical and UV spectroscopy.  Five more sources are readily identified as stellar and are fit only as naked stars but have $\chi^2/pt \ge 3.00$.  S293 is one of these and is classified as a B1 star by \citet{hutchings91}.  They also report one B2 star corresponding to  source 270 (see Section~\ref{270}), one O6-O8 star corresponding to S235, and three B stars that have insufficient IR photometry for our SED fitting.  A typical stellar SED, for K348, is shown in Figure~\ref{SEDK348}.

Nine readily identifiable stellar sources (S049, S213, S235, S293, S394, S406, S411, S456, and S486) can be decisively fit.  S049, S411, and S486 lie outside the {\it ACS} FOV.  S049 and S486 display characteristic diffraction patterns in IRAC imaging and are fit by naked stellar templates with $3.0 < {\chi^2/pt}_{min} < 4.0$.  S411 is well-fit but may be fit as a naked star, a YSO, or an AGB.  The rest fall within the {\it ACS} FOV and are clearly stellar.  S293, S394, and S406 have typical but imperfect photospheric SEDs.

Three stellar sources are poorly fit because of their evident mid-IR excesses.  S456 lies on the edge of the {\it ACS} FOV  in Figure~\ref{dist}) and has no reliable optical photometry.  Its SED does not drop as quickly in the IR as would be expected for a naked star.  S213 lies near cluster center, and several optical sources fall within the IRAC aperture.  \citet{hutchings91} identify the brightest of these as an O9 star, though there is significant confusion even in our high-resolution optical imaging.  Its SED falls fairly smoothly through [5.8] but then rises sharply through 8.0 to 24~$\mu$m.  S235 lies at the bright tip of the MS on our optical CMD (Figure~\ref{massCMD}).  \citet{hutchings91} report that it is (optically) the brightest source in NGC~602 and classify it as an O6 or O8 type bright, young MS star based on optical and UV spectroscopy respectively.  Spectra are shown in their figures 3, 4, and 5 for their star 8.  Looking at this source in IR wavelengths shows something different.  The SED drops steadily through [8.0], but the MIPS 24~$\mu$m flux is far higher than would be expected (Figure~\ref{SEDS235}).  The flux points for optical through IRAC can be well-fit as a MS star, but the 8.0~$\mu$m measurement is slightly brighter than MS fits would suggest.  This is likely the result of the O star heating the nearby interstellar medium.  Alternately, this star could be the first to emerge from an embedded cluster \citep[cf.][for similar examples in 30~Doradus and N~66]{walborn99,walborn02,h-m10} the O star may exhibit a mid-IR excess due to free-free emission in the stellar wind \citep[as described in ][]{bonanos09}, or it could be a chance superposition.

\subsubsection{Galaxies and Unidentified Sources}
\label{gals}
We visually identify 41 of the 497 IRAC sources with at least one band of data (Table~\ref{t:iracphot}) as galaxies in ACS images.  Eight of these are fitter sources, six of which are readily identifiable as ellipticals or spirals and one as an irregular.  The eighth galaxy (G211) is outside of the ACS FOV but appears markedly elongated in mosaiced IRAC images.  All eight are best fit as YSOs, some very well.  Only seven of the eight are plotted in the IRAC CMD (Figure~\ref{iracCMD}); G150 has $\sim 1.5$~mag error estimate in [8.0].  Only two of the eight have 5.8~$\mu$m flux measurements.  In Figure~\ref{galseds}, we show the SEDs for the two most obvious galaxies in the regions, elliptical G133 and face-on grand design spiral G372.  These are the only two that are not well-fit as YSOs, and we show their best-fit galaxy templates.

\section{THE YSO MASS FUNCTION AND STAR FORMATION RATE}
\label{mf}
The total mass of our well-fit YSOs is $\sim 300~\Msun$.  We plot a histogram of YSO masses (Figure~\ref{mf_plot}, also Table~\ref{t:ysoresult}) and assume approximate completeness at the peak of the distribution ($\sim6-10 ~\Msun$) to estimate a lower limit for the total mass.  We consider a two-part MF with $\xi \propto {\rm M}_{star}^{-1.3}$ for M$_{star} < 0.5~\Msun$ and $\xi \propto {\rm M}_{star}^{-2.3}$ for M$_{star} > 0.5~\Msun$, where $\xi$ is the number of sources with mass $M_{star}$ \citep[as in][for complete IMF discussion]{whitney08,kroupa01}.  Requiring the two functions to be equal at M$_{star} = 0.5~\Msun$ and scaling to the peak of the observed mass function, we  integrate over the mass range $0.08-50~\Msun$, determining a total YSO mass of $\sim 2250~\Msun$.  We may then estimate a star formation rate (SFR) assuming constant star formation over a given time and considering the physical size $\sim 9.16 \times 10^{-3} $~kpc$^2$ of the studied $3\arcmin$ radius region.  
We consider two possible time scales:
\begin{enumerate} 
\item{1 Myr:} As seen in Figure~\ref{OptCMD}~(b) the PMS stars related to YSOs are characteristically younger than $\sim 1$~Myr.  Applying this time scale, we calculate an SFR of $\sim 2.2 \times 10^{-3} ~\Msun $~yr$^{-1}$ ($\sim 0.24 ~\Msun $~yr$^{-1}~$kpc$^{-2}$).
\item{$2 \times 10^5 $~yr:}  \citet{whitney08} apply a time of $2 \times 10^5$~yr for the formation of current YSOs in the LMC, based on the relative number of Stage~{\sc i} YSOs in their sample and the timescale calculations of \citet{lada99}.  For NGC~602, this results in an SFR of  $\sim11.1 \times 10^{-3} ~\Msun $~yr$^{-1}$ ($\sim 1.22 ~\Msun $~yr$^{-1} $~kpc$^{-2}$).
\end{enumerate}

In our earlier paper \citep{cignoni09}, we calculated an SFR for the observed optical population of $3 - 7 \times 10^{-4}~\Msun $~yr$^{-1}$.  This estimate considers the mass range $0.45-120 ~\Msun$ and covers only the {\it ACS} FOV (the white square in Figure~\ref{FOV}), which has a physical size of 58~pc~$\times$~58~pc ($3.36 \times 10^{-3}$~kpc$^2$).  This results in an optical SFR of $0.089-0.208~ \Msun $~yr$^{-1} $~kpc$^{-2}$, assuming star formation has been constant over the past 2.5~Myr.  In order to directly compare this optical estimate with our calculated YSO SFR, we integrate the IR MF over the same $0.45-120~\Msun$ mass range and obtain a total YSO mass of $\sim 1300~\Msun$.  This results in a YSO SFR of $0.14-0.71~\Msun$~yr$^{-1} $~kpc$^{-2}$.  Comparing the optical value from \citet{cignoni09} (as well as with the similar optical mass function determined by \citet{schmalzl08}) with this IR calculation implies that the  region's SFR has been approximately constant from $\sim2.5$~Myr ago to the present, possibly with a slight increase in the last 0.5~Myr.

As a point of comparison, the Orion Nebula is of a similar physical size and age to NGC~602.  Orion's physical extent is $\sim 25$~pc radius \citep[cf.][]{hillenbrand97}; in our current study, we consider an area of radius $\sim 53$~pc around NGC~602, more than encompassing the main cluster.  We estimate NGC~602 young stellar populations to have ages in the range $\sim 10^5$~yr to a few~$\times 10^6$~yr.  \citet{hillenbrand97} quote exactly the same age range for Orion and see a similar PMS population.  In the same paper, the SFR is calculated as $> 10^{-4} ~\Msun $~yr$^{-1}$ over the central active star forming area of the Orion Nebula, considering optical sources down to a mass of $\sim 1~ \Msun$ within a FOV of $\sim 36\arcmin \times 34\arcmin$.  If we assume a distance of $\sim 400$~pc \citep[e.g.,][]{hirota07}, the physical area is $\sim 4$~pc$~ \times ~4$~pc or $1.6 \times 10^{-5}$~kpc$^2$, and the SFR is $\sim 6.25 ~ \Msun $yr$^{-1} $kpc$^{-2}$.  Our SFR of $0.14-0.71~\Msun $~yr$^{-1} $~kpc$^{-2}$ for NGC~602 is a factor of ten to fifty less than that given for Orion, unsurprising considering the inclusion of the more diffuse star-forming environs of NGC~602's outskirts and the higher molecular gas densities in Orion.

\section{SPATIAL AND TEMPORAL DISTRIBUTION OF YOUNG STELLAR POPULATIONS}
\label{SpaceTime}

\subsection{Spatial Distribution of Optical and Infrared Populations}
\label{gridsect}

\begin{figure}
\centering
\subfigure[]{\label{isoCMD}
        \includegraphics[width=0.48\textwidth]{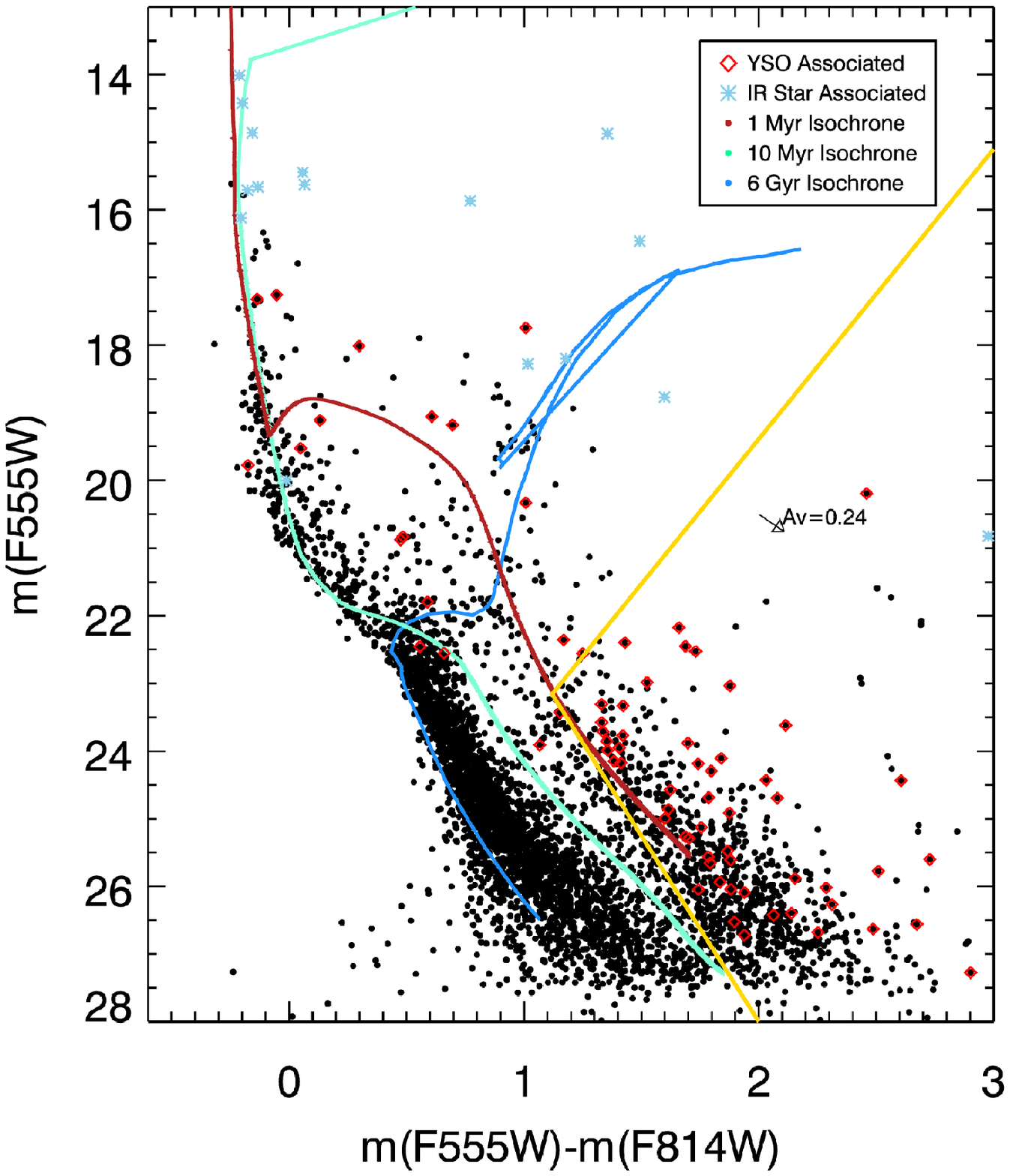}}
\hspace{0.01in}
\subfigure[]{\label{massCMD}
        \includegraphics[width=0.48\textwidth]{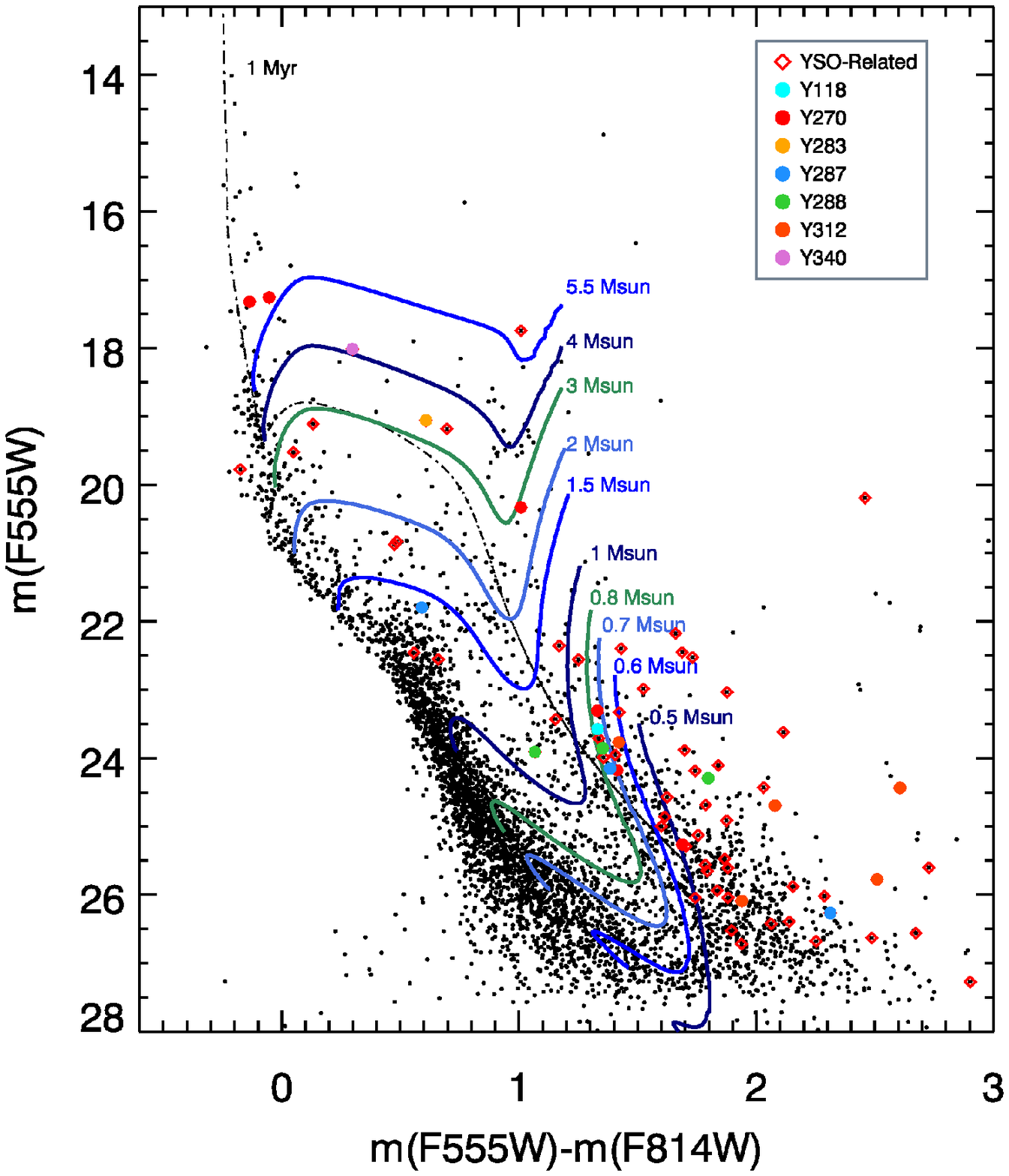}}
\caption{\label{OptCMD} (a) Optical Color-Magnitude Diagram of the entire {\it ACS} field of view, using apparent magnitudes in the {\it HST/ACS/WFC} bands F555W and F814W (roughly {\it V} and {\it I}, respectively).  The PMS lies to the right (red) of the lower MS and is a few times 10$^5$~yr.  We also mark optical sources associated with our stellar and proto-stellar fitter sources.  Blue asterisks are spatially related to IR sources fit as (non-YSO) stars.  Red diamonds are spatially related to fit YSO sources; many ``YSOs" encompass multiple optical sources, mostly PMS.  The 1 and 10 Myr FRANEC isochrones from \citet[ $Z=0.004$; ][]{cignoni09}, based on \citet{siess00}, are shown as representative age boundaries.  The 6 Gyr isochrone is based on \citet{bertelli94} with $Z=0.001$.  The yellow line shows Equation~\ref{pmseq}; points to the right are PMS candidates.  The arrow indicates the reddening vector for $A_V = 0.24$ ($E(B-V)=0.08$).  (b) CMD used to determine masses of YSOs' optical counterparts.  Blues show the evolutionary tracks for various stellar masses from 0.5 to 5.5~$\Msun$.  The dashed black line shows the 1~Myr isochrone.  Colored points are related to fit YSO candidates.}
\end{figure}

\begin{figure}
\centering
\subfigure[]{\label{iracCMD}
        \includegraphics[width=0.5\textwidth]{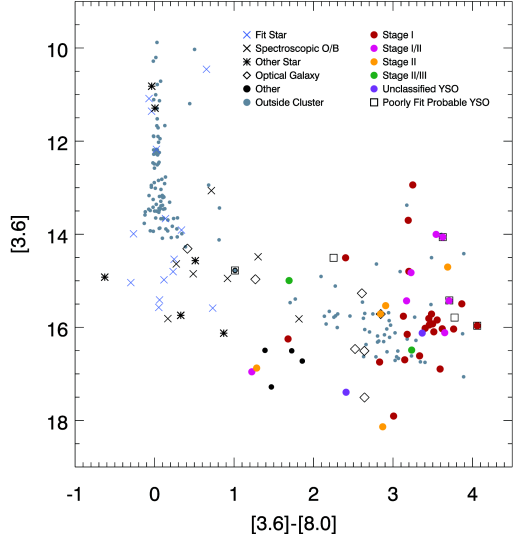}}
\hspace{0.1in}
\subfigure[]{\label{iracCCD}
        \includegraphics[width=0.5\textwidth]{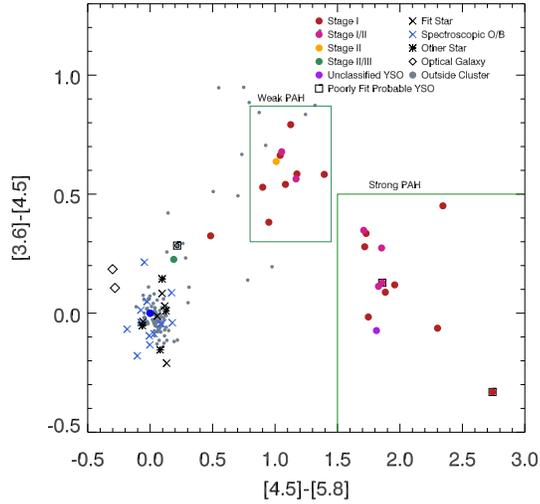}}
\caption{\label{iracc}  (a) IRAC CMD with object types.  Colored points are for the 3\arcmin\ cluster region with YSO candidates marked as dots, color-coded by evolutionary stage.  
Black dots are sources of unknown nature within the 3$\arcmin$ region.  Grey points are for sources in an area of $18\arcmin \times 18\arcmin$ outside the main $3\arcmin$ radius cluster region. (b)  Color-Color diagram with object types and PAH color-color space indicated.  Boxes outline empirically determined regions with significant PAH contribution by selecting sources that are both relatively blue in [3.6]-[4.5] and relatively red in [4.5]-[5.8].  The most prominent 4.5~$\mu$m dip in the SEDs correspond to the color-color region $([3.6]-[4.5]) < 0.5$ and $1.5 < ([4.5]-[5.8])$.  Sources with shallow 4.5~$\mu$m SED dips (slight PAH contribution) are found in the color region $0.3 < ([3.6]-[4.5]) < 0.8$ and $0.8 < ([4.5]-[5.8]) < 1.5$.  (For clarity, we have allowed a slight space between the two boxes on the plot.)}
\end{figure}

\begin{figure}
\centering
\subfigure[]{\label{mini90}
        \includegraphics[width=0.15\textwidth]{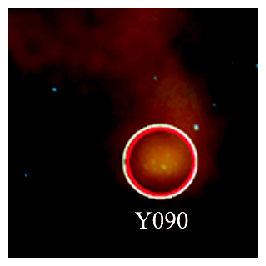}}
\hspace{0.2in}
\subfigure[]{\label{mini96}
        \includegraphics[width=0.15\textwidth]{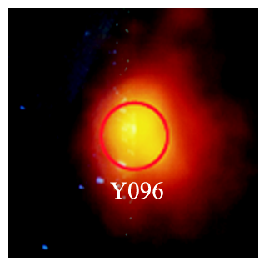}}
\hspace{0.2in}
\subfigure[]{\label{mini118}
        \includegraphics[width=0.15\textwidth]{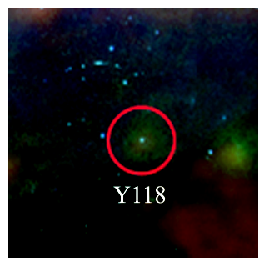}}
\hspace{0.2in}
\subfigure[]{\label{mini142}
        \includegraphics[width=0.15\textwidth]{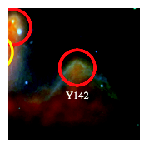}}
\hspace{0.2in}
\subfigure[]{\label{mini143}
        \includegraphics[width=0.15\textwidth]{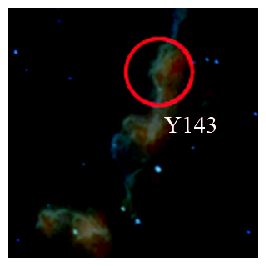}}
\vspace{0.2in}
\subfigure[]{\label{mini148}
        \includegraphics[width=0.15\textwidth]{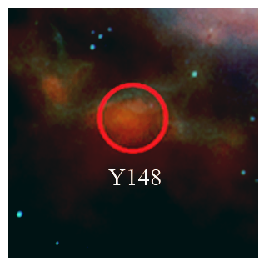}}
\hspace{0.2in}
\subfigure[]{\label{mini149163}
        \includegraphics[width=0.15\textwidth]{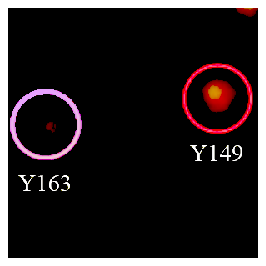}}
\hspace{0.2in}
\subfigure[]{\label{mini162170171174}
        \includegraphics[width=0.15\textwidth]{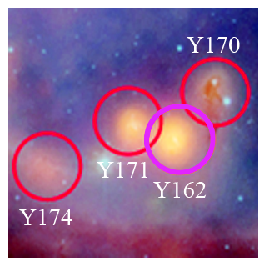}}
\hspace{0.2in}
\subfigure[]{\label{mini179196197206}
       \includegraphics[width=0.15\textwidth]{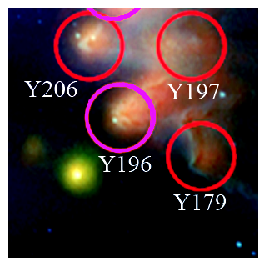}}
\hspace{0.2in}
\subfigure[]{\label{mini198}
       \includegraphics[width=0.15\textwidth]{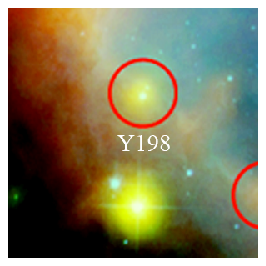}}
\vspace{0.2in}
\subfigure[]{\label{mini217237}
        \includegraphics[width=0.15\textwidth]{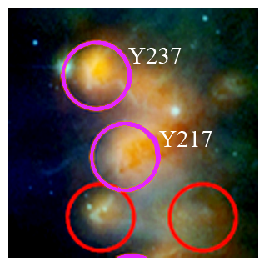}}
\hspace{0.2in}
\subfigure[]{\label{mini223227240}
        \includegraphics[width=0.15\textwidth]{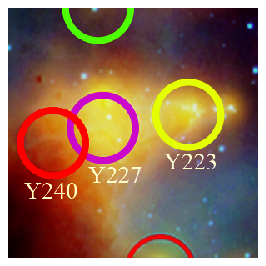}}
\hspace{0.2in}
\subfigure[]{\label{mini251255264270}
        \includegraphics[width=0.15\textwidth]{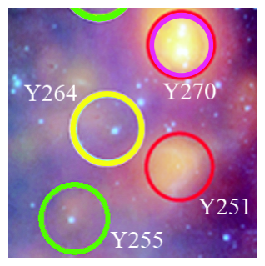}}
\hspace{0.2in}
\subfigure[]{\label{mini270271}
        \includegraphics[width=0.15\textwidth]{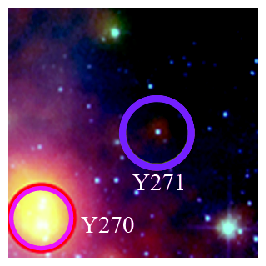}}
\hspace{0.2in}
\subfigure[]{\label{mini283}
        \includegraphics[width=0.15\textwidth]{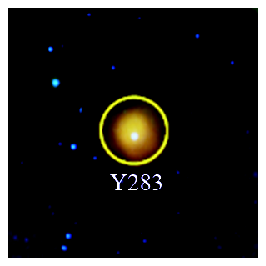}}
\vspace{0.2in}
\subfigure[]{\label{mini285}
        \includegraphics[width=0.15\textwidth]{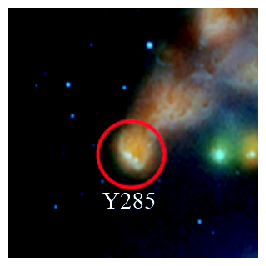}}
\hspace{0.2in}
\subfigure[]{\label{mini287288}
        \includegraphics[width=0.15\textwidth]{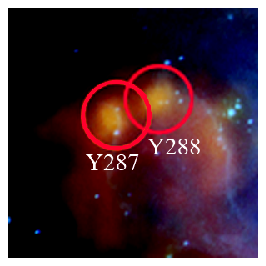}}
\hspace{0.2in}
\subfigure[]{\label{mini290312}
       \includegraphics[width=0.15\textwidth]{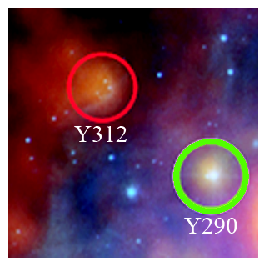}}
\hspace{0.2in}
\subfigure[]{\label{mini326}
        \includegraphics[width=0.15\textwidth]{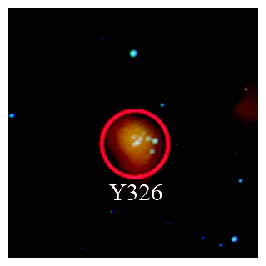}}
\hspace{0.2in}
\subfigure[]{\label{mini327}
        \includegraphics[width=0.15\textwidth]{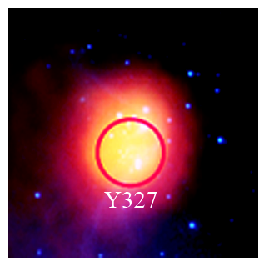}}
\vspace{0.2in}
\subfigure[]{\label{mini340}
        \includegraphics[width=0.15\textwidth]{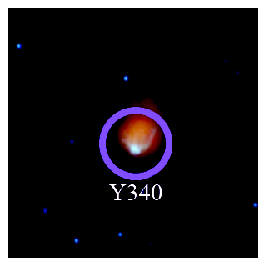}}
\hspace{0.2in}
\subfigure[]{\label{mini358}
        \includegraphics[width=0.15\textwidth]{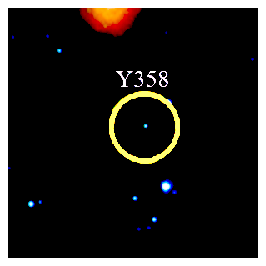}}
\hspace{0.2in}
\subfigure[]{\label{mini396}
        \includegraphics[width=0.15\textwidth]{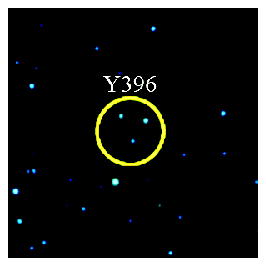}}
\hspace{0.2in}
\subfigure[]{\label{mini493}
        \includegraphics[width=0.15\textwidth]{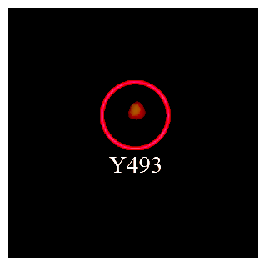}}
\hspace{0.2in}
\subfigure[]{\label{mini700}
        \includegraphics[width=0.15\textwidth]{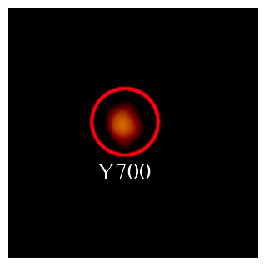}}
\caption{\label{minis} YSOs in NGC~602.  Each square image is approximately $15\arcsec$ (4.5~pc) on a side.  Circles are $1\farcs 6$ in radius, representing the IRAC photometric apertures.  Circle colors are Red (Stage {\sc i}), Orange (Stage {\sc i}/{\sc ii}), Yellow (Stage {\sc ii}), Green (Stage~{\sc ii}/{\sc iii}, Lavender (Unclassified), and White (Poorly fit).  The underlying images are 8-color: Blue= Optical (F555W[{\it V}]~+~F658N[H$\alpha$]~+~F814W[{\it I}] from {\it HST/ACS}).  Green to Yellow= Infrared IRAC  3.6~+~4.5~+~5.8~$\mu$m.  Red= IRAC 8.0~$\mu$m.  Magenta - MIPS~24$\mu m$.  The regions shown in \ref{mini149163}, \ref{mini493}, and \ref{mini700} are outside the {\it ACS} FOV.}   
\end{figure}

\begin{figure}
\centering
\subfigure[]{\label{SED142}
        \includegraphics[width=0.45\textwidth]{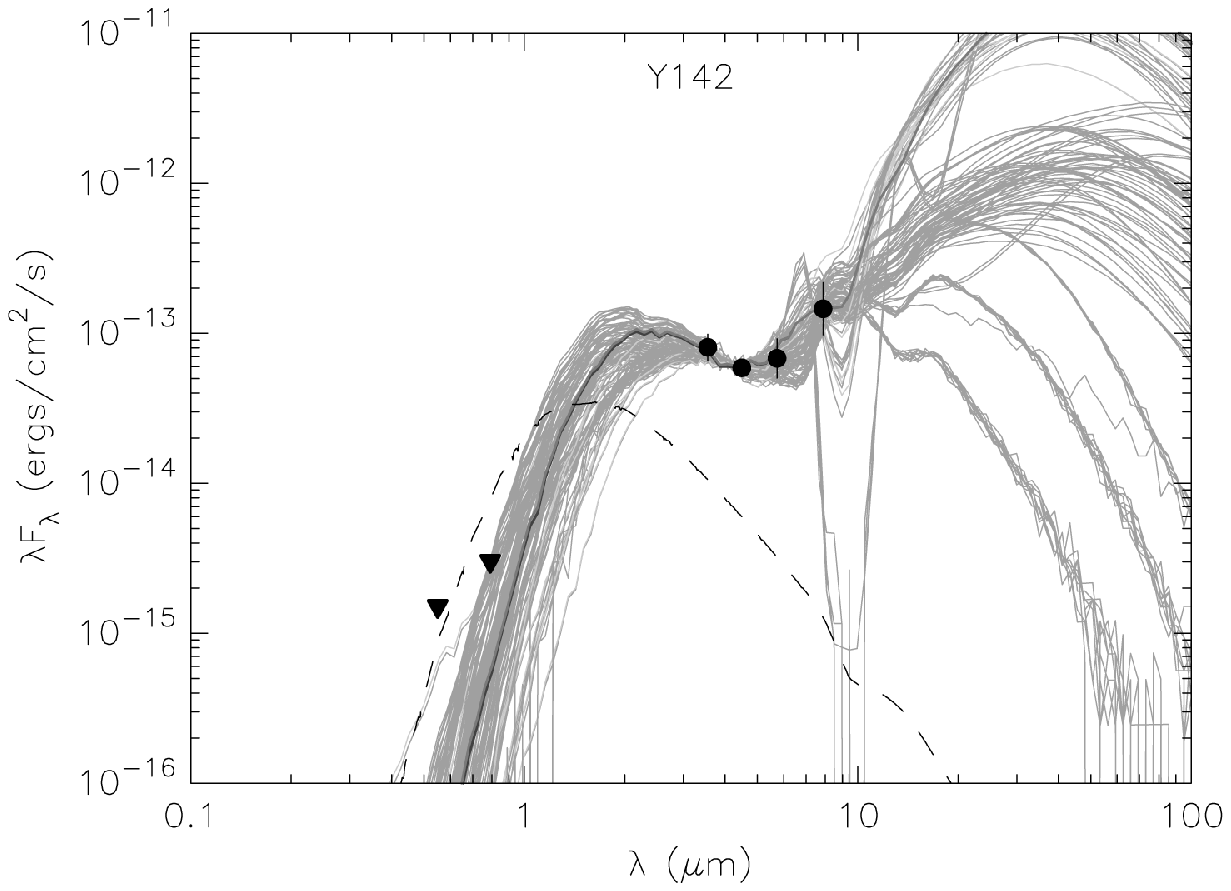}}
\hspace{0.2in}
\subfigure[]{\label{SED217}
        \includegraphics[width=0.45\textwidth]{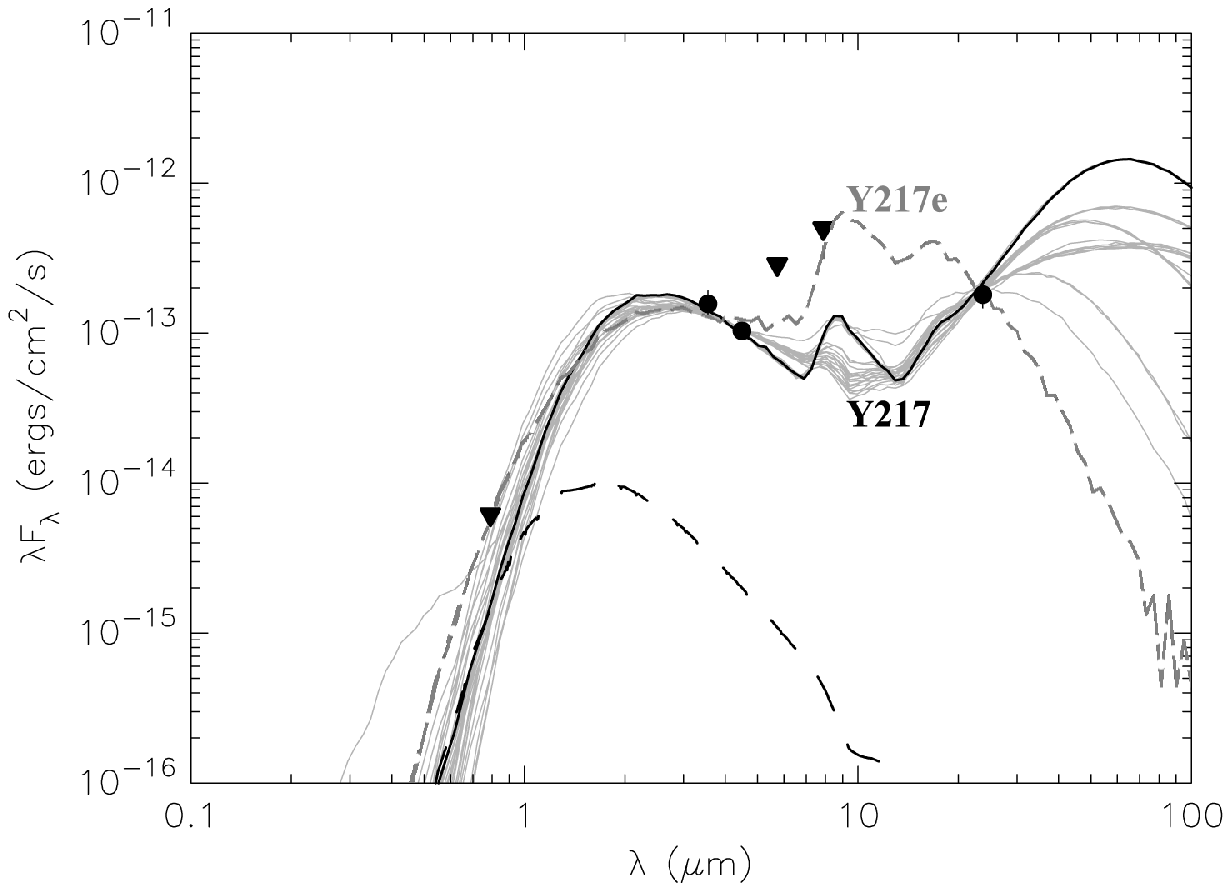}}
\caption{\label{noopt} Example SEDs for YSO candidates with no optical counterpart.  Both SEDs are fit with optical data used as upper limits.  Images of the two sources are shown in Figures~\ref{mini142} and \ref{mini217237}.  (a) SED for Stage~{\sc i} source Y142. The fit indicates a mass of $\sim 7.4 ~\Msun$.  With no 24~$\mu$m measurement, the fit is poorly constrained.  (b) SED of Stage~{\sc i}/{\sc ii} source Y217 with mass $\sim 5.7 ~\Msun$.  The deep dip at 4.5~$\mu$m indicates significant PAH contamination, and in the presence of the 24~$\mu$m measurements, 5.8 and 8.0~$\mu$m fluxes are treated as upper limits.  The dashed grey line (Y217e) shows the best fit using the error bar method of correcting for PAHs and shows quite a different shape from the fit using the upper limit method.  Filled circles - photometric data points with applied error bars.  Triangles - photometric data points used as upper limits in SED fitting as described in the text.  All fits within 0.5~$\chi^2/pt$ are shown in grey with the best fit in black.  The dashed black lines indicate the photosphere corresponding to the best fit.  All SEDs are available online.}
\end{figure}

\begin{figure}
\epsscale{0.5}
\plotone{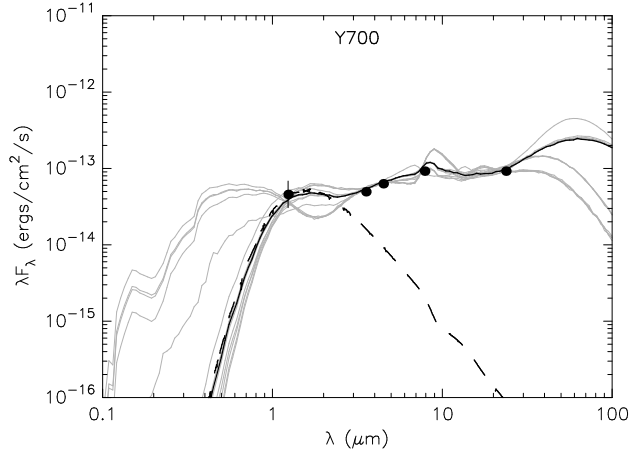}
\caption{\label{sed700} Y700 lies outside the {\it ACS} FOV along a ridge of dust east of the main clusters, highlighted in 8.0~$\mu$m in Figure~\ref{8col}.  It is fit as a Stage {\sc i} source with a flat SED.  Based on this fit, the source's mass is $\sim 4.3 ~\Msun$.  See also Figure~\ref{mini700}.  Symbols as in Figure~\ref{noopt}.}
\end{figure}

\begin{figure}
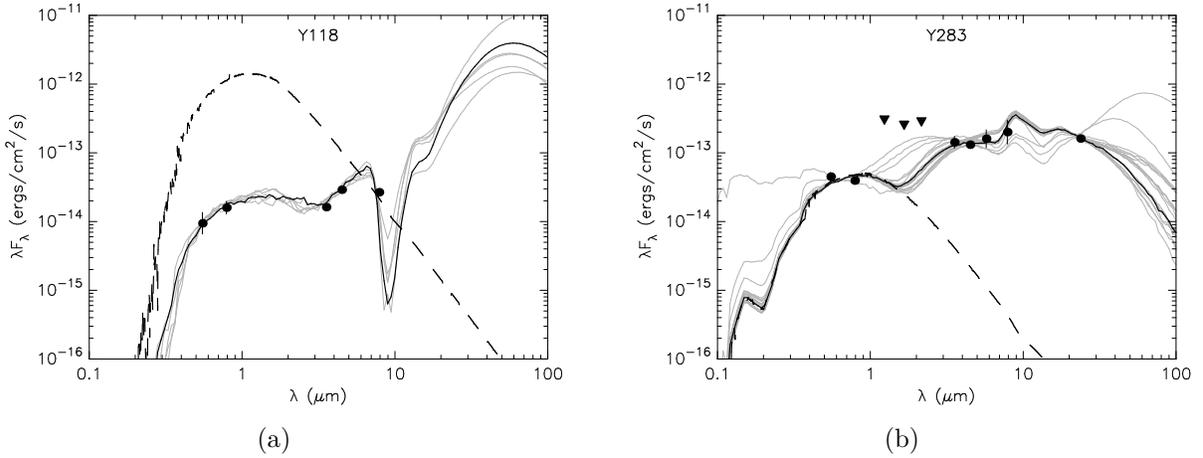

\centering
\subfigure[]{\label{SED118}
        \includegraphics[width=0.45\textwidth]{f8a_g.eps}}
\hspace{0.2in}
\subfigure[]{\label{SED283}
        \includegraphics[width=0.45\textwidth]{f8b_g.eps}}
\caption{\label{oneopt} Example SEDs for YSO candidates with single optical counterparts.  Images are Figures~\ref{mini118} and \ref{mini283}.  (a) Y118 a relatively isolated Stage~{\sc i} source with a mass estimate of $\sim 7.4 ~\Msun$ based on this SED but approximately corresponding to a single PMS star (cyan in Figure~\ref{massCMD}) with mass $\sim 0.8 ~\Msun$.  (b) Y283 is a bright Stage~{\sc ii} source of mass $\sim 4.5 ~\Msun$; it is plotted in yellow in Figure~\ref{massCMD}.  Symbols as in Figure~\ref{noopt}.}
\end{figure}

\begin{figure}
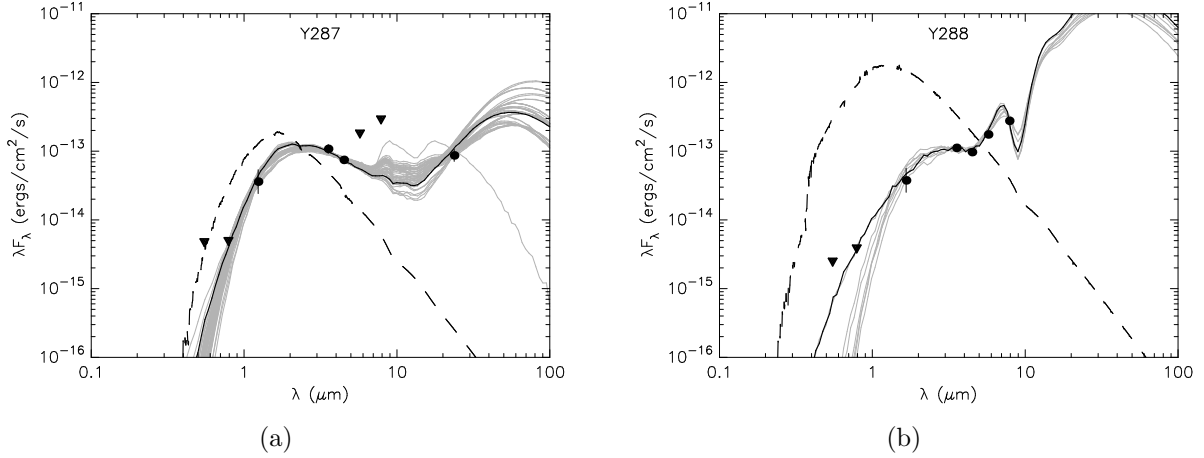

\centering
\subfigure[]{\label{SED287}
        \includegraphics[width=0.45\textwidth]{f9a_g.eps}}
\hspace{0.2in}
\subfigure[]{\label{SED288}
        \includegraphics[width=0.45\textwidth]{f9b_g.eps}}
\caption{\label{seds28788} Example SEDs for YSO candidates with multiple optical counterparts.  Y287 and Y288 are fit as Stage~{\sc ii} and {\sc i}, respectively, and lie near each other on the plane of the sky.  In Figure~\ref{mini287288}, they appear to lie along adjacent molecular ridges.  Corresponding optical sources are marked in green in Figure~\ref{massCMD}.  Symbols as in Figure~\ref{noopt}.} 
\end{figure}

\begin{figure}
\centering
\subfigure[]{\label{SED312}
        \includegraphics[width=0.5\textwidth]{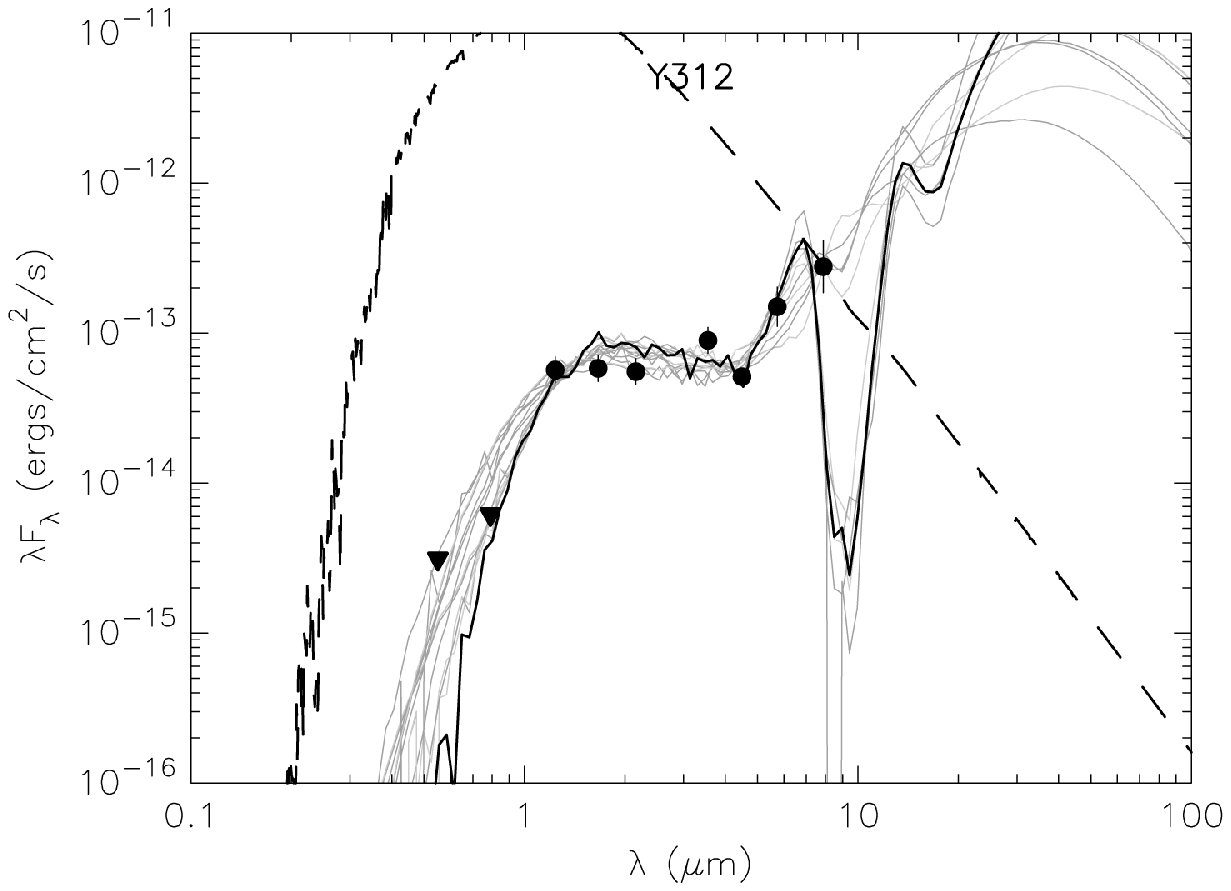}}
\hspace{0.2in}
\subfigure[]{\label{312_ACS}
        \includegraphics[width=0.25\textwidth]{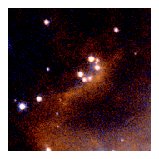}}
\caption{\label{sed312} Y312 appears as an arc of Pre-Main Sequence stars (also Figure~\ref{mini290312}, five orange optical sources in CMD Figure~\ref{massCMD}) fit as a Stage {\sc i} source.  %The deep dip around 10~$\mu$m indicates silicate absorption.  
SED symbols as in Figure~\ref{noopt}.}
\end{figure}

\begin{figure}
\centering
\subfigure[Optical ACS: F555W + F658N + F814W]{\label{340ACS}
        \includegraphics[width=0.25\textwidth]{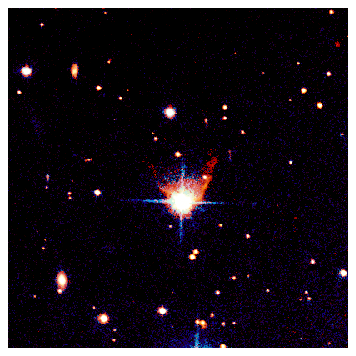}}
\hspace{0.2in}
\subfigure[ACS: H$\alpha$]{\label{340Halpha}
        \includegraphics[width=0.25\textwidth]{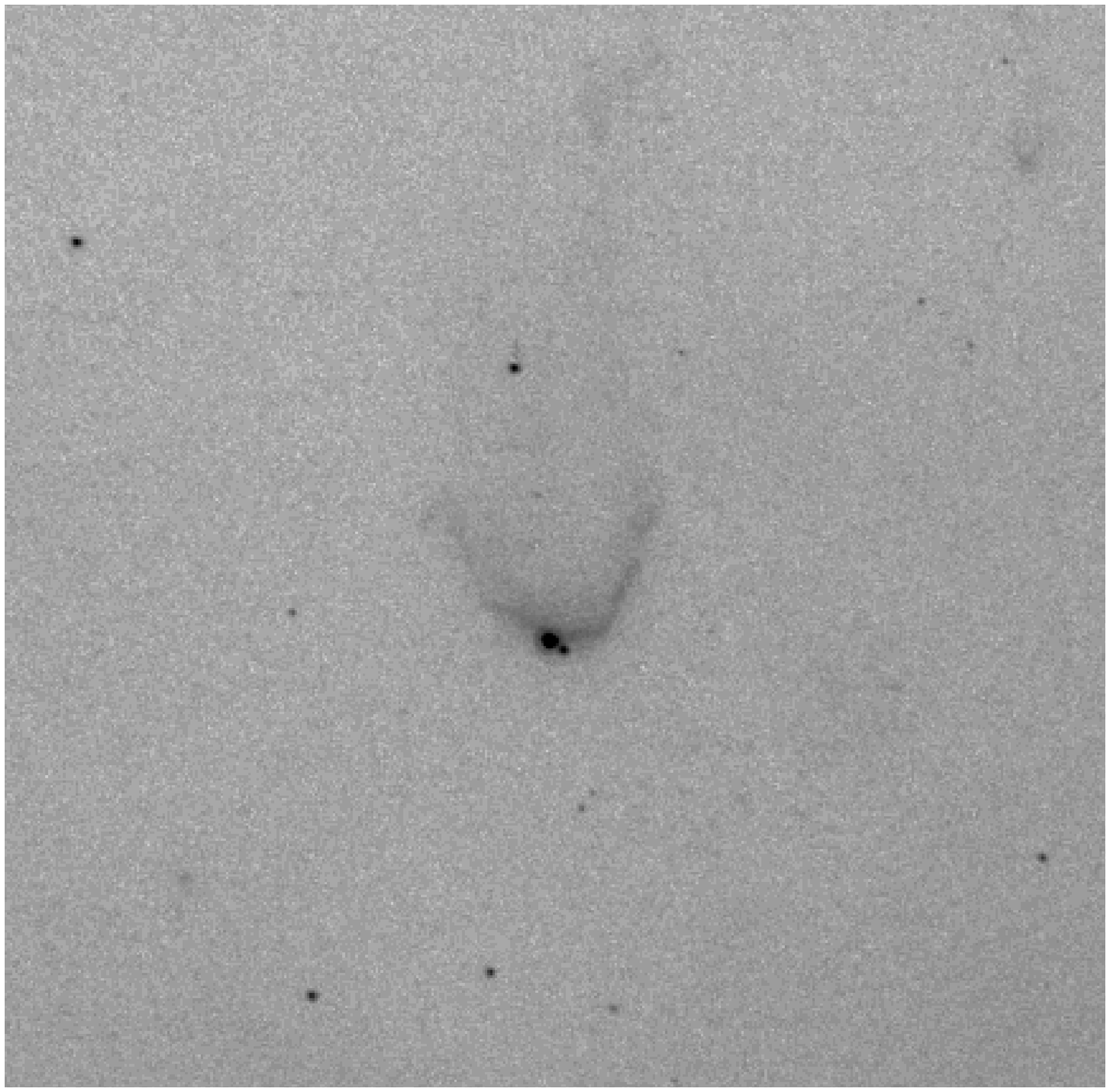}}
\hspace{0.2in}
\subfigure[Eight-color]{\label{340_8col}
        \includegraphics[width=0.25\textwidth]{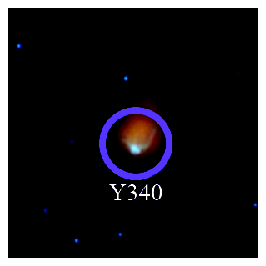}} 
\vspace{0.15in}
\subfigure[Optical stellar fit]{\label{SED_K340}
        \includegraphics[width=0.4\textwidth]{f11d_g.eps}}
\hspace{0.2in}
\subfigure[IR YSO fit]{\label{SED_Y340}
        \includegraphics[width=0.4\textwidth]{f11e_g.eps}}
\vspace{0.15in}
\subfigure[All points fit]{\label{SED_A340}
        \includegraphics[width=0.4\textwidth]{f11f_g.eps}}
\hspace{0.2in}
\subfigure[Fits overlaid]{\label{SED_C340}
        \includegraphics[width=0.4\textwidth]{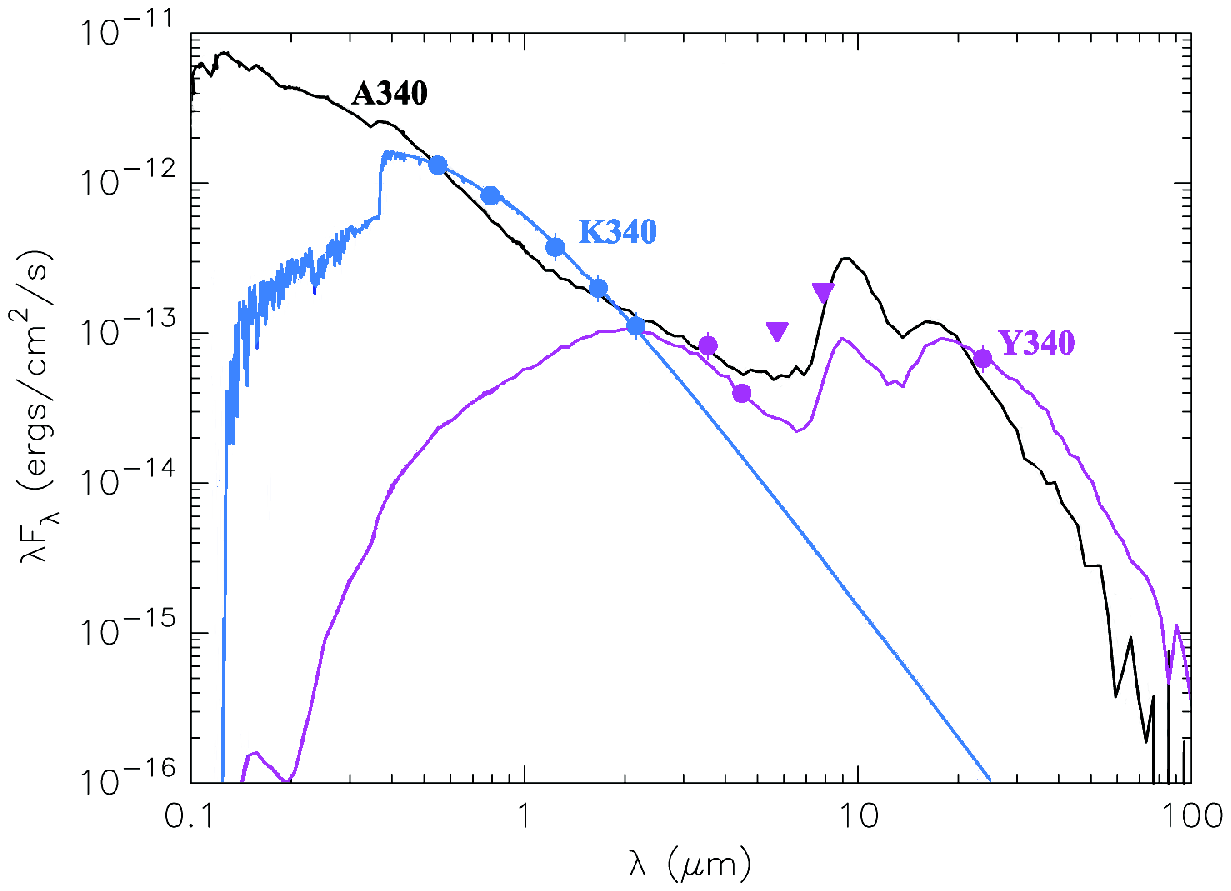}}
\caption{\label{340set} \small{Source 340.  An optically bright source dominates (a) the three-color optical image, but (b) the narrow band H$\alpha$ reveals a second, fainter optical point source.  (c) In eight colors, the infrared source looks orange (strong emission in IRAC bands); the approximate photometric aperture is shown.  Images (a), (b), and (c) are approximately 22\arcsec\ on each side.  The three black and white SEDs are (d) the K340 stellar optical fit, (e) the Y340 IR fit, and (f) the A340 fit to all photometric data.  (g) The color SED shows the three fits overlaid together with A340 in black, K340 in blue, and Y340 in violet.  SED symbols as in Figure~\ref{noopt}.}}
\end{figure}

\begin{figure}
\centering
\subfigure[Optical ACS: F555W + F658N + F814W]{\label{270ACS}
        \includegraphics[width=0.25\textwidth]{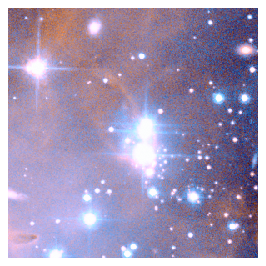}}
\hspace{0.2in}
\subfigure[ACS: H$\alpha$]{\label{270Halpha}
        \includegraphics[width=0.25\textwidth]{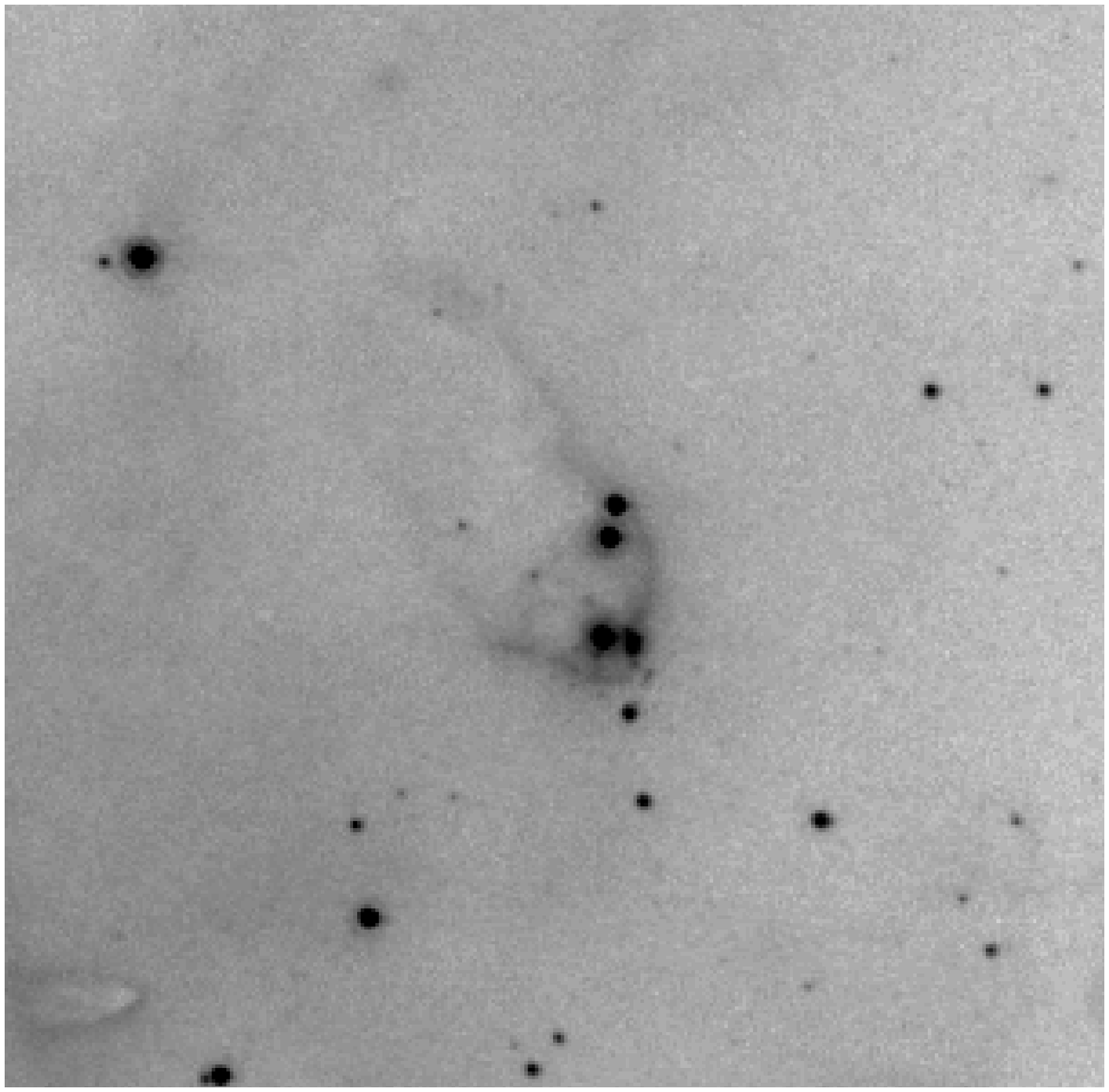}}
\hspace{0.2in}
\subfigure[Eight-color]{\label{270_8col}
        \includegraphics[width=0.25\textwidth]{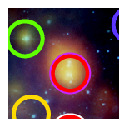}} 
\vspace{0.15in}
\subfigure[Optical YSO fit]{\label{SED_Y270o}
        \includegraphics[width=0.4\textwidth]{f12d_g.eps}}
\hspace{0.2in}
\subfigure[IR YSO fit]{\label{SED_Y270i}
        \includegraphics[width=0.4\textwidth]{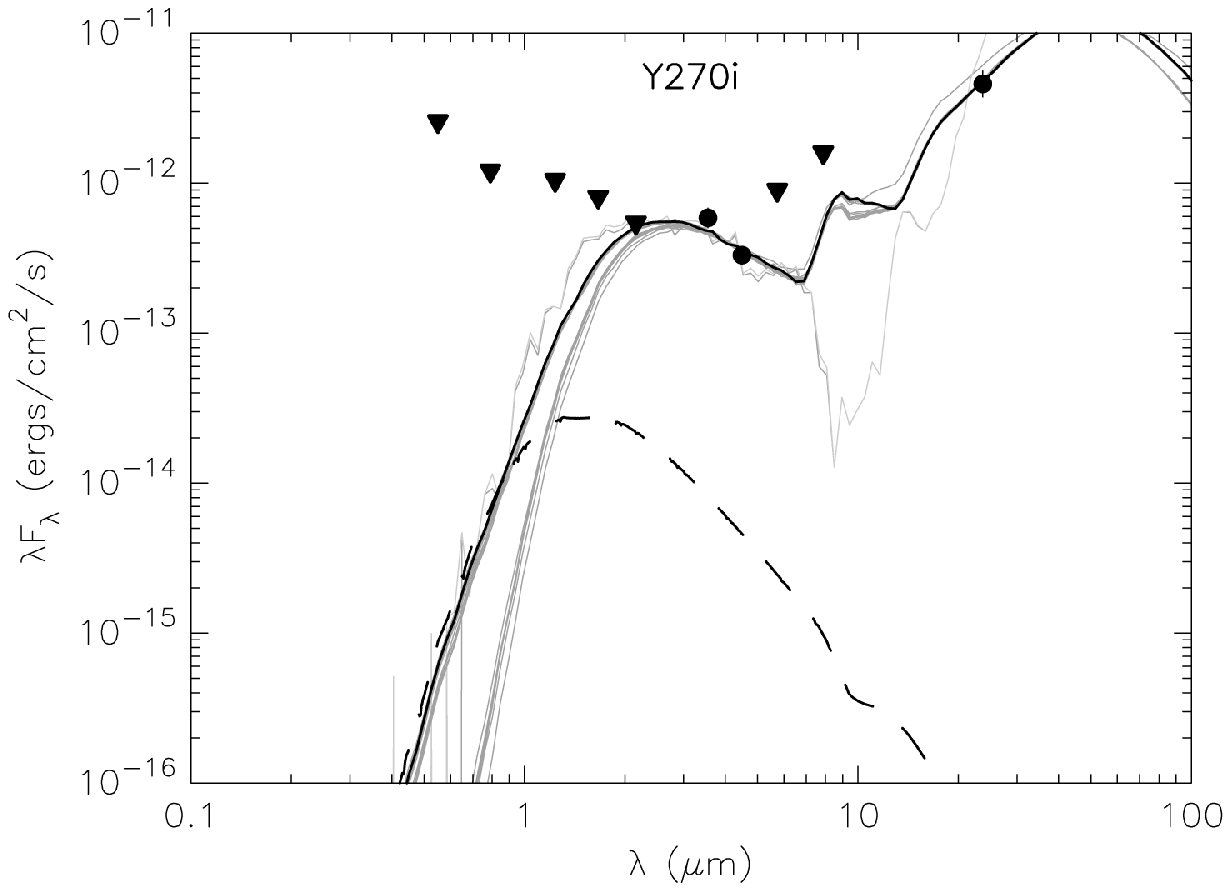}}
\vspace{0.15in}
\subfigure[All points fit]{\label{SED_A270}
        \includegraphics[width=0.4\textwidth]{f12f_g.eps}}
\hspace{0.2in}
\subfigure[Fits overlaid]{\label{SED_C270}
        \includegraphics[width=0.4\textwidth]{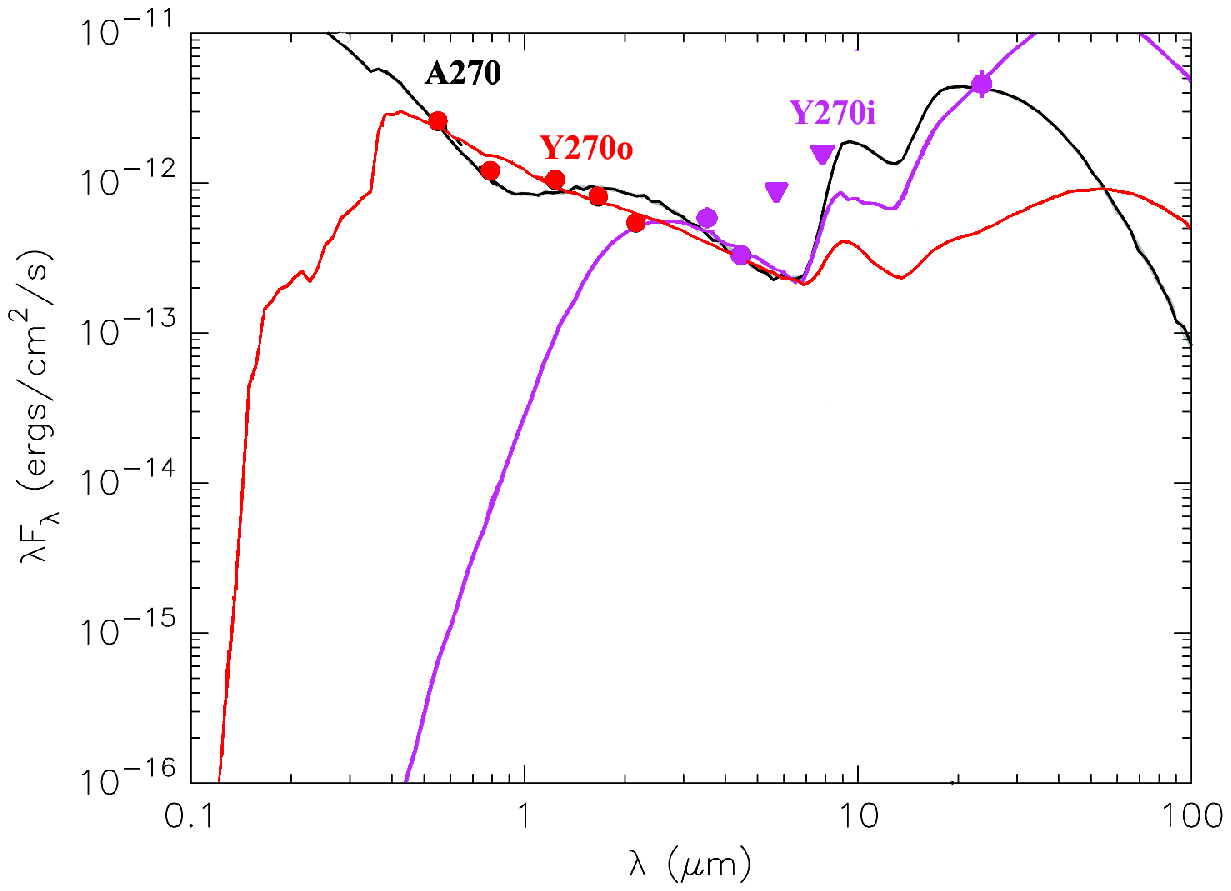}}
\caption{\label{270set} Images and SEDs for source 270.  Images (a), (b), and (c) are approximately 15\arcsec\ on each side.  The brightest optical source is a B2 star \citep{hutchings91}.  We fit the photometric data best as two Stage~{\sc i} YSOs.  The photometric optical sources are plotted in red in Figure~\ref{massCMD}.  SED symbols as in Figure~\ref{noopt}.} 
\end{figure}

\begin{figure}
\centering
\subfigure[]{\label{SEDK348}
        \includegraphics[width=0.4\textwidth]{f13a_g.eps}}
\hspace{0.2in}
\subfigure[]{\label{SEDS235}
        \includegraphics[width=0.4\textwidth]{f13b_g.eps}}
\vspace{0.2in}
\subfigure[]{\label{K348_8col}
        \includegraphics[width=0.25\textwidth]{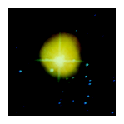}}
\hspace{0.2in}
\subfigure[]{\label{S235_8col}
        \includegraphics[width=0.25\textwidth]{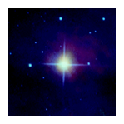}}
\caption{\label{starseds} Two stellar SEDs.  K348 is a well-fit naked star.  According to \citet{hutchings91} S235 is an O6 star, optically the brightest in NGC~602.  Our data is well fit as a naked star {\it if} we ignore the MIPS 24~$\mu$m flux, which is likely the result of the O star heating the nearby interstellar medium.  Symbols as in Figure~\ref{noopt}.}
\end{figure}

\begin{figure}
\centering
\subfigure[]{\label{SEDG133}
        \includegraphics[width=0.4\textwidth]{f14a_g.eps}}
\hspace{0.2in}
\subfigure[]{\label{SEDG372}
        \includegraphics[width=0.4\textwidth]{f14b_g.eps}}
\vspace{0.2in}
\subfigure[]{\label{G133ACS}
        \includegraphics[width=0.25\textwidth]{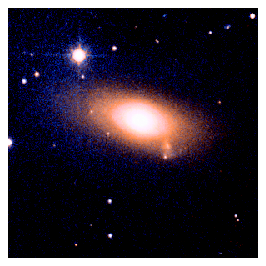}}
\hspace{0.2in}
\subfigure[]{\label{G372ACS}
        \includegraphics[width=0.25\textwidth]{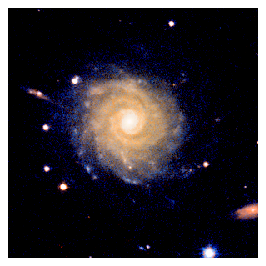}}
\caption{\label{galseds}  Example SEDs for known galaxies.  G133 is a readily identifiable as a bright elliptical and appears bright green in the west of Figure~\ref{8col}.  It is fit best (though poorly fit) by AGB templates; here, we show the best fit galaxy template.  G372 is the face-on grand design spiral just north-east of the main nebula in Figure~\ref{8col}.  It is best (though not well) fit by YSO models; we show the best galaxy template.  The other 6 galaxies in our 77 fitter sources are all well-fit as YSOs.  None of these galaxies are well-fit by galaxy templates currently incorporated in the fitter.  All galaxy SEDs in online figures show the best-fit galaxy templates.  Symbols as in Figure~\ref{noopt}.}  
\end{figure}

\begin{figure}
\plotone{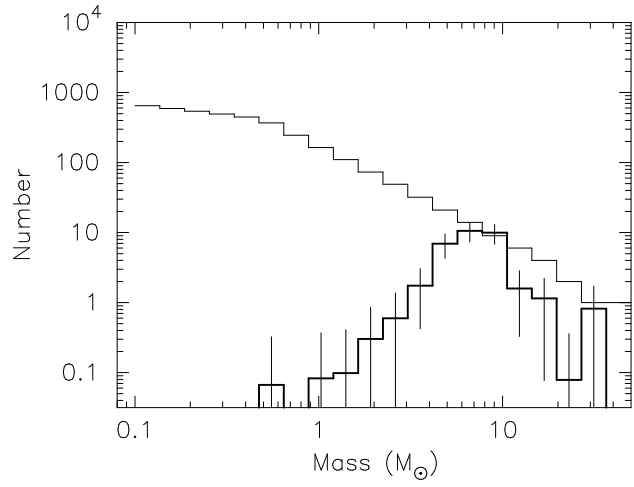}
\caption{\label{mf_plot} Mass Function of YSO candidates.  The total mass of the 41 fit YSO candidates is $\sim 300~ \Msun$.  The total integrated mass, assuming a two-part mass function \citep[grey line;][]{kroupa01} from 0.08 to $50~\Msun$, is $\sim 2250 ~\Msun$.}
\end{figure}

Inspection of the optical CMD (Figure~\ref{OptCMD}) together with our analysis of combined optical and infrared data reveals that different stellar populations coexist in the area:  

\begin{enumerate}
\item{\it Young Stellar Objects.}  YSOs are the least evolved stellar sources.  We identify Stage~{\sc i} and Stage~{\sc ii} sources with masses $2.5 \la {\rm M}/\Msun \la 13 $ (except for Y206 with ${\rm M}/\Msun \sim 26$).  They are heavily concentrated in the periphery of the cluster, along dusty ridges.  Most correspond to multiple optical sources with ages less than 1~Myr.
\item{\it Pre-Main Sequence Stars.} Low mass ($0.6< {\rm M}/\Msun<3 $) PMS represent the most remarkable feature in the optical CMD.  They are characterized by red colors ($m_{\rm F555W}-m_{\rm F814W}\ga 1.0$) and faint magnitudes ($m_{\rm F555W}\ga 22.3$).  Their ages are generally less than $\sim$5~Myr, certainly less than 10~Myr.  They are concentrated near cluster center, but clumps also appear in the dusty outskirts of the cluster, corresponding to YSO candidates.  A total of 494 PMS candidates are included in our final optical photometry list (Table~\ref{t:optphot}).  
\item {\it Young Stars.} A bright ($14.5\la m_{\rm F555W}\la 22.3$), blue ($m_{\rm F555W}-m_{\rm F814W}\la 0.3$) and well-defined MS is visible in the upper left of the optical CMD. The majority of these stars belong to NGC~602, and the population is strongly concentrated toward the center of the cluster.
\item{\it Old stars.} Old SMC wing stars populate the lower part of the MS ($m_{\rm F555W}\ga 22.0$).  A few red giant branch (RGB) and red clump stars are visible in the upper right of the CMD.  These are primarily a background population.  They are most visible around the edges of our analyzed region, outside the nebula.  Our work in \citet{cignoni09} indicates that this population is consistent with the nearby field population and at a larger distance modulus than the young population.
\end{enumerate}

\begin{figure}
\epsscale{1.0}
\plotone{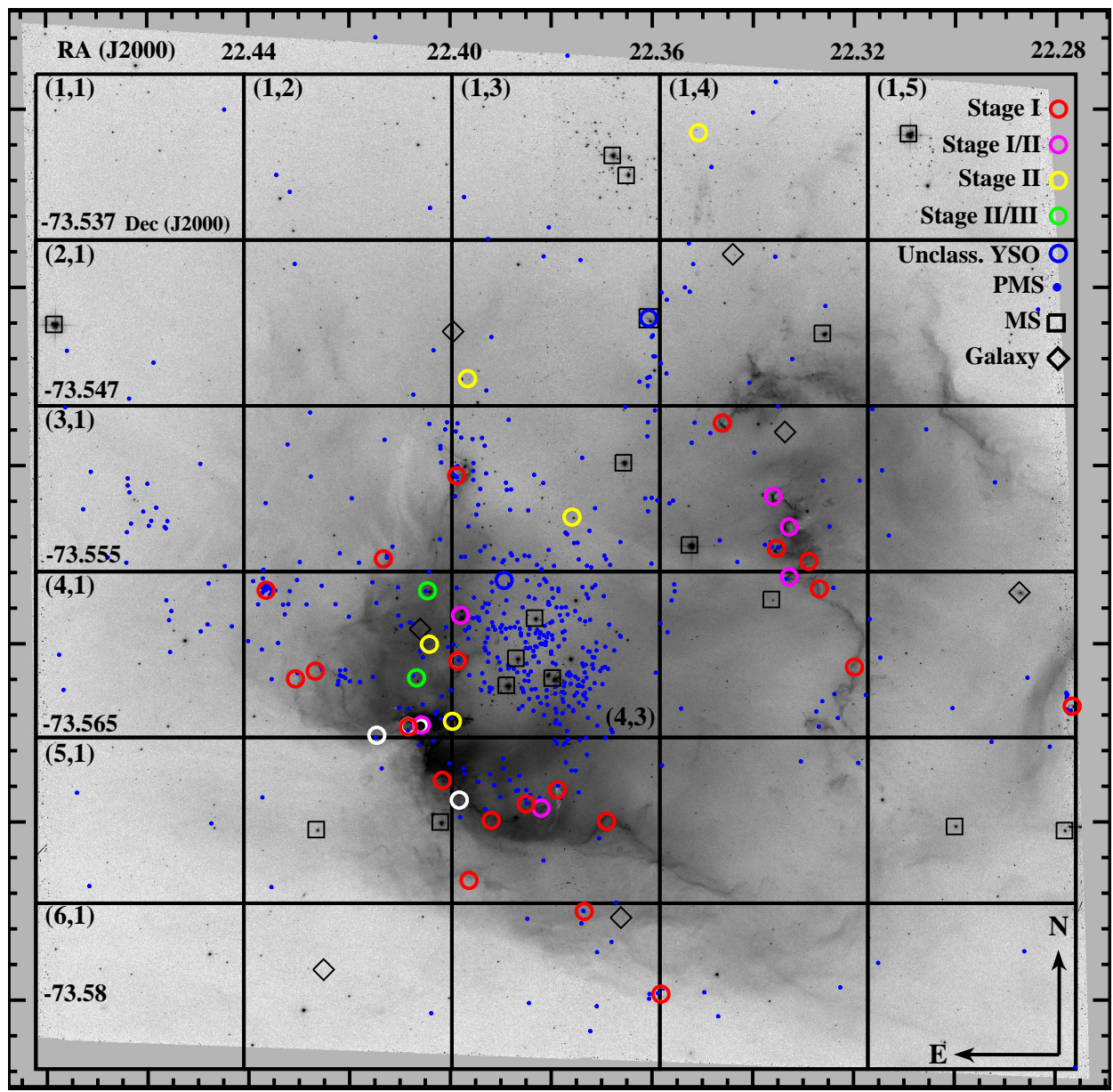}
\caption{\label{dist} Distribution of young sources on an inverted F658N (H$\alpha$) {\it ACS} image.  Circles indicate IRAC YSO candidates and are shown with the IRAC photometric aperture of $1\farcs 6$ radius and are color-coded by object type as in legend.  White circles indicate probable YSOs (based on their IR colors and environments) that do not have enough bands for SED fits.  Sources for which we can estimate stages but which are not well-fit are indicated with white in addition to their appropriate stage colors.  Boxes indicate fit stellar sources and O/B stars from \citet{hutchings91}.  All 565 PMS candidates identified in the optical are plotted.}  
\end{figure}

\begin{figure}
\epsscale{1.0}
\plotone{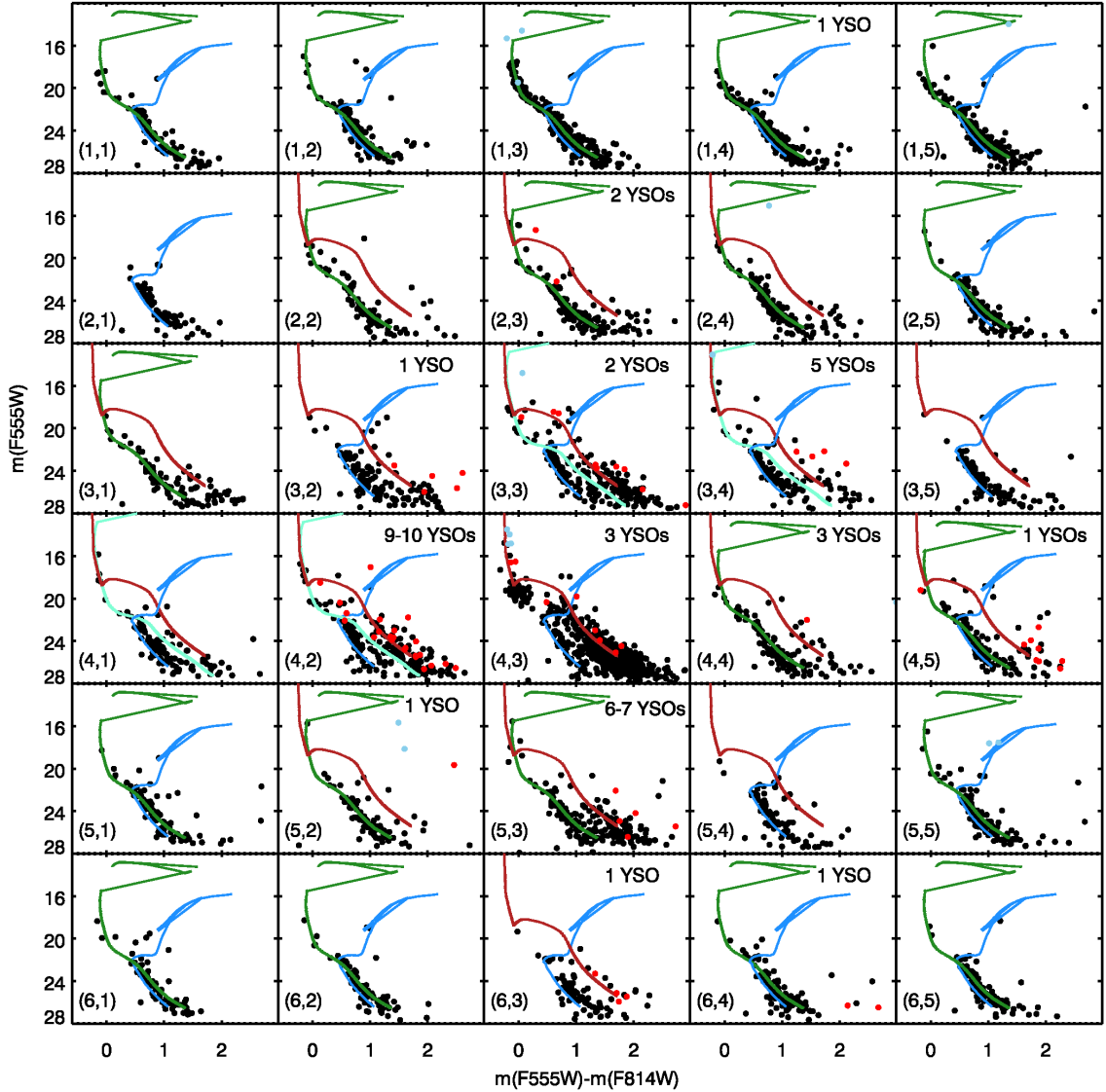}
\caption{\label{gridcmd}  Optical CMDs for grid sections in Figure~\ref{dist}.  The youngest optical sources (both PMS and upper MS) are concentrated near cluster center.  YSOs are concentrated farther from cluster center, especially along the ``broken ring" of dusty ridges.  The older population is more diffuse and appears to be largely background.  The number of \spit-identified YSO candidates within each grid region is noted.  Optical sources related to YSOs are marked in red; optical sources corresponding to fit IR stars are marked in blue.  Isochrones are for 1Myr (dark red), 10 Myr (light green in sections (3,3), (3,4), (4,1), and (4,2)), 50 Myr (true green), and 6 Gyr (blue).}
\end{figure}

Figure~\ref{dist} outlines the spatial distribution of these stellar populations with a grid on the {\it ACS} H$\alpha$.  PMS stars are marked as blue dots and YSOs as circles with colors indicating their determined stages.  Stellar sources are marked as squares.  Background galaxies are marked as diamonds.  Figure~\ref{gridcmd} shows optical CMDs corresponding to each of the grid regions delineated in Figure~\ref{dist} with isochrones from Figure~\ref{isoCMD} for reference.  The number of \spit-identified YSO candidates is noted in the top right-hand corner of each CMD.  PMS candidates associated with \spit\ YSOs are shown in red (as in Figure~\ref{isoCMD}).

The concentration of PMS stars versus YSOs is striking.  Of the candidate PMS stars, 38\% lie within grid-section (4,3).  The majority of YSO candidates lie farther from cluster center, notably along ridges in the East/South-East and West, and four YSOs (Y149, Y163, Y493, and Y700) lie outside the optical FOV.  Grid section (5,3) contains 6 of the 36 well-fit YSOs but less than 7\% of the PMS stars.  Grid-section (4,2) includes 6 well-fit YSOs (seven if  Y270o and Y270i are counted separately), 3 YSO candidates that are not well-fit, and another probable YSO with insufficient photometry for our fitting requirements.  Although these (4,2) YSOs correspond to many PMS candidates, less than 12\% of all PMS candidates fall within this grid-section.  Also notable are the evident ages of the PMS sources in the central versus outer grid-sections.  In (4,3), the PMS population aligns well with the 1~Myr isochrone (red) with some older and some younger.  Other grid-sections (e.g., (3,2), (3,3), (3,4), (4,2), (4,4), (4,5), (5,3), and (6,3)) suggest younger PMS populations that lie redward of the 1~Myr isochrone (in red).  While some of this is likely the result of increased reddening in these dustier areas, extinction is insufficient to account for the measurable trend.  (Figure~\ref{isoCMD} shows the appropriate reddening vector.)

We consider two components of the MS, old and young populations.  Many of the edge regions contain only old MS and RGBs.  Virtually all of the optical sources in grid-sections (1,3) and (1,5), which contain \citet{cignoni09} sub-clusters NGC~602-B and NGC~602-B2 \citep[alternately B~164 and Cluster~A in ][]{schmalzl08}, are old stars.  The young MS (m$_{F555W}\la 17$) population is concentrated at cluster center, particularly in grid-section (4,3) with the PMS concentration.

\subsection{Progression of Star Formation in NGC~602}
\label{ysotime}

The distances from cluster center to PMS and YSO candidates of different stages reveal a gradient.  We define the cluster center as the center of the distribution of PMS candidates (RA, Dec: 22$^\circ$.3855, -73$^\circ$.5584 or $1^h29^m32^s.5$, $-73^{\circ}33\arcmin30\farcs24$), which is approximately the center of the main O/B association.  We measure the projected distance to PMS and YSO candidates.  As a correction for the three-dimensional distance, we make the simple assumption that the actual physical distances are $\sqrt2$ times these projected distances. 

\begin{figure}
\epsscale{1.0}
\plotone{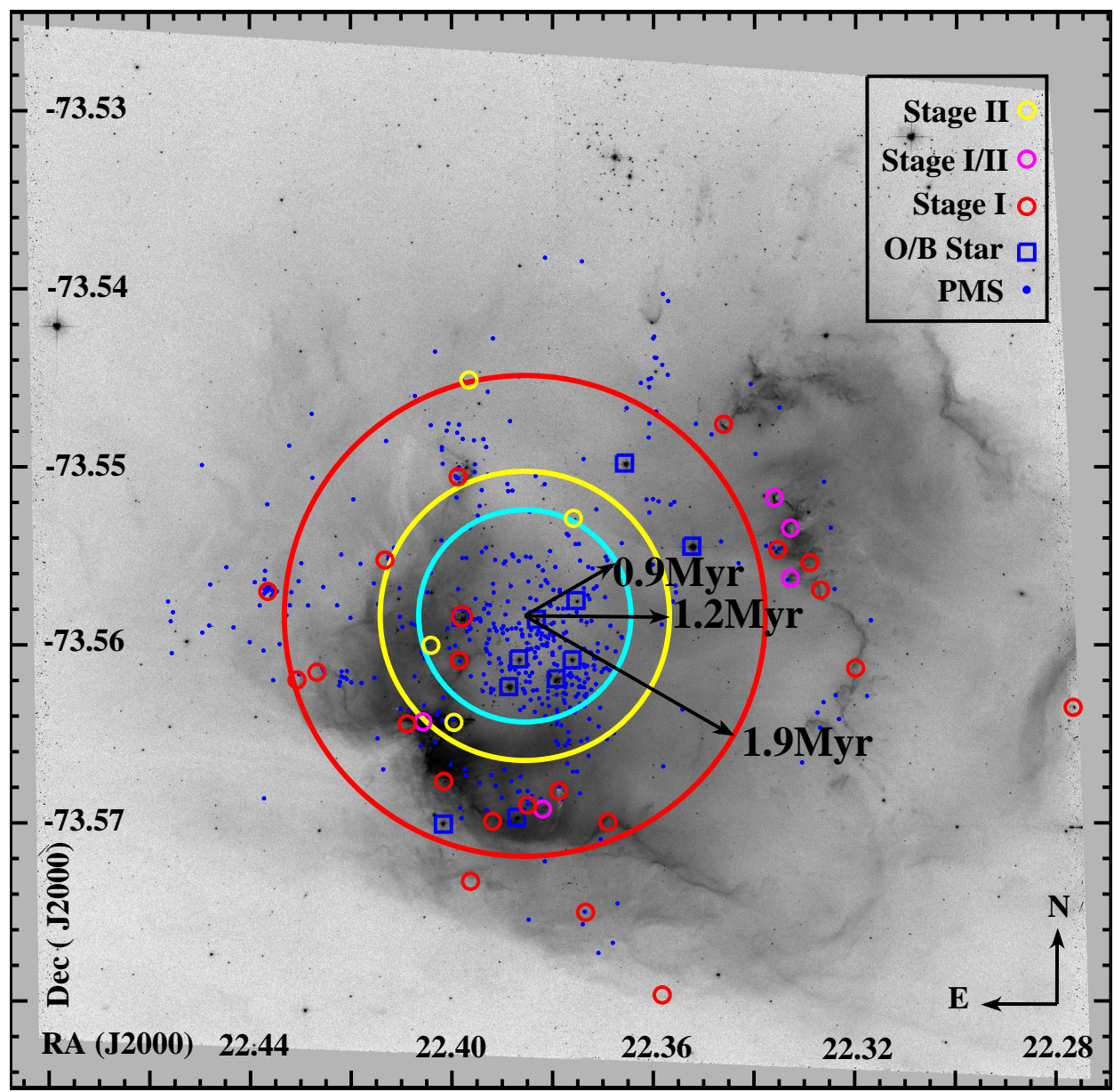}
\caption{\label{timescale} Relative timescales for triggered formation of YSO Stages.  We assume a sound speed $c_s\sim 10~$km~s$^{-1}$ and calculate the time from cluster center (the center of the PMS distribution) to the average positions of YSO Stages, multiplying by a factor $\sqrt 2$ as a partial correction from line-of-sight distance from cluster center.  The large cyan circle encloses 50\% of the PMS distribution within $\sim 30$~pc of cluster center (blue dots). The red and yellow circles show the average radial distances to Stage~{\sc i} and Stage~{\sc ii} YSOs.  Background image: {\it ACS} H$\alpha$.}
\end{figure}

We have a list of 565 PMS candidates, 484 of which lie within 30~pc of cluster center.  We consider these 484 to be strong candidates, as their positions are consistent with a common triggering process, and their clustering near the O/B association makes their nature more certain.  The primary locus of the PMS population is clear from image examination (See Figures~\ref{dist} and \ref{timescale}), and the distribution is approximately gaussian.  We define the PMS distance as the radius from cluster center that will include 50\% of these strong PMS candidates and use error bars defined by the radii including 40\% and 60\% of the strong candidates.  The result is r$_{PMS} = 9.0~\pm~_{1.3}^{2.5}$~pc.

For the average distances to YSOs, we remove sources outside the ACS FOV because their nature is less certain, and their formation may be triggered in a different manner.  Y700, for example, lies on its own semi-circular ridge of 8.0~$\mu$m emission.  The  average distances from cluster center to YSO stages are: Stage~{\sc ii} $\sim 12.0~ \pm ~5.5$~pc, Stage~{\sc i}/{\sc ii} $\sim 16.6 \pm 5.9$~pc, and Stage~{\sc i} $\sim19.4 \pm 9.2$~pc, where the error bars are the standard deviations of the mean.

We consider a simple picture of NGC~602 in which the region consists of a central O/B association surrounded by concentric shells of star formation.  We make the assumption that star formation began at cluster center (the center of the PMS distribution) approximately 3~Myr ago and is propagating outward.  This picture is shown schematically in Figure~\ref{timescale}; for simplicity, we only include shells for PMS, Stage~{\sc ii}, and Stage~{\sc i} sources.  As mentioned in Section~\ref{intro}, \citet{nigra08} have shown that the nebular expansion velocities are negligible, so collapse/star formation (beginning YSO Stage 0) is most likely being triggered along ionization fronts.  We assume a local sound speed of $\sim 10$~km/s (as appropriate for an ionized region at T$\sim 10^4$K) for the speed of star formation propagation.  Dividing the distances by this sound speed, we derive timescales for the populations.  The PMS population formed within $\sim~0.9$~Myr of the O/B association formation. The average time between the formation of the O/B association and the YSOs increase with less evolved YSO stages as follows:  Stage~{\sc ii} $\sim 1.2 \pm 0.55$~Myr,  Stage~{\sc i}/{\sc ii} $\sim 1.66~\pm 5.9$~Myr;   and Stage~{\sc i}  $\sim 1.94 \pm 0.92$~Myr.  The  Stage~{\sc i}/{\sc ii} shell forms between Stage~{\sc i} and Stage~{\sc ii}, representing a mix of Stage {\sc i} and {\sc ii} sources.  These timescales associated with the YSO stages represent samples of the continuous propagation of star formation radially outwards in the NGC 602 region; there is, however, overlap between all of the shells.  Sources of uncertainty include imprecise distances due to three-dimensional structure and low-number statistics.

The relative times between the triggering of these Stages can also be used to estimate Stage lifetimes.  PMS stars and what are now Stage~{\sc ii} sources are triggered $3 \pm~6 \times~10^5$~yr apart, indicating that the Stage II evolution lasts on the order of  $\sim 3\times~10^5$~yr.  Likewise, Stage~{\sc i} sources have a lifetime $\sim 3 \times~10^5$~yr, from the time difference between Stage~{\sc i} and {\sc ii} shells.  We are unable to account for the probable effects of mass on the range of YSO evolutionary timescales.  The smallest value for the Stage~{\sc i} shell radius (i.e., the best estimate minus the uncertainty) is $\sim 10$~pc (1~Myr), indicating that triggering occurred as much as 2~Myr, which may be taken as a firm upper limit to Stage~{\sc i} lifetimes.  Massive sources are expected to evolve more quickly than low mass sources; most of our candidates are protoclusters/multiple sources, leading us to expect a multiplicity of masses and thus stages within individual YSO candidate.  The uncertainty in these numbers is large, particularly because we are unable to analyze the expected correlation between mass and stage lifetime with such low number statistics and the current resolution.  However, the results are interesting; the temporal differences between the triggering of PMS and Stage~{\sc ii}, Stage~{\sc ii} and Stage~{\sc i}/{\sc ii}, and Stage~{\sc i}/{\sc ii} and Stage~{\sc i} are all approximately equal (a few $10^5$~yr).  Our Stage~{\sc i} lifetime is consistent with the lifetimes measured for YSOs in the galactic star-forming region, M17, by \citet{povich2010}. 

We construct a time-ordered scenario for the progression of star formation in NGC~602.  Approximately 7 Myr ago, two expanding H{\sc i} shells began to interact, creating an over-density in the region that is now N90.  After $\sim 3$~Myr, turbulence subsided sufficiently for star formation to begin in earnest, $\sim 4$~Myr ago \citep{nigra08}.  The current population of bright main sequence (O and B) stars and low-mass PMS stars formed near the center of the over-dense region approximately 2-3~Myr ago.  The radiation from the massive stars began to erode the surrounding nebula, creating a photodissociation region and triggering further star formation around the edges.  The formation of the current Stage~{\sc ii} YSOs was triggered $\sim 1.8$~Myr ago, the formation of the current Stage~{\sc i} YSOs $\sim 0.7$~Myr later (or $\sim 1.1$~Myr).

\section{CONCLUSIONS}

We present a multi-wavelength analysis of photometry and imaging of the NGC~602 active star-forming region in the SMC, covering {\it HST} 0.55 through \spit\ 24~$\mu$m.  From these data, we define stellar and proto-stellar populations and their spatial distribution.  We estimate the present-day star formation rate and derive time scales for the formation of Stage~{\sc i} and Stage~{\sc ii} YSOs.  We provide full mutli-wavelength photometric catalogs online and present approximate physical and evolutionary parameters for all of our YSO candidates.

Our primary modes of source analysis are CMD examination and SED fitting.  Optical $\it HST$ CMDs reveal $\sim$565 PMS candidates, essentially low mass Stage~{\sc iii} YSOs, through isochrone fitting.  Through multi-wavelength SED fitting, we identify 41 YSO candidates, including 24 Stage~{\sc i}, 8 Stage~{\sc i}/{\sc ii}, 5 Stage~{\sc ii}, 2 Stage~{\sc ii}/{\sc iii}, and 2 unclassified candidates.  High-resoultion {\it HST} imaging shows that $\ga 70\%$ of the YSO candidates include multiple sources or are protoclusters, and most of these optical sources are PMS candidates.  Efforts to construct YSO protocluster models and incorporate them into the SED fitter are underway but are beyond the scope of this paper.  For the $\sim$20 YSO candidates, we are able to estimate masses for one or more optical counter-parts via comparison with CMD evolutionary tracks, and we find consistency between these lower limit optical masses and the YSO SED fitter masses.  We also construct a mass function from YSO SED fitter masses and derive a present-day star formation rate of 0.2-1~$\Msun$~yr$^{-1}$~kpc$^{-2}$. 

Finally, we present a quantitative analysis of the spatial distribution of the YSO population with respect to the central cluster and PMS population.  We find that star  formation has progressed from cluster center to the edge of the star forming dust cloud in NGC~602.  The PMS stars are heavily concentrated near cluster center and that the YSO population distribution can be represented as concentric shells with Stage~{\sc ii} sources preferentially closer to cluster center and Stage~{\sc i} sources farther away.  Previous observations \citep{nigra08} have shown that there is no significant expansion of the dust shell in which most of our YSO candidates are located and that the photo-ionization front is the prime mover of the star formation activity.  We therefore correlate average distances of the Stage~{\sc i} and {\sc ii} YSOs from cluster center with the times at which their formation is triggered; we divide the distances by the sound speed.  Relating the timescales, we find the lifetimes of each YSO Stage to be a few $10^5$~yr, comparable to timescale estimates in the literature which apply independent techniques to galactic star-formation regions.

%% In a manner similar to \objectname authors can provide links to dataset
%% hosted at participating data centers via the \dataset{} command.  The
%% second curly bracket argument is printed in the text while the first
%% parentheses argument serves as the valid data set identifier.  Large
%% lists of data set are best provided in a table (see Table 3 for an example).
%% Valid data set identifiers should be obtained from the data center that
%% is currently hosting the data.

\acknowledgments
Special thanks to Joana Oliveira and Tom Robitaille for discussion and advise.  HST funding came from STScI GO grant GO-10248.07-A.  24$\mu$m data are taken from the first epoch of \spit\ MIPS observations under the SAGE-SMC \spit\ Legacy Program (Program 40245, PI Gordon).  Near Infrared observations made with \spit/IRAC come from Program 125, PI Fazio.  The Spitzer Space Telescope is operated by the Jet Propulsion Laboratory, California Institute of Technology, under NASA contract 1407.  Support for this work was provided by NASA through contract 1256790 issued by JPL/Caltech.  
{\it Facilities: \facility{HST (ACS)}, \facility{Spitzer (IRAC, MIPS)}}

\clearpage

\clearpage

\clearpage

\clearpage

\clearpage

\clearpage

\clearpage

\clearpage

\clearpage

\clearpage
 
\begin{deluxetable}{lccccc}
\tabletypesize{\small}
\tablecaption{\label{t:obs} NGC~602 log-book of optical observations. }
\tablehead{\colhead{Image Name} & \colhead{Filter} & \colhead{Exp. Time} & \colhead{R.A.} & \colhead{Dec.} & \colhead{Instrument} \\
& & (sec.) & (J2000) & (J2000) & \\}
\startdata
J92F05LIQ & F555W & 3 & 01:29:27.57 & -73:33:17.1 & HST/ACS/WFC \\ % 
J92FA6R7Q & F555W & 3 & 01:29:27.57 & -73:33:17.1 & HST/ACS/WFC \\ % 
J92F05LJQ & F555W & 430 & 01:29:27.57 & -73:33:17.1 & HST/ACS/WFC \\ %  
J92F05MOQ & F555W & 430 & 01:29:27.25 & -73:33:17.4 & HST/ACS/WFC \\ % 
J92F05LQQ & F555W & 430 & 01:29:27.89 & -73:33:16.9 & HST/ACS/WFC \\ % 
J92F05LSQ & F555W & 430 & 01:29:28.21 & -73:33:16.7 & HST/ACS/WFC \\ %  
J92F05LUQ & F555W & 430 &  01:29:28.52 & -73:33:16.4 & HST/ACS/WFC \\ %
J92F05MSQ & F814W & 2 & 01:29:27.57 & -73:33:17.1 & HST/ACS/WFC \\ % 
J92FA6QZQ & F814W & 2 & 01:29:27.57 & -73:33:17.1 & HST/ACS/WFC \\ % 
J92F05MTQ & F814W & 453 &01:29:27.25 & -73:33:17.4 & HST/ACS/WFC \\ %  
J92F05N9Q & F814W & 453 & 01:29:27.57 & -73:33:17.1 & HST/ACS/WFC \\ %  
J92F05N7Q & F814W & 453 & 01:29:27.89 & -73:33:16.9  & HST/ACS/WFC \\ % 
J92F05N3Q & F814W & 453 & 01:29:28.21 & -73:33:16.7 & HST/ACS/WFC \\ % 
J92F05MXQ & F814W & 453 & 01:29:28.52 & -73:33:16.4 & HST/ACS/WFC \\ %  
J92FA6R0Q & F658N & 636 & 01:29:19.35 & -73:33:18.7 &  HST/ACS/WFC \\ %
J92FA6R2Q & F658N & 636 & 01:29:18.71 & -73:33:19.0 &  HST/ACS/WFC \\ %
J92FA6R4Q & F658N & 636 & 01:29:18.08 & -73:33:19.3 &  HST/ACS/WFC \\ % 
\tableline
%  rccd061014.056 & B & 60 & 01:29:24.87 & -73:33:16.0 & SMARTS/ANDICAM \\
% rccd061014.057 & B & 60 & 01:29:24.81 & -73:33:15.6 & SMARTS/ANDICAM \\
%rccd061014.058 & B & 60 & 01:29:24.79 & -73:33:15.8 & SMARTS/ANDICAM \\
rccd061014.059 & V & 45 & 01:29:24.78 & -73:33:15.6 & SMARTS/ANDICAM \\ % 
rccd061014.060 & V & 45 & 01:29:24.80 & -73:33:15.6 & SMARTS/ANDICAM \\ %
rccd061014.061 & V & 45 & 01:29:24.81 & -73:33:15.4 & SMARTS/ANDICAM \\ %
rccd061014.062 & I & 30 & 01:29:24.82 & -73:33:15.5 & SMARTS/ANDICAM \\ %
rccd061014.063 & I & 30 & 01:29:24.83 & -73:33:15.7 & SMARTS/ANDICAM \\ %
rccd061014.064 & I & 30 & 01:29:24.85 & -73:33:15.5 & SMARTS/ANDICAM \\ % 
%\tableline
%SST & [3.6] & SST & SST & SST & IRAC \\
%SST & [4.5] & SST & SST & SST & IRAC \\
%SST & [5.8] & SST & SST & SST & IRAC \\
%SST & [8.0] & SST & SST & SST & IRAC \\
\enddata
\end{deluxetable}
     
\clearpage

\begin{deluxetable}{rrccccccr}
\tabletypesize{\scriptsize}
\tablecaption{\label{t:optphot} Optical photometry table.  The full 4535-source table is available online and includes matches to IRAC sources, and optical \citet{schmalzl08} stars and \citet{hutchings91} O/B stars (matched within 0$\farcs$2).  The ACS filters F555W and F814W are comparable to {\it V} and {\it I} , respectively.  Flags indicate the photometric data used: 2 ACS, 6 ground-based (SMARTS), 12 ACS/SMARTS combination.  Designations begin ``HSTNGC602 ."  }
\tablehead{\colhead{ID} & \colhead{Designation} & \colhead{R.A.} & \colhead{Dec.} & \colhead{F555W} & \colhead{dF555W} & \colhead{F814W} & \colhead{dF814W} & \colhead{Flag} \\
& & ($^{\circ}$, J2000) & ($^{\circ}$, J2000) & (mag) & (mag) & (mag) & (mag) & \\}
\startdata
      1   & J012935.538-733330.75 & 22.39806  &  -73.55852  &   17.257   &     0.024	 &  17.310   & 0.076   &     6 \\
      2   & J012950.969-733139.21 & 22.46245  &  -73.52756  &   \nodata  &    \nodata &  17.060   & 0.093   &     6 \\
      3   & J012934.757-733247.31 & 22.39484  &  -73.54645  &   17.603   &     0.014	 &  17.594   & 0.036   &     6 \\
      4   & J012944.118-733206.11 & 22.43388  &  -73.53502  &   17.695   &     0.014	 &  17.853   & 0.089   &     6 \\
      5   & J012937.031-733325.52 & 22.40429  &  -73.55707  &   17.742   &     0.015	 &  16.734   & 0.014   &    12 \\
      6   & J012935.517-733329.37 & 22.39798  &  -73.55814  &   \nodata  &    \nodata &  17.978   & 0.029   &     6 \\
      7   & J012921.403-733251.69 & 22.33920  &  -73.54766  &   17.897   &     0.016	 &  17.342   & 0.095   &     6 \\
      8   & J012926.544-733230.95 & 22.36064  &  -73.54190  &   18.012   &     0.017	 &  17.714   & 0.030   &     6 \\
      9   & J012930.638-733406.40 & 22.37761  &  -73.56842  &   18.069   &     0.021	 &  18.042   & 0.026   &     6 \\
     20  & J012935.361-733300.21 & 22.39735  &  -73.55004  &   18.202   &     0.038	 &  17.958   & 0.031   &     2 \\

\enddata
\end{deluxetable}

\clearpage

\begin{deluxetable}{cccccccccccc}
\tabletypesize{\scriptsize}
\tablecaption{\label{t:iracphot} IRAC photometry table.  The full 497-source table is available online and notes matches to optical sources from Table~\ref{t:optphot} as well as matches to \citet{gouliermis07} YSO candidates.  Designations begin with the prefix ``SSTNGC602 ."}
\tablehead{\colhead{ID} & \colhead{Designation} & \colhead{R.A.} & \colhead{Dec} & \colhead{[3.6]} & \colhead{d[3.6]} & \colhead{[4.5]} & \colhead{d[4.5]} & \colhead{[5.8]} & \colhead{d[5.8]} & \colhead{[8.0]} & \colhead{d[8.0]} \\
 & & ($^{\circ}$, J2000) & ($^{\circ}$, J2000) & (mag) &(mag) & (mag) &(mag) & (mag) &(mag) & (mag) &(mag)\\}
\startdata
194	& J012920.73-733327.1 &22.3364 & 	-73.5575 &	15.040 &	0.016 &	15.027 &	0.027 &	15.101  &	0.108    &	15.339  &	0.262   \\
195	& J012938.80-733411.4 &22.4117 &        -73.5698 &	16.142 &	0.087 &	15.926 &	0.084 &	\nodata &	\nodata &	\nodata &	\nodata \\
196	& J012919.88-733322.5 &22.3328 &         -73.5563 &	16.116 &	0.033 &	15.768 &	0.046 &	14.058  &	0.044    &	12.466  &	0.021    \\
197	& J012918.95-733319.4 &22.3290 &	-73.5554 &	15.927 &	0.027 &	15.604 &	0.037 &	\nodata &	\nodata &	12.429  &	0.020    \\   
198	& J012936.38-733403.6 &22.4016 &	-73.5677 &	14.507 &	0.021 &	14.182 &	0.030 &	13.698  &	0.047    &	12.104  &	0.034    \\
199	& J012933.75-733357.0 &22.3907 &	-73.5658 &	17.298 &	0.234 &	16.053 &	0.124 &	\nodata &	\nodata &	\nodata &	\nodata \\
201	& J012921.57-733325.4 &22.3400 &	-73.5571 &	15.040 &	0.016 &	15.518 &	0.039 &	\nodata &	\nodata &	\nodata &	\nodata \\
203	& J012925.36-733331.1 &22.3557 &	-73.5587 &	17.720 &	0.116 &	17.106 &	0.138 &	\nodata &	\nodata &	\nodata &	\nodata \\
204	& J012922.91-733324.7 &22.3455 &	-73.5569 &	17.733 &	0.108 &	17.247 &	0.152 &	\nodata &	\nodata &	\nodata &	\nodata \\
205	& J012927.67-733334.9 &22.3653 &	-73.5597 &	15.240 &	0.018 &	15.264 &	0.034 &	15.712  &	0.174    &	\nodata &	\nodata \\
206	& J012920.48-733316.8 &22.3354 &	-73.5547 &	16.097 &	0.031 &	15.646 &	0.039 &	13.302  &	0.025    &	12.584  &	0.022   \\
\enddata
\end{deluxetable}

\clearpage

\begin{deluxetable}{lllllllllllll}
\rotate
\tabletypesize{\tiny}
\tablewidth{-2pt}
\tablecaption{\label{t:flux} Fluxes of all 77 fitter sources, including those fit in multiple ways.  Uncertainties are included in the full online photometry table and are on the order of 10 to 20\%.}
\tablehead{\colhead{Name} & \colhead{R.A.} & \colhead{Dec.} & \colhead{Flux$_V$} & \colhead{Flux$_I$} &  \colhead{Flux$_J$} & \colhead{Flux$_H$} & \colhead{Flux$_{K_s}$} & \colhead{Flux$_{3.6\mu{\rm m}}$} & \colhead{Flux$_{4.5\mu{\rm m}}$} & \colhead{Flux$_{5.8\mu{\rm m}}$} & \colhead{Flux$_{8\mu{\rm m}}$} & \colhead{Flux$_{24\mu{\rm m}}$} \\
 & ($^{\circ}$, J2000) & ($^{\circ}$, J2000) & (mJy) & (mJy) & (mJy) & (mJy) & (mJ) & (mJy) & (mJy) & (mJy) & (mJy) & (mJy) \\}
\startdata
Y090 & 22.35823 & -73.57969 & 0.0003\tablenotemark{95} & 0.0012\tablenotemark{95} & 0.017 &  \nodata &  \nodata &  0.12 &  0.054 &  0.43  &  1.1 & \nodata \\
Y096   & 22.27672 & -73.56351  &  0.0075\tablenotemark{95} &   0.012\tablenotemark{95} &  0.18\tablenotemark{90} &  0.35\tablenotemark{90} &   0.30\tablenotemark{90} & 0.93 &  1.0 &1.9 & 4.0 &   35 \\ 
Y118   & 22.37345 & -73.57505  &  0.0018  &   0.0043 & \nodata & \nodata &  \nodata & 0.019 &  0.044 &   \nodata & 0.070 &	\nodata \\
Y142   & 22.36902 & -73.56999  &  0.0003\tablenotemark{95} &   0.0008\tablenotemark{95} & \nodata & \nodata &  \nodata & 0.097 &  0.089 &    0.14 & 0.41 &   \nodata \\
Y143   & 22.31988 & -73.56133  &  0.0039 &   0.0048 & \nodata & \nodata &  \nodata & 0.049 &  0.045 &   \nodata & 0.31 &   \nodata \\
Y148   & 22.39630 & -73.57331  & \nodata &   0.0001 & \nodata & \nodata &  \nodata & 0.11 &  0.075 &    0.27 & 0.69 &   \nodata \\
Y149   & 22.23401 & -73.54512  & \nodata &  \nodata &  0.036 &  0.058 &   0.058 & 0.056 &  0.057 &   \nodata & 0.17 &   \nodata \\
Y162   & 22.38200 & -73.56925  &  0.0010 &   0.0019 &  0.032\tablenotemark{90} &  0.047\tablenotemark{90} &	0.068\tablenotemark{90} & 0.33 &  0.39 & 0.66 & 1.5 &    5.3 \\
Y163   & 22.24721 & -73.54578  & \nodata &  \nodata &  0.034 &  0.047 &   0.053 & 0.046 &  0.042 &   \nodata & 0.033 &   \nodata \\
Y170   & 22.37877 & -73.56824  &  0.015\tablenotemark{95} &   0.017\tablenotemark{95} &  0.022 &  0.044 &  \nodata & 0.15 &  0.15 &    0.27 & 0.82 &   \nodata \\
Y171   & 22.38497 & -73.56902  &  0.0012 &   0.0015 &  0.026 & \nodata &  \nodata & 0.34 &  0.40 &    0.66 & 1.5 &   \nodata \\
Y174   & 22.39191 & -73.56994  &  0.0003\tablenotemark{95} &   0.0008\tablenotemark{95} &  1.0\tablenotemark{90} &  0.99\tablenotemark{90} &   0.63\tablenotemark{90} & 0.14 &  0.14 &0.21 & 0.57 &   \nodata \\
Y179   & 22.32686 & -73.55694  &  0.0037 &   0.0033 & \nodata & \nodata &  \nodata & 0.064 &  0.052 &   \nodata & 0.31&   \nodata \\
Y196   & 22.33283 & -73.55628  &  0.0018 &   0.0022 &  0.029 &  0.033 &  \nodata & 0.10 &  0.089 &    0.27 & 0.66 &   \nodata \\
Y197   & 22.32897 & -73.55540  &  0.016 &   0.0044 & \nodata & \nodata &  \nodata & 0.12 &  0.10 &   \nodata & 0.68 &   \nodata \\
Y198   & 22.40160 & -73.56769  &  0.037 &   0.20 &  0.52 &  0.69 &   0.51 & 0.44 &  0.38 &    0.38 & 0.92 &   \nodata \\
Y206   & 22.33535 & -73.55467  &  0.0028\tablenotemark{95} &   0.0025\tablenotemark{95} &  0.028\tablenotemark{90} &  0.043\tablenotemark{90} &  \nodata & 0.10 &  0.099 &    0.55 & 0.59 &   \nodata \\
Y217   & 22.33283 & -73.55347  & 0.000005\tablenotemark{95}&   0.00008  & \nodata & \nodata &  \nodata & 0.19 &  0.16 &	 0.5489 & 1.3 &    1.5 \\
Y223   & 22.39958 & -73.56438  &  0.0019  &   0.0053   &  0.066\tablenotemark{90} &  0.090\tablenotemark{90} &   0.094\tablenotemark{90} & 0.37 & \nodata &    1.0 & 2.5 &	\nodata \\
Y227   & 22.40573 & -73.56458  &  0.0043\tablenotemark{95} &   0.0039\tablenotemark{95} & \nodata & \nodata &	 \nodata & 0.67 &  0.48 &	 1.7 & 4.3 &   11 \\
Y237   & 22.33603 & -73.55176  &  0.0017\tablenotemark{95} &   0.0018\tablenotemark{95} &  0.35\tablenotemark{90} &  0.21\tablenotemark{90} &   0.14\tablenotemark{90} & 0.19 &  0.20 &    0.38 & 0.80 & 1.9 \\
Y240   & 22.40832 & -73.56469  &  0.031 &   0.018  & \nodata & \nodata  &   \nodata & 0.67 &  0.48 &   \nodata & 4.3 &   \nodata \\
Y251   & 22.39853 & -73.56098  &  0.0009 &   0.0015  &  0.012\tablenotemark{90} &  0.020\tablenotemark{90} &  \nodata & 0.18 &  0.12 &	0.57 & 1.4 &   \nodata \\
Y255   & 22.40668 & -73.56194  &  0.079 &   0.066  & \nodata & \nodata &  \nodata & 0.071 &  0.089 &   \nodata & 0.32 &	\nodata \\
Y264   & 22.40414 & -73.56005  &  0.13 &   0.10  & \nodata & \nodata &  \nodata & 0.15 &  0.15 &   \nodata & 0.46 &	\nodata \\
Y270o  & 22.39799 & -73.55843  &  0.47\tablenotemark{95} &   0.32\tablenotemark{95} &  0.43\tablenotemark{90} &  0.45\tablenotemark{90} &   0.39\tablenotemark{90} & 0.70 &  0.50 &    1.7 & 4.2 &   37 \\
Y270i  & 22.39799 & -73.55843  &  0.47  &   0.32 &  0.43 &  0.45&   0.40 & 0.70\tablenotemark{95} &  0.50\tablenotemark{95} &	1.7\tablenotemark{95} & 4.2\tablenotemark{95} &	37\tablenotemark{95}\\
A270   & 22.39799 & -73.55843  &  0.47  &   0.32 &  0.43 &  0.45 &   0.39 & 0.70 &  0.50  &    1.7 &	  4.2 &   37 \\
Y271   & 22.38932 & -73.55647  &  0.015  &   0.019 & \nodata & \nodata &  \nodata & 0.031 &  0.030 &   \nodata &	  0.065 &   \nodata \\
Y283   & 22.37587 & -73.55292  &  0.0083  &   0.011 &  0.13\tablenotemark{90} &  0.15\tablenotemark{90} &   0.21\tablenotemark{90} & 0.17 &  0.20  &    0.32 &	0.57 &    1.3 \\
Y285   & 22.34611 & -73.54764  &  0.0024  &   0.0041 &  0.052 &  0.082 &   0.081 & 0.12 &  0.097  &    0.30 &	  0.65 &   \nodata \\
Y287   & 22.43070 & -73.56199  &  0.0009\tablenotemark{95} &   0.0013\tablenotemark{95} &  0.016 & \nodata &  \nodata & 0.13 &  0.12  &	0.35 &     0.78 &	 0.70 \\
Y288   & 22.17680 & -73.01200  &  0.0005\tablenotemark{95} &   0.0010\tablenotemark{95} & \nodata &  0.023 &  \nodata & 0.13 &  0.15  &	0.34 &     0.73 &	\nodata \\
Y290   & 22.40450 & -73.55705  &  0.026		   &   0.042 &  0.68\tablenotemark{90} &  0.74\tablenotemark{90} &   0.47\tablenotemark{90} & 0.28 &  0.22  &    0.17 &	   0.31 &   \nodata \\
Y312   & 22.41324 & -73.55526  &  0.0006\tablenotemark{95} &   0.0016\tablenotemark{95} &  0.024  &  0.033 &   0.041 & 0.11 &  0.077  &	 0.30 &      0.79 &   \nodata \\
Y326   & 22.43648 & -73.55702  &  0.0016\tablenotemark{95} &   0.0037\tablenotemark{95} &  0.023\tablenotemark{90} &  0.039\tablenotemark{90} &  \nodata & 0.11 &  0.069  &    0.22 &	 0.58 &   \nodata \\
Y327   & 22.39873 & -73.55059  &  0.015\tablenotemark{95} &   0.019\tablenotemark{95} &  0.18\tablenotemark{90} &  0.18\tablenotemark{90} &   0.37\tablenotemark{90} & 1.9 &  2.5  &    4.5 &     8.5 &   39 \\
Y340   & 22.36068 & -73.54176  & \nodata  &  \nodata &  \nodata &  \nodata &  \nodata &  0.10 &  0.060 &    0.20 &  0.51 &    0.55 \\
A340   & 22.36068 & -73.54176  &  0.24  &   0.22 &   0.16 &   0.11 &   0.082 &  0.10 &  0.060 &    0.20 &  0.51 &    0.55 \\
Y358   & 22.39656 & -73.54515  &  0.0035  &   0.0050 &  \nodata &  \nodata &  \nodata &  0.016 &  0.021 &   \nodata &  0.050 &   \nodata \\
Y396   & 22.35085 & -73.53134  &  0.042  &   0.04 &   0.016 &  \nodata &  \nodata &  0.050 &  0.053 &   \nodata &  0.037 &   \nodata \\
Y493   & 22.31646 & -73.50489  & \nodata  &  \nodata &   0.036 &   0.080 &   0.11 &  0.089 &  0.075 &   \nodata &  0.095 &   \nodata \\
Y700   & 22.52856 & -73.54969  & \nodata  &  \nodata &   0.020 &  \nodata &  \nodata &  0.059 &  0.095 &   \nodata &  0.24 &    0.73 \\
\tableline
S049   &  22.26367 & -73.57036  & \nodata  &  \nodata &   4.0 &  41 &  28 & 13 &  8.0 &	4.9 &  2.9 &    0.71 \\
K050   &  22.38150 & -73.59051  & \nodata  &  \nodata &   2.2 &   2.1 &	1.5 &  0.71  &  0.42 &	0.28 &  0.13 &   \nodata \\
K063   &  22.27820 & -73.57049  &  0.22  &   0.41 &   0.86 &   1.3 &	0.83 &  0.43 &  0.23 &	0.14 &  0.12 &   \nodata \\
K086   &  22.30000 & -73.57027  &  0.20  &   0.33 &   0.52 &   0.64 &	 0.39 &  0.19 &  0.12 &	0.062 &  0.050 &   \nodata \\
K181   &  22.40200 & -73.57003  &  1.1  &   2.7 &   3.9 &   5.3 &	3.3 &  1.7 &  1.1 &	0.71 &  0.74 &   \nodata \\
K194   &  22.33638 & -73.55754  &  0.0014  &   0.016 &   0.47 &   0.55 &	0.43 &  0.27 &  0.18 &	0.10 &  0.047 &   \nodata \\
K210   &  22.42660 & -73.57045  &  0.13  &   0.35 &   0.72 &   1.0 &	0.66 &  0.34 &  0.20 &	0.13 &  0.095 &   \nodata \\
S213   &  22.37980 & -73.56195  &  4.3  &   2.6 &   1.2 &   0.79 &	0.51 &  0.45 &  0.30 &	0.21 &  0.34\tablenotemark{95} &   12\tablenotemark{95}\\
K225   &  22.38890 & -73.56236  &  6.4  &   3.8 &   1.9 &   1.1 &	0.65 &  0.29 &  0.16 &	0.11 &  0.16\tablenotemark{95} &   \nodata \\
K232   &  22.38700 & -73.56084  &  2.0  &   1.3 &   0.64 &   0.37 &	0.26 &  0.13 &  0.062 &   \nodata &  0.16\tablenotemark{95} &   \nodata \\
S235   &  22.35260 & -73.55448  &  9.3  &   5.5 &   2.66 &   1.5 &	0.94 &  0.39 &  0.24 &	0.15 &  0.12 &    9.0\tablenotemark{95}\\
K248   &  22.38330 & -73.55859  &  1.9  &   1.2 &   0.60 &   0.34 &	0.22 &  0.13 &  0.065 &   \nodata &  0.035 &   \nodata \\
K261   &  22.25457 & -73.53501  & \nodata  &  \nodata &   1.9 &   2.30 &	1.66 &  0.96 &  0.64 &	0.40 &  0.25 &   \nodata \\
S293   &  22.36570 & -73.54987  &  2.1  &   1.6 &   1.0 &   0.68 &	0.51 &  0.32 &  0.22 &	0.16 &  0.12 &   \nodata \\
K296   &  22.32630 & -73.54260  &  1.8  &   2.4 &   2.6 &   2.4 &	1.6 &  0.77 &  0.47 &	0.32 &  0.24 &   \nodata \\
K340   &  22.36068 & -73.54176  &  0.24  &   0.22 &   0.16 &   0.11 &	0.082 &  0.10\tablenotemark{95} &	\nodata &   \nodata & \nodata &   \nodata \\
K348   &  22.30930 & -73.53139  &  4.7  &  10 &  18 &  26 &  16 &  8.0 &  4.6 &	2.9 &  1.8 &   \nodata \\
S394   &  22.36515 & -73.53374  &  0.94  &   0.58 &   0.41 &   0.23 &	0.15 &  0.14 &  0.092 &   \nodata &  0.044 &   \nodata \\
S406   &  22.36790 & -73.53262  &  2.5  &   1.9 &   1.2 &   0.78 &	0.57 &  0.42 &  0.31 &	0.21 &  0.15 &   \nodata \\
S411   &  22.32367 & -73.52343  & \nodata  &  \nodata &  \nodata &   0.016 &  \nodata &  0.10 &  0.076 &   \nodata &  0.051 &   \nodata \\
K421   &  22.31828 & -73.52143  & \nodata  &  \nodata &  36 &  35 &  22 & 10 &  6.38 &	3.8 &  2.2 &   \nodata \\
K441   &  22.29871 & -73.51362  & \nodata  &  \nodata &  15 &  30 &  26 & 18 & 13 &	9.6 &  7.7 &    0.68 \\
K442   &  22.49467 & -73.54764  & \nodata  &  \nodata &   0.066 &   0.067 &	0.12 &  0.29 &  0.18 &	0.13 &  0.073 &   \nodata \\
K444   &  22.34401 & -73.52120  & \nodata  &  \nodata &   0.058 &   0.11 &	0.096 &  0.16 &  0.13 &	0.078 &  0.073 &   \nodata \\
S456   &  22.47845 & -73.54208  & \nodata  &  \nodata &  19 &  26 &  17 &  8.5 &  4.7 &	3.3 &  2.0 &    1.3 \\
K481   &  22.49347 & -73.53891  & \nodata  &  \nodata &   0.026 &   0.060 &  \nodata &  0.17 &  0.10 &   \nodata &  0.040 &   \nodata \\
S486   &  22.39100 & -73.51968  & \nodata  &  \nodata &   1.0 &   1.0 &	0.64 &  0.30 &  0.19 &	0.14 &  0.040 &   \nodata \\
K701   &  22.39838 & -73.51712  & \nodata  &  \nodata &  10 &  12 &	8.6 &  3.8 &  2.32 &	1.6 &  0.88 &   \nodata \\
\tableline
G109   & 22.36616 & -73.57540  & \nodata  &  \nodata &   0.016 &    \nodata &  \nodata &  0.070 &	0.052 &   \nodata &	 0.18 &   \nodata \\
G133   & 22.28723 & -73.55713  &  0.0080  &   0.024 &   0.28\tablenotemark{90} & 0.52\tablenotemark{90} &   0.53\tablenotemark{90} & 0.53 & 0.37 & 0.18 &  0.18  & \nodata \\
G150   & 22.42516 & -73.57831  &  0.0003  &   0.0026 &   0.019 &    \nodata &  \nodata &     0.0548 &     0.0346 &   \nodata &     0.0075 &   \nodata \\
G211   & 22.27448 & -73.54372  & \nodata  &  \nodata &   0.20 &     0.29 &   0.27 &     0.29 &     0.22 &    0.11 &     0.21 &   \nodata \\
G262   & 22.33369 & -73.54814  &  0.0016  &   0.0093 &   0.014 &     0.029 &  \nodata &     0.073 &     0.082 &   \nodata &     0.17 &   \nodata \\
G275   & 22.40597 & -73.55920  &  0.0033  &   0.013 &   0.025 &     0.054 &   0.053 &     0.15 &     0.15 &   \nodata &     0.46 &   \nodata \\
G343   & 22.34403 & -73.53817  &  0.0073  &   0.018 &   0.022 &     0.036 &  \nodata &     0.028 &     0.031 &   \nodata &     0.073 &   \nodata \\
G372   & 22.39943 & -73.54250  &  0.014  &   0.036 &   0.056\tablenotemark{90} & 0.062\tablenotemark{90} &   0.065\tablenotemark{90} & 0.22 & 0.15  &   \nodata & 0.55 &   \nodata  \\
\tableline
U364   & 22.28734 & -73.52486  & \nodata  &  \nodata &  \nodata  &     0.044 &   0.27 &     0.057 &     0.052 &   \nodata &     0.072 &	 0.52\\
U703   & 22.29610 & -73.51102  & \nodata  &  \nodata &   0.87  &     0.65 &   0.52 &     0.34 &     0.29 &	0.22 &     0.20 &	 0.53\\
																		      
\enddata
\tablenotetext{90}{Flux considered upper limit with 90\% certainty}
\tablenotetext{95}{Flux considered upper limit with 95\% certainty}
\end{deluxetable}

\clearpage

\begin{deluxetable}{lrcrrcrrrrrrrrc}
\rotate
\tablewidth{-2pt}
\tabletypesize{\scriptsize}
\tablecaption{\label{t:ysoresult} Young Stellar Object Parameters.  For those sources with optical counterparts, we list the number of visually identifiable optical sources.  Where possible, we determine masses of the optical sources; others either have no good optical photometry or are so young or of such low mass (M~$\la 0.5~\Msun$) that their masses cannot be determined from our evolutionary tracks (See Figure~\ref{OptCMD}~(b)).  M$_{opt}$ is the sum of the optical masses which we can determine.  Sources with Stage listed as U are unclassified probable YSOs.}
\tablehead{\colhead{Name} & \colhead{G07\tablenotemark{a}} & \colhead{$N_{opt}$\tablenotemark{b}} & \colhead{$\chi^2$} & \colhead{$n_{fits}$\tablenotemark{d}}& \colhead{$M_{opt}$} & \colhead{$M_{\star}^{ave}$} & \colhead{$\sigma M_{\star}$\tablenotemark{e}} & \colhead{$L_{\star}^{ave}$} & \colhead{$\sigma L_{\star}$} & \colhead{$\dot{M}_{env}^{ave}$} & \colhead{$\sigma \dot{M}_{env}$} & \colhead{$M_{disk}^{ave}$} & \colhead{$\sigma M_{disk}$} & \colhead{Evol.} \\
& & & (cpd\tablenotemark{c}) & & ($\Msun$) & ($\Msun$) & ($\Msun$) & ($10^3 \Lsun$) & ($10^3 \Lsun$) & ($10^{-4}\Msun/yr$) & ($10^{-4}\Msun/yr$) & ($\Msun$) & ($\Msun$) & Stage  \\ }
\startdata

Y090\tablenotemark{f,g,h} &       2  & $>10$\tablenotemark{i} &  5.39  &    2  &  0.5\tablenotemark{j}	 &   7.56  &   0.23  &  1.223  &   0.496   &   20.56   &   8.535   &  0.1407	&    0.0637   &  \sc{i} \\   
Y096\tablenotemark{g}     &      51  & $>10$\tablenotemark{i} &  0.31  &   69  &    1\tablenotemark{j}	 &   8.04  &   0.01  &  2.475  &   0.013   &   1.271   &   0.035   &  0.2276    &    0.0091   &  \sc{i} \\   
Y118             	  & \nodata  & 1\tablenotemark{i}     &  0.22  &    6  &  0.8			 &   7.37  &   0.09  &  0.729  &   0.090   &  33.770   &   3.291   &  0.7915    &    0.1824   &  \sc{i} \\   
Y142\tablenotemark{k}	  &      36  & 1	              &	0.02   &  120  & \nodata                 &   7.40  &   0.03  &  2.016  &   0.023   &  14.830   &   0.212   &  0.2022    &    0.0040   &  \sc{i} \\   
Y143\tablenotemark{k}	  & \nodata  & 1	              &	0.39   &    1  & \nodata                 &   7.45  &   0.00  &  2.257  &   0.000   &   2.214   &   0.000   &  0.1467    &    0.0000   &  \sc{i} \\   
Y148\tablenotemark{h,k}	  & \nodata  & 0	              &	1.19   &    9  & \nodata                 &   8.17  &   0.08  &  2.205  &   0.010   &   3.711   &   0.175   &  0.7449    &    0.1132   &  \sc{i} \\   
Y149             	  & \nodata  & N/A\tablenotemark{l}   &  0.03  &   37  & \nodata 		 &   4.19  &   0.06  &  0.442  &   0.017   &   3.572   &   0.193   &  0.1828    &    0.0087   &  \sc{i} \\   
Y162			  &      53  & 3	              &	0.73   &   15  & \nodata                 &   6.67  &   0.18  &  2.581  &   0.219   &   0.147   &   0.025   &  0.3671    &    0.0425   &  \sc{i}/\sc{ii} \\ 
Y163             	  & \nodata  & N/A\tablenotemark{l}   &	0.04   &  128  & \nodata                 &   3.64  &   0.01  &  0.143  &   0.008   &   3.280   &   0.190   &  0.1486	&    0.0019   &  \sc{i}/\sc{ii} \\ 
Y170\tablenotemark{g}     &      46  & 4\tablenotemark{i}     &	0.39   &    6  & \nodata                 &   8.23  &   0.09  &  2.541  &   0.166   &   4.744   &   0.325   &  0.1472    &    0.0172   &  \sc{i} \\   
Y171             	  & \nodata  & 7\tablenotemark{i}     &	0.43   &   35  &  0.4\tablenotemark{j}   &  10.95  &   0.11  & 11.840  &   0.353   &  33.170   &   1.948   &  0.2325    &    0.0148   &  \sc{i} \\   
Y174\tablenotemark{g,k}   &      38  & 0	              &	0.01   &  263  & \nodata                 &   7.63  &   0.01  &  2.439  &   0.009   &   9.728   &   0.095   &  0.4575    &    0.0039   &  \sc{i} \\   
Y179\tablenotemark{k}	  & \nodata  & 0	              &	0.92   &    1  & \nodata                 &   7.45  &   0.00  &  2.257  &   0.000   &   2.214   &   0.000   &  0.1467	&    0.0000   &  \sc{i} \\   
Y196\tablenotemark{h,k}   & \nodata  & 2\tablenotemark{i}     &	2.09   &    9  &  0.6\tablenotemark{j}   &   7.74  &   0.07  &  2.741  &   0.070   &   0.829   &   0.132   &  0.2822	&    0.0246   &  \sc{i}/\sc{ii} \\ 
Y197\tablenotemark{k}	  & \nodata  & 0	              &	0.87   &    3  & \nodata                 &   8.11  &   0.16  &  2.556  &   0.137   &   2.346   &   0.144   &  0.5644	&    0.1899   &  \sc{i} \\ 
Y198             	  & \nodata  & 2\tablenotemark{i}     &	0.59   &    5  & \nodata                 &   6.85  &   0.09  &  0.409  &   0.018   &  18.702   &   5.690   &  0.7289	&    0.1602   &  \sc{i} \\ 
Y206\tablenotemark{g,k}   &      49  & 2\tablenotemark{i}     &	0.90   &   11  &  1.4\tablenotemark{j}   &  25.87  &   0.38  & 23.680  &   0.354   & 103.900   &   7.538   &  0.5077	&    0.0098   &  \sc{i} \\   
Y223\tablenotemark{k}	  & \nodata  & $>10$       	      &  0.24  &   18  & \nodata                 &   6.37  &   0.02  &  1.202  &   0.015   &   0.002   &   0.003   &  0.1762	&    0.0110   &  \sc{ii} \\ 
Y237\tablenotemark{g,k}   & \nodata  & 2	              &  0.49  &   42  & \nodata                 &   5.80  &   0.04  &  1.422  &   0.074   &   0.110   &   0.009   &  0.1266	&    0.0052   &  \sc{i}/\sc{ii} \\ 
Y240\tablenotemark{f}	  & \nodata  & 3           	      &  5.38  &    2  &    1\tablenotemark{j}   &   8.78  &   0.38  &  4.037  &   0.609   &   0.533   &   0.137   &  0.0031	&    0.0013   &  \sc{i} \\  
Y251\tablenotemark{f,h,k} &       1  & 1\tablenotemark{i}     &  4.38  &    4  & \nodata                 &  11.12  &   0.90  & 14.550  &   4.471   &  64.900   &  23.480   &  0.1505	&    0.0254   &  \sc{i} \\ 
Y255             	  & \nodata  & 2\tablenotemark{i}     &  0.77  &   27  & 3.55\tablenotemark{j}   &   6.50  &   0.03  &  1.316  &   0.013   & $< 0.001$ & $< 0.001$ & $< 0.0001$ & $< 0.0001$  &  \sc{ii}/\sc{iii} \\ 
Y264             	  & \nodata  & 7\tablenotemark{i}     &  0.32  &   10  & \nodata                 &   6.57  &   0.00  &  1.295  & $< 0.001$ &   0.000   &   0.000   & $< 0.0001$ &    0.0000   &  \sc{ii} \\  
Y271             	  & \nodata  & 3\tablenotemark{i}     &  0.05  &   31  &   2	                 &   4.58  &   0.02  &  0.367  &   0.005   &   0.065   &   0.012   &  0.0001	& $< 0.0001$  &  U  \\ 
Y283			  &      52  & 1		      &  0.63  &   20  &   3                     &   4.51  &   0.02  &  0.349  &   0.006   &   0.001   & $< 0.001$ &  0.1072	&    0.0043   &  \sc{ii} \\ 
Y285\tablenotemark{g,h,k} &      34  & 7\tablenotemark{i}     &	 1.76  &   12  & \nodata                 &   8.56  &   0.06  &  3.837  &   0.079   &   1.859   &   0.165   &  0.4557	&    0.0511   &  \sc{i} \\
Y288\tablenotemark{g}	  & \nodata  & 6   		      &  0.23  &    8  &  1.8\tablenotemark{j}	 &  11.25  &   0.06  &  3.516  &   0.086   &   6.695   &   0.253   &  0.3750	&    0.0367   &  \sc{i} \\
Y290             	  & \nodata  & 2\tablenotemark{i}     &  0.05  &   46  &  5.5\tablenotemark{j}	 &   6.75  &   0.01  &  1.494  &   0.012   & $< 0.001$ & $< 0.001$ & $< 0.0001$ & $< 0.0001$  &  \sc{ii}/\sc{iii} \\
Y312\tablenotemark{g,h,k} &      24  & 8\tablenotemark{i}     &	 2.35  &   11  &  0.65\tablenotemark{j}  &  12.41  &   0.36  &  6.367  &   0.429   &  68.470   &   7.649   &  0.4087	&    0.0733   &  \sc{i} \\
Y326\tablenotemark{g,h}   &      22  & $>10$\tablenotemark{i} &  1.46  &   12  &  3.85\tablenotemark{j}  &   8.01  &   0.08  &  1.548  &   0.086   &   2.534   &   0.114   &  0.7441    &    0.0793   &  \sc{i} \\
Y327\tablenotemark{g}     &   54/57  & $>10$\tablenotemark{i} &  0.14  &   67  &  8.2\tablenotemark{j}	 &   8.78  &   0.02  &  4.175  &   0.028   &   2.969   &   0.143   &  0.5354    &    0.0130   &  \sc{i} \\
Y358		          & \nodata  & 3\tablenotemark{i}     &  0.07  &   38  &  1.2\tablenotemark{j}	 &   2.58  &   0.01  &  0.042  & $< 0.001$ &   0.001   & $< 0.001$ &  0.0408	&    0.0017   &  \sc{ii} \\  
Y396			  & \nodata  & 4	              &	 1.49  &    6  & \nodata                 &   2.69  &   0.00  &  0.056  &   0.000   &   0.000   &   0.000   &  0.0159	&    0.0000   &  \sc{ii} \\  
Y493			  & \nodata  & N/A\tablenotemark{l}   &  0.07  &  248  & \nodata 		 &   4.44  &   0.00  &  0.283  &   0.001   &   3.914   &   0.034   &  0.1174	&    0.0015   &  \sc{i} \\ 
Y700			  & \nodata  & N/A\tablenotemark{l}   &  0.24  &   14  & \nodata 		 &   4.31  &   0.07  &  0.205  &   0.003   &   0.737   &   0.169   &  0.0449	&    0.0043   &  \sc{i} \\ 
\tableline
Y217\tablenotemark{k}     &      26  &  0                     &  0.03  &   28  & \nodata                 &   5.68  &   0.01  &  0.455  &   0.008   &   0.514   &   0.026   &  0.2646	&    0.0196   & \sc{i}/\sc{ii} \\ 
Y217e\tablenotemark{h}    &          & 0	              &  2.66  &    2  & \nodata                 &   9.74  &   0.15  &  6.020  &   0.427   &   0.000   &   0.000   &  0.6202	&    0.3101   &  \sc{ii}/\sc{iii} \\ 
Y227\tablenotemark{f,g,k} &      50  & 3                      &  0.74  &   17  & \nodata                 &   6.86  &   0.03  &  1.260  &   0.015   &   4.161   &   0.696   &  0.4697	&    0.0748   & \sc{i}/\sc{ii} \\ 
Y227e\tablenotemark{h}    &          & 3	              &  4.32  &    6  &    2\tablenotemark{j}   &   8.87  &   0.32  &  4.894  &   0.436   &   0.001   &   0.001   &  0.3339	&    0.0997   &  \sc{ii} \\
Y287\tablenotemark{k}	  &      45  & 3                      &  0.07  &   43  &  2\tablenotemark{j}     &   5.36  &   0.01  &  0.225  &   0.002   &   4.895   &   0.294   &  0.2185	&    0.0060   & \sc{i} \\ 
Y287e\tablenotemark{h} 	  &          & 3   		      &	 2.59  &    1  &  2.2\tablenotemark{j}	 &  12.04  &   0.00  & 11.550  &   0.000   &   0.000   &   0.000   &  0.4850	&    0.0000   &  \sc{ii} \\ 
\tableline
Y270i                     &     29   &   $>10$	              &  1.05  &   22  &    15\tablenotemark{j}  &   9.27  &   0.14  &  7.400  &   0.772   &  23.900   &   4.978   &  0.0025	&    0.0005   & \sc{i}/\sc{ii} \\
Y270ie\tablenotemark{h}   &          & $>10$		      &  2.90  &   10  &   15\tablenotemark{j}   &   8.17  &   0.11  &  1.207  &   0.051   &  41.680   &   4.025   &  0.4940	&    0.0498   &  \sc{i} \\  
Y270o             	  &          & $>10$		      &	1.79   &   18  &   15\tablenotemark{j}   &   5.64  &   0.01  &  0.472  &   0.004   &   0.174   &   0.006   &  0.0666	&    0.0042   &  \sc{i} \\  
A270                      &          &   $>10$	              &  3.36  &    8  &  15\tablenotemark{j}	 &   8.45  &   0.00  &  3.463  & $< 0.001$ &   0.000   &   0.000   & $< 0.0001$ & $< 0.0001$  & \sc{ii} \\ 
A270e\tablenotemark{h}    &          & $>10$		      &	4.95   &    9  &   15\tablenotemark{j}   &   8.45  &   0.00  &  3.463  & $< 0.001$ &   0.000   &   0.000   & $< 0.0001$ & $< 0.0001$  &  \sc{ii} \\  
\tableline 
Y340                      &     5    & 2                      &  0.62  &   29  & \nodata &   4.72  &   0.04  &  0.409  &   0.012   &   0.739   &   0.049   &  0.04732   &    0.0048   & U \\
Y340e\tablenotemark{h}    &          & 2		      &  2.84  &    5  & \nodata &   7.72  &   0.00  &  2.459  &   0.000   &   0.000   &   0.000   & $< 0.0001$ &    0.0000   &  \sc{iii} \\ 
K340		          &          & 2		      &  0.03  & 4138  &    4\tablenotemark{j}	 & \nodata & \nodata & \nodata & \nodata   &  \nodata  &   \nodata &  \nodata   &    \nodata  &  Stellar  \\  
A340		          &          & 2	 	      &  3.51  &    6  &    4\tablenotemark{j}	 &   7.72  &   0.00  &  2.459  &   0.000   &   0.000   &   0.000   & $< 0.0001$ & $< 0.0001$  &  \sc{iii} \\ 
A340e\tablenotemark{h}    &          & 2	&  3.61  &    6  &    4\tablenotemark{j}	 &   7.72  &   0.00  &  2.459  &   0.000   &   0.000   &   0.000   & $< 0.0001$ & $< 0.0001$  &  \sc{iii} \\ 
\enddata

\tablenotetext{a}{ID from \citet{gouliermis07}.}
\tablenotetext{b}{$N_{opt}$ is the approximate number of optical sources within the $1\farcs 6$ IRAC aperture radius.}
\tablenotetext{c}{$\chi_{min}^2$ (cpd) per datapoint}
\tablenotetext{d}{$n_{fits}$ is the number of fits with cpd between $cpd_{min}$ and $cpd_{min}+0.5$}
\tablenotetext{e}{$\sigma$ values represent the standard deviation of the mean.}
\tablenotetext{f}{These sources are likely YSOs based on CMD examination and physical location, but they are not considered well-fit.}
\tablenotetext{g}{These sources are fit with optical fluxes set as upper limits.}
\tablenotetext{h}{Fit using the error bar method of correcting for PAH emission.}
\tablenotetext{i}{Corresponds to one or more PMS star.}
\tablenotetext{j}{These optical masses neglect one or more optical source.}
\tablenotetext{k}{These are in environments of illuminated dust, mostly along ridges or in the tips of ``pillars."}
\tablenotetext{l}{Outside optical FOV.}

\end{deluxetable}

\clearpage

%% Here we use \plottwo to present two versions of the same figure,
%% one in black and white for print the other in RGB color
%% for online presentation. Note that the caption indicates
%% that a color version of the figure will be available online.
%%

%% This figure uses \includegraphics to scale and rotate the still frame

%% If you are not including electonic art with your submission, you may
%% mark up your captions using the \figcaption command. See the
%% No more than seven \figcaption commands are allowed per page,
%% so if you have more than seven captions, insert a \clearpage
%% after every seventh one.\\\
%% Tables should be submitted one per page, so put a \clearpage before

%% This table also includes a table comment indicating that the full
%% version will be available in machine-readable format in the electronic

%% Tables may also be prepared as separate files. See the accompanying
%% sample file table.tex for an example of an external table file.
%% To include an external file in your main document, use the \input
%% command. Uncomment the line below to include table.tex in this
%% sample file. (Note that you will need to comment out the \documentclass,
%% \begin{document}, and \end{document} commands from table.tex if you want
%% to include it in this document.)


%% The following command ends your manuscript. LaTeX will ignore any text


\begin{thebibliography}{}

\bibitem[Bernasconi \& Maeder(1996)]{bernasconi96}
Bernasconi,~P.~A., \& Maeder,~A. 1996, A\&A, 307, 829 

\bibitem[Bertelli et al.(1994)]{bertelli94}
Bertelli,~G., Bressan,~A., Chiosi,~C., Fagotto,~F., \& Nasi,~E. 1994, A\&AS, 106, 275

\bibitem[Bonanos et al.(2009)]{bonanos09}
Bonanos,~A.~Z., et al. 2009, \aj, 138, 1003

\bibitem[Bot et al.(2007)]{bot07} 
Bot, C., Boulanger, F., Rubio, M., \& Rantakyro, F. 2007, \aap, 471, 103 

\bibitem[Carlson et al.(2007)]{carlson07}
Carlson,~L.~R., et al. 2007, \apjl, 665, 109

\bibitem[Chieffi(1989)]{chieffi89}
Chieffi,~A., \& Straniero O., 1989, \apj S, 71, 47

\bibitem[Churchwell et al. (2004)]{churchwell04}
Churchwell, ~E. et al. 2004, \apjs, 154, 322

\bibitem[Cignoni et al.(2009)]{cignoni09}
Cignoni, M., et al. 2009, \aj, 137, 3668

\bibitem[Crawford(1975)]{crawford75}
Crawford, D. L. 1975, \aj 80, 955

\bibitem[Cohen, Wheaton, \& Megeath(2003)]{cohen03}
Cohen,~M., Wheaton,~Wm.~A., \& Megeath,~S.~T. 2003, \aj, 126, 1090

\bibitem[Degl'Innocenti et al.(2008)]{deglin08}
Degl'Innocenti,~S., Prada~Moroni,~P.~G., Marconi,~M. \& Ruoppo,~A. 2008, Astrophysics \& Space
Science 316, 25

\bibitem[Diolaiti et al.(2000)]{diolaiti00}
Diolaiti, E., Bendinelli, O., Bonaccini, D., Close, L., Currie, D., \& Parmeggiani, G. 2000, in ASP Conf. Ser., Vol. 216, Astronomical Data Analysis Software and Systems IX, eds. N. Manset, C. Veillet, D. Crabtree (San Francisco: ASP), 623 

\bibitem[Fagotto et al.(1994)]{fagotto94}
Fagotto, F., Bressaqn, A. Bertelli, G., \& Chiosi, C. 1994, A\&AS, 105, 29

\bibitem[Fazio et al.(2004)]{fazio04}
Fazio, G., et al. 2004, ApJS, 154, 10

\bibitem[Gordon et al.(2010)]{gordon10}
Gordon, K., et al. 2010 (in prep)

\bibitem[Gouliermis et al.(2007)]{gouliermis07}
Gouliermis, D.A., Quanz, S.P., \& Henning, T. 2007, \apj, 665, 306 

\bibitem[Hatzidimitriou et al.(2005)]{hatzidimitriou05}
Hatzidimitriou, D., Stanimirovic, S., Maragoudaki, F., Staveley-Smith, L., Dapergolas, A., \& Bratsolis, E. 2005 \mnras, 360, 1171

\bibitem[Henize(1956)]{henize56}
Henize, K. 1956, ApJS, 2, 315

\bibitem[Heyderi-Malayeri \& Selier(2010)]{h-m10}
Heyderi-Malayeri,~M. \& Selier,~R. 2010, A\&A, 517, 39

\bibitem[Hilditch et al.(2005)]{hilditch05}
Hilditch, R.W., Howarth, I.D., \& Harries, T.J. 2005, \mnras, 125, 336

\bibitem[Hillenbrand(1997)]{hillenbrand97}
Hillenbrand,~L.~A. 1997, \aj, 113,173

\bibitem[Hirota et al.(2007)]{hirota07}
Hirota,~T., et al. 2007, \pasj, 59, 857

\bibitem[Hutchings et al.(1991)]{hutchings91}
Hutchings, J. B., Cartledge, S., Pazder, J., \& Thompson, I. B.  1991, \aj, 101, 933

\bibitem[Kato et al.(2007)]{kato07}
Kato, D., et al.  2007, \pasj, 59, 615

\bibitem[Kroupa(2001)]{kroupa01}
Kroupa,~P. 2001, \mnras, 322, 231

\bibitem[Kurucz(1993)]{kurucz93}
Kurucz,~R. 1993, ATLAS9 Stellar Atmosphere Programs and 2 km s$^1$ Grid. Kurucz CD-ROMNo. 13 (Cambridge,MA: Smithsonian Astrophysical Obs.) 13

\bibitem[Lada(1987)]{lada87}
Lada, C. L. 1987, in IAU Symp. 115, Star Forrming Regions, ed. M. Piembert \& J. Jugaku (Dordrecht: Reidel), 1

\bibitem[Lada(1999)]{lada99}
Lada,~C.~J. 1999, in NATO ASIC Proc. 540, The Origin of Stars and Planetary Systems, ed. C. J. Lada, \& N. D. Kylafis, (Dordrecht: Kluwer), 143

\bibitem[Lee et al.(2005)]{lee05}
Lee,~H., Jackson,~D.~C., Skillman,~E.~D., Cannon,~J.~M., Gehrz,~R.~D., Polomski,~E., \& Woodward,~C.~E. 2005, AAS meeting, 207, 113.11

\bibitem[Meixner et al.(2006)]{meixner06}
Meixner, M., et al, 2006, \aj, 132, 2268

\bibitem[Nigra et al.(2008)]{nigra08}
Nigra, L., Gallagher, J. L., III, Smith, L. J., Stanimirovi\'c, S., Nota, A., \& Sabbi, E.  2008, PASP, 120, 972

\bibitem[Povich \& Whitney(2010)]{povich2010} 
Povich, M.~S., \& Whitney, B.~A.\ 2010, \apjl, 714, L285 

\bibitem[Reach et al. (2005)]{reach05}
Reach, W. T., et al. 2005, PASP, 117, 978

\bibitem[Rieke at al.(2004)]{rieke04}
Rieke,~G.~H., et al. 2004 \apjs, 154, 25

\bibitem[Robitaille et al. (2007)]{robitaille07}
Robitaille, T. P., Whitney, B. A., Indebetouw, R., \& Wood, K. 2007, \apjs, 169, 328

\bibitem[Robitaille et al. (2006)]{robitaille06}
Robitaille, T. P., Whitney, B. A., Indebetouw, R., Wood, K., \& Denzmore, P.  2006, \apjs, 167, 256

\bibitem[Rolleston et al.(1999)]{rolleston99}
Rolleston, W.R.J., Dufton, P.L., McErlean, N.D., \& Venn, K.A. 1999, A\&A, 348, 728

\bibitem[Sabbi et al.(2007)]{sabbi07}
Sabbi, E., et al. 2007, \aj, 133, 44

\bibitem[Salpeter(1955)]{salpeter55}
Salpeter,~E.~E. 1955, \apj, 121, 161

\bibitem[Schmalzl et al.(2008)]{schmalzl08}
Schmalzl, M., Gouliermis,~D.~A., Dolphin,~A.~E., \& Henning, T. 2008, \apj, 681, 290

\bibitem[Schuster et al.(2006)]{schuster06}
Schuster,~M.~T., Marengo, M., \& Patten,~B.~M. 2006, Proc. SPIE, 6270, 65

\bibitem[Siess et al.(2000)]{siess00}
Siess, L., Dufour, E., \& Forestini, M. 2000, A\&A, 358, 593

\bibitem[Sewilo et al. (2010)]{sewilo10}
Sewilo,~M., et al. 2010 A\&A 518, 73

\bibitem[Sirianni et al.(2002)]{sirianni02}
Sirianni, M., Nota, A., De Marchi, G., Leitherer, C., \& Clampin, M. 2002, \apj, 579, 275

\bibitem[Sirianni et al.(2005)]{sirianni05}
Sirianni, M., et al. 2005, \pasp, 117, 1049

\bibitem[Skrutskie et al.(2006)]{skrut06}
Skrutskie, M. F., et al. 2006, \aj, 131, 1163

\bibitem[Stanimirovic et al.(2000)]{stanimirovic00}
Stanimirovic, S., Staveley-Smith, L., van~der~Hulst, J.~M., Bontekoe,~T.~J.~R,
Kester, D.~J.~M., \& Jones, P.~A. 2000, \mnras, 315, 791

\bibitem[Stetson(1987)]{stetson87}
Stetson, P. B. 1987 \pasp\ 99, 191

\bibitem[Thompson, Smith, \& Hester(2002)]{thompson02}
Thompson,~R.~I., Smith,~B.~A.,\& Hester,~J.~J.\ 2002, \apj, 570, 749

\bibitem[Walborn et al.(1999)]{walborn99}
Walborn,~N.~R. et al. 1999, \aj, 117, 225

\bibitem[Walborn et al.(2002)]{walborn02}
Walborn,~N.~R., Ma$\acute{i}$z-Apell$\acute{a}$niz,~J., \& Barb$\acute{a}$,~R.~H. 2002, \aj, 124, 1601

\bibitem[Whitney et al. (2004)]{whitney04}
Whitney,~B.~A., Indebetouw,~R., Bjorkman,~J.~E., \& Wood,~K. 2004, \apj, 617, 1177

\bibitem[Whitney et al.(2008)]{whitney08}
Whitney,~B.~A., et al. 2008, \aj, 136, 18

\bibitem[Whitney et al. (2003a)]{whitney03a}
Whitney,~B.~A., Wood, K., Bjorkman,~J.~E., \& Cohen, M. 2003a, \apj, 598, 1079

\bibitem[Whitney et al. (2003b)]{whitney03b}
Whitney, B. A., Wood, K., Bjorkman, J. E., \& Wolff, M. J. 2003b, \apj, 591, 1049

\bibitem[Zaritsky et al.(2000)]{zaritsky00}
Zaritsky,~D., Harris,~J., Grebel,~E.~K., \& Thompson,~I.~B. 2000, \apj, 534, L53

\bibitem[Zaritsky \& Harris(2004)]{zaritsky04}
Zaritsky, D., \& Harris, J. 2004, \apj, 604, 167

\end{thebibliography}
\end{document}